\documentclass[12pt]{article}
\pdfoutput=1
\usepackage{amssymb}
\usepackage{amsmath}
\usepackage{latexsym}
\usepackage[usenames]{color}
\usepackage{fancybox}
\usepackage{simplewick}
\usepackage{comment}
\usepackage{cite}
\usepackage{framed}
\definecolor{shadecolor}{rgb}{0.95,0.95,0.97}
\usepackage{setspace}
\usepackage[usenames,dvipsnames]{xcolor}
\usepackage{bm}
\usepackage{longtable}
\definecolor{darkgreen}{rgb}{0,0.5,0}
\definecolor{darkblue}{cmyk}{0.9,0.9,0,0}
\definecolor{darkred}{rgb}{0.6,0,0.3}

\usepackage{graphicx}
\usepackage[whole]{bxcjkjatype}
\usepackage[unicode,setpagesize=false,pagebackref=false, linktocpage, bookmarksopen=true, colorlinks=true, linkcolor=RoyalBlue,citecolor=Maroon,urlcolor=Maroon]{hyperref}
\usepackage{hyperref}

\newcommand{\id}{{\bf 1}}
\newcommand{\tr}{{\rm tr}}

\renewcommand{\thefootnote}{\arabic{footnote}}

\def\del{\partial}

\def\fn#1{\footnote{#1}}
\def\nn{\nonumber}
\def\eqref#1{(\ref{#1})}
\def\comma{\,,}
\def\period{\,.}

\def\beq{\begin{equation}}
\def\eeq{\end{equation}}
\def\pmatrix#1#2{\left(
\begin{array}{#1}
#2\end{array}
\right)}
\def\red#1{\textcolor[rgb]{1, 0, 0}{#1}}

\newcommand{\rd}{\mathrm{d}}

\newcommand{\rK}{\mathrm{K}}

\newcommand{\rP}{\mathrm{P}}

\numberwithin{equation}{section}
\begin{document}
\thispagestyle{empty}

\renewcommand{\thefootnote}{\fnsymbol{footnote}}
\setcounter{page}{1}
\setcounter{footnote}{0}
\setcounter{figure}{0}
%%%%%%%%%%%%%%%%%%%%%%%%%%%%%%%%%%%%%%%%%%%%%%%%%%%%%%%%%%%%%%%%%%%%%%%%%%%%%%%%%%%%%%%%%%%%%%%%%%%
\begin{flushright}
CJQS-2022-001\\
CERN-TH-2021-042\\
USTC-ICTS/PCFT-21-14
\end{flushright}
\begin{center}
$$$$
{\large\textbf{\mathversion{bold}
Three-Point Functions in ABJM and Bethe Ansatz
}\par}

\vspace{1.3cm}
\textrm{Peihe Yang$^{\text{一}}$, Yunfeng Jiang$^{\text{二，三，四}}$, Shota Komatsu$^{\text{四，五}}$, Jun-Bao Wu$^{\text{一、六，七}}$\footnote{Corresponding author.}\footnote{The unusual ordering of authors instead of the standard alphabetical one in hep-th community is for students to get proper recognition of contribution under the current out-dated practice in China.}}
\\ \vspace{0.5cm}
\begin{flushleft}
\footnotesize{\textit{
\!\!$^{\text{一}}$Center for Joint Quantum Studies and Department of Physics, School of Science, Tianjin University,\\\hspace{10pt} 135 Yaguan Road, Tianjin 300350, P. R. China\\
$^{\text{二}}$School of physics, Southeast University, Nanjing 211189, P.R. China\\
$^{\text{三}}$Shing-Tung Yau Center of Southeast University, Nanjing 210096, P.R. China\\
$^{\text{四}}$Department of Theoretical Physics, CERN,
1 Esplanade des Particules, 1211 Meyrin, Switzerland\\
$^{\text{五}}$School of Natural Sciences, Institute for Advanced Study, 1 Einstein Dr. Princeton, NJ 08540, USA\\
$^{\text{六}}$Peng Huangwu Center for Fundamental Theory, Hefei, Anhui 230026, P. R. China \\
$^{\text{七}}$Center for High Energy Physics, Peking University, 5 Yiheyuan Rd, Beijing 100871, P. R. China
}\\
}
\end{flushleft}
\textrm{\footnotesize{E-mail: {\tt peihe\_yang@tju.edu.cn, Yunfeng.Jiang@cern.ch, shota.komatsu@cern.ch, junbao.wu@tju.edu.cn}}}

%\textrm{Peihe Yang$^{\text{一}}$, Yunfeng Jiang$^{\text{二,六}}$, Shota Komatsu$^{\text{二,三}}$, Jun-Bao Wu$^{\text{一,四,五}}$}
%\\ \vspace{0.5cm}
%\begin{flushleft}
%\footnotesize{\textit{
%\!\!$^{\text{一}}$Center for Joint Quantum Studies and Department of Physics, School of Science, Tianjin University,\\\hspace{10pt} 135 Yaguan Road, Tianjin 300350, P. R. China\\
%$^{\text{二}}$Department of %Theoretical Physics, CERN,
%1 Esplanade des Particules, 1211 %Meyrin, Switzerland\\
%$^{\text{三}}$School of Natural Sciences, Institute for Advanced %Study, 1 Einstein Dr. Princeton, NJ 08540, USA\\
%$^{\text{四}}$Peng Huangwu Center for Fundamental Theory, Hefei, Anhui 230026, P. R. China \\
%$^{\text{五}}$Center for High Energy Physics, Peking University, 5 Yiheyuan Rd, Beijing 100871, P. R. China\\
%$^{\text{六}}$Shing-Tung Yau Center and School of physics, Southeast University, Nanjing 210096, China
%}\\
%}
%\end{flushleft}
%\textrm{\footnotesize{E-mail: {\tt peihe\_yang在tju.edu.cn, Yunfeng.Jiang在cern.ch, shota.komatsu在cern.ch, junbao.wu在tju.edu.cn}}}

\par\vspace{1.5cm}

\textbf{Abstract}\vspace{2mm}
\end{center}
We develop an integrability-based framework to compute structure constants of two sub-determinant operators and a single-trace non-BPS operator in ABJM theory in the planar limit. In this first paper, we study them at weak coupling using a relation to an integrable spin chain. We first develop a nested Bethe ansatz for an alternating SU(4) spin chain that describes single-trace operators made out of scalar fields. We then apply it to the computation of the structure constants and show that they are given by overlaps between a Bethe eigenstate and a matrix product state. We conjecture that the determinant operator corresponds to an integrable matrix product state and present a closed-form expression for the overlap, which resembles the so-called Gaudin determinant. We also provide evidence for the integrability of general sub-determinant operators. The techniques developed in this paper can be applied to other quantities in ABJM theory including three-point functions of single-trace operators.
%\date{today}

\noindent

\setcounter{page}{1}
\renewcommand{\thefootnote}{\arabic{footnote}}
\setcounter{footnote}{0}
\setcounter{tocdepth}{2}
\newpage
\tableofcontents

\parskip 5pt plus 1pt   \jot = 1.5ex

%%%%%%%%%%%%%%%%
\newpage

	\section{Introduction}
	Over the last fifteen years, there have been breathtaking developments in solving $\mathcal{N}=4$ supersymmetric Yang-Mills theory (SYM) in four dimensions in the large $N_c$ limit. This was mainly thanks to the successful application of {\it integrability}, a method originally invented to solve special two-dimensional systems. A key finding which triggered this burst of activity was made by Minahan and Zarembo \cite{Minahan:2002ve}, who recognized that the action of the dilatation operator on a subclass of operators, called single-trace operators, can be mapped to a Hamiltonian of an integrable spin chain. Subsequently the idea was generalized and refined in various directions, and by now the integrability was applied not just to the spectrum of single-trace operators \cite{Gromov:2009tv,Bombardelli:2009ns,Arutyunov:2009ur,Gromov:2013pga,Gromov:2014caa} but also to their correlation functions \cite{Basso:2015zoa,Fleury:2016ykk,Basso:2015eqa,Basso:2017khq,Basso:2017muf,Eden:2015ija,Fleury:2017eph,Eden:2016xvg,Coronado:2018ypq,Belitsky:2019fan,Belitsky:2020qrm,Kostov:2019stn,Kostov:2019auq,Fleury:2020ykw,deLeeuw:2019qvz,Bargheer:2019kxb,Bargheer:2019exp} and even to their $1/N_c$ corrections\footnote{See also recent interesting work on non-planar anomalous dimensions \cite{McLoughlin:2020siu}.} \cite{Bargheer:2017nne,Eden:2017ozn,Ben-Israel:2018ckc,Bargheer:2018jvq}. All these progresses indicate that we are close to a complete solution to an interacting gauge theory in four dimensions; a feat never achieved in the long history of theoretical physics.
	
	More recently the integrability approach was applied to quantities that are far beyond single-trace operators: the paper \cite{Jiang:2019xdz,Jiang:2019zig} showed that the correlation function of two determinant operators and one single-trace operator can be computed exactly using the integrability machinery. This came as a surprise since the quantum numbers of determinant operators scale with $N_c$, and even performing the perturbative computation is a nontrivial task. Nevertheless, by a judicious rewriting one can map these observables to overlaps of states in an integrable two-dimensional system (a spin chain at weak coupling and a string worldsheet at strong coupling), and compute them exactly \cite{Jiang:2019xdz,Jiang:2019zig}.
	
	In this series of papers \cite{secondpaper,thirdpaper}, we generalize the analysis of \cite{Jiang:2019xdz,Jiang:2019zig} to $\mathcal{N}=6$ supersymmetric Chern-Simons-matter theory in three dimensions, constructed by Aharony, Bergman, Jafferis and Maldacena \cite{Aharony:2008ug}; {\it ABJM theory} for short. Soon after the construction of ABJM theory, it was realized in \cite{Minahan:2008hf,Gaiotto:2008cg,Nishioka:2008gz,Bak:2008cp} that ABJM theory is also integrable in the large $N_c$ limit as far as the spectrum of single-trace operators\fn{The full solution to the spectral problem was obtained by Quantum Spectral Curve \cite{Cavaglia:2014exa,Bombardelli:2017vhk}.} is concerned. However, unlike $\mathcal{N}=4$ SYM, not much is known beyond the spectrum. In particular, we are still lacking an integrability-based framework to compute correlation functions at finite 't Hooft coupling. The goal of this series of papers is to propose and establish the first of such frameworks. In this first paper, we focus on the computation at weak coupling and show that the tree-level structure constants of two determinant operators and one single-trace operator can be computed by overlaps between a matrix product state and a Bethe eigenstate in an integrable spin chain. Furthermore, we present evidence that the relevant matrix product state preserves integrability and conjecture a closed-form expression for the overlaps in terms of Gaudin-like determinants. Similar expressions were found in various different contexts, ranging from the study of quench dynamics \cite{Brockmann,Pozsgay:2014,Foda:2015nfk,Piroli:2017sei,Pozsgay:2018,deLeeuw:2019ebw,Pozsgay:2019,Jiang:2020sdw,Chen:2020xel} to the defect one-point functions in $\mathcal{N}=4$ SYM \cite{deLeeuw:2015hxa,Komatsu:2020sup,Buhl-Mortensen:2015gfd,deLeeuw:2016umh,Buhl-Mortensen:2017ind,deLeeuw:2018mkd,Gombor:2020kgu,Gombor:2020auk,Kristjansen:2020mhn,Kristjansen:2020vbe}.
	
	We should also mention an important difference from $\mathcal{N}=4$ SYM. In $\mathcal{N}=4$ SYM, the result in \cite{Chen:2019gsb} suggests that only the determinant operator corresponds to an integrable boundary state and all the sub-determinant operators are not integrable\fn{We should however note that the (non-)integrability of the sub-determinant operators is not fully settled even for $\mathcal{N}=4$ SYM. For instance, there are results on the spectrum of open string attached to the sub-determinant operators which suggest the sub-determinant operators might actually correspond to integrable boundary states \cite{Berenstein:2006qk,Ciavarella:2010tp}. It would be interesting to revisit this question in view of recent discovery of a new integrable boundary condition in $AdS_5\times S^{5}$ \cite{Linardopoulos:2021rfq}, which was missed in the classification in \cite{Dekel:2011ja}.}. By contrast, the analysis of this paper suggests that the sub-determinant operators are also integrable in ABJM theory, at least at tree level.  The possibility of having such a family of integrable boundary states motivates further study of these operators in ABJM theory.
	
	Before describing the contents of the paper, let us explain a couple of more motivations. Two rather obvious reasons were already mentioned; 1.~we can test the formalism developed in \cite{Jiang:2019xdz,Jiang:2019zig} in other theories, and 2.~we establish the first integrability-based framework to compute correlation functions in ABJM theory.
	Yet another motivation comes from the fact that the structure constants in ABJM theory receive quantum corrections even when all the operators are BPS\cite{Young:2014lka,Young:2014sia,Bianchi:2020cfn}. For the purpose of checking the AdS/CFT correspondence, this is more like a curse than a blessing since it inhibits a direct comparison between the results at weak coupling and the results in supergravity. On the other hand, this feature makes ABJM theory an ideal testing ground for the integrability approach to the correlation functions: In the integrability description, the BPS operator corresponds to the vacuum state of the spin chain and provides the simplest setup for the computation. Unfortunately, this setup was ``too simple'' for $\mathcal{N}=4$ SYM since the quantum corrections all vanish \cite{Baggio:2012rr}. By contrast, in ABJM theory the setup provides a simple yet nontrivial test of the formalism because of the dependence on the coupling constant. In addition, structure constants of BPS single-trace operators in planar ABJM theory can be computed by supersymmetric localization \cite{Dedushenko:2016jxl,Dedushenko:2017avn,Dedushenko:2018icp,Mezei:2017kmw,Gaiotto:2020vqj,Chester:2018aca,Binder:2019mpb,Chester:2020jay,KW:2021}. At present, localization computation has not been generalized to sub-determinant operators, but if we succeed in doing so, we will be able to compare two rather different approaches and deepen our understanding on the relation between them.

	The rest of the paper is organized as follows. In section \ref{sec:setup}, we explain the setup to be discussed in this paper: In ABJM theory, even the three-point functions of BPS operators can have several $R$-symmetry tensor structures and therefore depend on a multitude of structure constants. In order to simplify the analysis, we focus on the {\it twisted-translated frame}. We show that the structure constant computed in the twisted-translated frame is a particular linear combination of the full structure constants\footnote{In principle, we can recover the full structure constants by acting the $R$-symmetry transformation to each operator, which in the spin-chain language corresponds to adding zero-momentum magnons.}.  In section \ref{sec:CBA}, we first review the basic properties of the $SU(4)$ invariant alternating spin chain, such as the Hamiltonian, the Bethe equations and the relation to the single-trace operator. We then develop the nested coordinate Bethe ansatz and write down wave functions of Bethe states. In section \ref{sec:MPS}, we derive a matrix product state representation of the structure constant of two giant gravitons and a single-trace BPS operator. Such analyses were performed in $\mathcal{N}=4$ SYM in \cite{Jiang:2019xdz,Jiang:2019zig,Chen:2019gsb} and in ABJ(M) theory in \cite{Chen:2019kgc}. The novelty of our analysis is that we derive an explicit expression for the matrix product state for the sub-determinant operators in the twisted-translated frame which can be readily used to evaluate the structure constants. We also evaluate the tree-level structure constants of two non-maximal sub-determinant operators and a single-trace BPS operator. In section \ref{sec:result}, we compute the structure constants using the results in the preceding sections and present our conjecture. As was the case with $\mathcal{N}=4$ SYM, the result for determinant operators exhibits a nontrivial selection rule on the rapidities and is given by a ratio of Gaudin-like determinants. These imply that the determinant operators in ABJM theory correspond to integrable boundary states. One notable difference from $\mathcal{N}=4$ SYM is that these features persist also for sub-determinant operators, indicating that ABJM theory admits a family of integrable boundary states. Our results also provide the first example of integrable matrix product states in the alternating spin chain. Finally in section \ref{sec:conclusion}, we conclude and discuss future directions. Several appendices are included to explain technical details.
	
	\section{Setup and Kinematics\label{sec:setup}}
	\subsection{Generalities}
	\paragraph{Basics.} ABJM theory is a $\mathcal{N}=6$ superconformal Chern-Simons matter theory in three dimensions with a product gauge group ${\rm U}(N)_{k}\times {\rm U}(N)_{-k}$ where $k$ is the Chern-Simons level. It consists of two sets of gauge fields and matter fields in the bi-fundamental representations. See  \cite{Aharony:2008ug,Benna:2008zy} for the explicit form of the Lagrangian. See also \cite{Klose:2010ki} for a review of the integrability properties of ABJM theory.
	
	A distinguishing feature of ABJM theory (as compared to $\mathcal{N}=4$ SYM) is that it admits two different large $N$ limits: The first limit is called the {\it M-theory limit} and can be defined by $N\to \infty$ with $k$ fixed. This limit has attracted much attention since it is dual to a M-theory on $AdS_4\times S^{7}/Z_{k}$ and provides one of the most concrete non-perturbative definitions of the M-theory currently available. Unfortunately this limit is difficult to study on the field theory side since it corresponds to a strong coupling limit (unless we focus on the BPS observables and use supersymmetric localization). We therefore study the second limit in this paper, namely the {\it planar limit}. The planar limit is defined by sending $N\to \infty$ while keeping the 't Hooft coupling
	\beq
	\lambda\equiv \frac{N}{k}\comma
	\eeq
	fixed. As is the case with the standard planar limit of Yang-Mills theories, the observables in the planar limit can be computed by the planar diagrams--- diagrams that can be drawn on a genus 0 Riemann surface. The holographic dual of this limit is given by type-IIA superstring theory in $AdS_4\times CP^{3}$.

	The main subject of this paper involves gauge invariant local operators in ABJM made out of scalar fields. ABJM theory has two sets of scalar fields; the one that transforms as $(\square, \overline{\square}; {\bf 4})$ and the other that transforms as $(\overline{\square}, \square; \bar{{\bf 4}})$ under the ${\rm U}(N)_k\times {\rm U}(N)_{-k}$ gauge groups and $SU(4)$ $R$-symmetry,
	\beq
	\begin{aligned}
	Y^{I}&:\quad (\square, \overline{\square};{\bf 4})\comma\quad \Delta^{0}=1/2\comma\\
	\bar{Y}_{I}&:\quad (\overline{\square}, \square;\bar{\bf 4})\comma\quad \Delta^{0}=1/2\comma
	\end{aligned}
	\eeq
	where $I=1,\ldots, 4$ and $\Delta^{0}$ is the mass dimension.
	The simplest gauge invariant operators constructed out of such fields are {\it single-trace operators}, which in general take the following form:
	\begin{align}\label{eq:generalsingletrace}
	    {\rm tr}\left(Y^{I_1}\bar{Y}_{J_1}\cdots Y^{I_L}\bar{Y}_{J_L}\right)+\cdots\period
	\end{align}
	As is clear from this expression, it consists of an alternating sequence of $Y^{I}$ fields and $\bar{Y}_{I}$ fields. Because of this feature, the spin chain that describes the anomalous dimension of such an operator is an {\it alternating} spin chain, in which spins living on odd sites are distinct from spins living on even sites. We will give a more detailed description of the spin chain and its Bethe ansatz in section \ref{sec:CBA}.
	\paragraph{BPS operators and giant gravitons.} General single-trace operators \eqref{eq:generalsingletrace} do not preserve any supersymmetry. However, for a special choice of the $R$-symmetry indices, they become BPS and invariant under $1/3$ of the supersymmetry transformations. Written explicitly they take the following form,
	\begin{align}
	    \mathcal{O}^{\circ}_{L}(x;n,\bar{n})\equiv {\rm tr}\left[\left((n\cdot Y)(\bar{n}\cdot \bar{Y})\right)^{L}\right]\comma
	\end{align}
	where the supercript $\circ$ is to indicate that the operator is BPS\footnote{Here we are following the notation used in \cite{Escobedo:2010xs}.} and we have
	\begin{align}
	    n\cdot Y\equiv\sum_{I=1}^{4}n_{I}Y^{I}\comma\qquad \bar{n}\cdot \bar{Y}\equiv \sum_{I=1}^{4}\bar{n}^{I}\bar{Y}_{I}\period
	\end{align}
	Here $n$ and $\bar{n}$ are four-component vectors specifying the $R$-symmetry polarizations of the operator and they need to satisfy
	\begin{align}
	    n\cdot \bar{n}=0\comma
	\end{align}
	in order for the operator to be BPS.
	
	Another class of BPS operators considered in this paper are {\it giant gravitons} \cite{Dey:2011ea,Chakrabortty:2011fd,Giovannoni:2011pn,Hirano:2012vz}. They are defined in terms of sub-determinants as
	\begin{align}
	    \mathcal{D}_M(x;n,\bar{n})\equiv \frac{1}{M!}\delta^{[b_1\cdots b_M]}_{[a_1\cdots a_M]}\left[(n\cdot Y)(\bar{n}\cdot \bar{Y})\right]^{a_1}_{b_1}\cdots \left[(n\cdot Y)(\bar{n}\cdot \bar{Y})\right]^{a_M}_{b_M}\comma
	\end{align}
	with $n\cdot\bar{n}=0$ and
	\begin{align}
	    \delta^{[b_1\cdots b_M]}_{[a_1\cdots a_M]}\equiv \sum_{\sigma\in S_M}(-1)^{|\sigma|}\delta^{b_1}_{a_{\sigma_1}}\cdots \delta^{b_M}_{b_{\sigma_M}}\period
	\end{align}
	The operator with a maximal $R$-charge ($M=N$) is called the {\it maximal giant gravitons} while others ($M< N$) are called {\it non-maximal giant gravitons}.
	
	Alternatively, they can be defined in terms of the antisymmetric Schur polynomial \cite{Corley:2001zk,Dey:2011ea,Chakrabortty:2011fd,Caputa:2012dg} as follows:
	\begin{align}\label{eq:defDM}
	    \mathcal{D}_M(x;n,\bar{n})=\frac{1}{M!}\sum_{\sigma\in S_M}\chi_{{\sf A}_M}(\sigma)\left[(n\cdot Y)(\bar{n}\cdot \bar{Y})\right]^{a_{\sigma_1}}_{a_1}\cdots \left[(n\cdot Y)(\bar{n}\cdot \bar{Y})\right]^{a_{\sigma_M}}_{a_M}\comma
	\end{align}
	where $\chi_{{\sf A}_M}$ is a Schur polynomial for the totally antisymmetric representation of size $M$. See e.g.~\cite{Chen:2019kgc} for explicit definitions. To study the correlation functions of giant gravitons, it is often convenient to consider a generating function $\mathcal{G}(x;n,\bar{n},t)$ defined by\fn{The second equality in \eqref{eq:defgenerating} follows from the invariance of the determinant under the addition and the subtraction of rows or columns,
	\begin{align}\nn
	    \det \pmatrix{cc}{{\bf 1}&-t (\bar{n}\cdot \bar{Y})\\ t (n\cdot Y)&{\bf 1}}=\det \pmatrix{cc}{{\bf 1}&{\bf 0}\\t(n\cdot Y)&{\bf 1}+t^2(n\cdot Y)(\bar{n}\cdot \bar{Y})}=\det\left[{\bf 1}+t^2(n\cdot Y)(\bar{n}\cdot \bar{Y})\right]\period
	\end{align}
	}
	\begin{align}\label{eq:defgenerating}
	    \mathcal{G} (x;n,\bar{n},t)\equiv \det \pmatrix{cc}{{\bf 1}&-t (\bar{n}\cdot \bar{Y})\\ t (n\cdot Y)&{\bf 1}}=\det \left[{\bf 1}+t^2(n\cdot Y)(\bar{n}\cdot \bar{Y})\right]\period
	\end{align}
	To extract the giant gravitons with a fixed charges, we perform the integral of $t$;
	\begin{align}
	    \mathcal{D}_M(x;n,\bar{n})=\oint \frac{\rd t}{2\pi i t^{1+2M}}\mathcal{G} (x;n,\bar{n},t)\period
	\end{align}
	\paragraph{Holographic dual.} Let us also briefly review the dual description of giant gravitons although it is not directly relevant for the analysis performed in this paper. There are two classes of giant gravitons in $AdS_4\times CP^3$ known in the literature which are conjectured to be dual to $1/3$ BPS operators\footnote{Some giant gravitions with more or less supersymmetries were also studied in the papers listed below in this paragraph.}.  The first class is the D2 branes extended in $S^2$ inside $AdS_4$ \cite{Berenstein:2008dc,Nishioka:2008ib,Hamilton:2009iv}. These branes are known to be dual to {\it symmetric} Schur polynomials and are analogs of the dual giant gravitons in $\mathcal{N}=4$ SYM. The other class is the D4 branes which are point-like in $AdS_4$ and extended in the $CP^3$ direction \cite{Giovannoni:2011pn,Berenstein:2008dc,Murugan:2011zd,Gutierrez:2010bb,Lozano:2011dd,Herrero:2011bk,Lozano:2013ota}. They are dual to antisymmetric Schur polynomials and are the subject of this paper. In the upcoming paper \cite{secondpaper}, we will study the correlation functions from these holographic perspectives.
	
	\subsection{Structures of two- and three-point functions}\label{subsec:structure}
	We now summarize the structures of two- and three-point functions and the constraints from symmetry, emphasizing the differences from $\mathcal{N}=4$ SYM.
	\paragraph{BPS two-point functions.} Let us first consider the two-point functions of $1/3$ BPS single-trace operators, $\langle \mathcal{O}_{L_1}^{\circ}\mathcal{O}_{L_2}^{\circ}\rangle$.
	Since $n_k$'s transform as the anti-fundamentals of $SU(4)$ while $\bar{n}_k$'s transform as the fundamentals of $SU(4)$, the $SU(4)$ symmetry determines the structure of the two-point function to be
	\begin{align}\label{eq:BPS2ptsingle}
	    \langle \mathcal{O}_{L_1}^{\circ}\mathcal{O}_{L_2}^{\circ}\rangle=\delta_{L_1,L_2}\mathcal{N}_{\mathcal{O}_{L_1}^{\circ}}(d_{12}d_{21})^{L_1}\comma
	\end{align}
	where $\mathcal{N}_{\mathcal{O}_{L}^{\circ}}$ is the normalization constant, which at weak coupling reads
	\begin{align}
	    \mathcal{N}_{\mathcal{O}_{L}^{\circ}}=L\lambda^{2L}\comma
	\end{align}
	while $d_{ij}$'s are defined by
	\begin{align}
	    d_{ij}\equiv \frac{n_i\cdot \bar{n}_j}{|x_{ij}|}\qquad \qquad |x_{ij}|\equiv |x_i-x_j|\period
	\end{align}
	Note that, unlike $\mathcal{N}=4$ SYM, $d_{ij}$ here is not symmetric under the exchange of indices: $d_{ij}\neq d_{ji}$.
	
	The symmetry is powerful enough to determine the two-point functions of (sub-)determinant operators as well:
	\begin{align}\label{eq:BPS2ptdet}
	    \langle \mathcal{D}_{M}(x_1,n_1,\bar{n}_1)\mathcal{D}_{M}(x_2,n_2,\bar{n}_2)\rangle=\mathcal{N}_{\mathcal{D}_M}(d_{12}d_{21})^{M}\period
	\end{align}
	Here again $\mathcal{N}_{\mathcal{D}_M}$ is the normalization constant.
	\paragraph{BPS three-point functions.} We then discuss the three-point functions of $1/3$-BPS operators. This is where ABJM theory shows significant differences from $\mathcal{N}=4$ SYM:
	\begin{itemize}
	    \item In $\mathcal{N}=4$ SYM, the $R$-symmetry structure of the BPS three-point functions is determined completely by the symmetry. On the other hand, the BPS three-point function in ABJM theory admits several different structures.
	    \item The structure constant of $1/2$-BPS operators in $\mathcal{N}=4$ SYM does not depend on the 't Hooft coupling. By contrast, the structure constant of $1/3$-BPS operators in ABJM theory is a nontrivial function of the 't Hooft coupling.
	\end{itemize}

	To see this explicitly, let us consider the three-point function of BPS single-trace operators. Imposing the $SU(4)$ symmetry, one finds that the three-point function has a multitude of allowed structures labelled by $p$ which is either an integer or a half-integer\fn{$p$ takes an integer-value when $\sum_k L_{k}$ is even while it takes a half-integer value when $\sum_{k}L_k$ is odd.} \cite{Liendo:2015cgi},
	\begin{align}\label{eq:structureBPS3pt}
	     \frac{\langle\mathcal{O}_{L_1}^{\circ}\mathcal{O}_{L_2}^{\circ}\mathcal{O}_{L_3}^{\circ}\rangle}{\sqrt{\mathcal{N}_{\mathcal{O}_{L_1}^{\circ}}\mathcal{N}_{\mathcal{O}_{L_2}^{\circ}}\mathcal{N}_{\mathcal{O}_{L_3}^{\circ}}}}=(d_{12}d_{21})^{L_{12|3}}(d_{23}d_{32})^{L_{23|1}}(d_{31}d_{13})^{L_{31|2}}\sum_{p=-\ell_{123}}^{\ell_{123}} C^{(p)}_{123}\,\,\xi^{p}
	    \comma
	\end{align}
	with
	\begin{align}
	    L_{ij|k}\equiv \frac{L_i+L_j-L_k}{2}\comma\qquad \ell_{123}\equiv {\rm min}\left[L_{12|3},L_{23|1},L_{31|2}\right]\period
	\end{align}
	Here $\xi$ is the {\it SU(4) cross ratio}
	\begin{align}
	    \xi\equiv \frac{d_{12}d_{23}d_{31}}{d_{21}d_{32}d_{13}}=\frac{(n_1\cdot \bar{n}_2)(n_2\cdot \bar{n}_3)(n_3\cdot \bar{n}_1)}{(n_2\cdot \bar{n}_1)(n_3\cdot \bar{n}_2)(n_1\cdot \bar{n}_3)}\comma
	\end{align}
	 and $C_{123}^{(p)}$'s are a set of structure constants. These structure constants can be readily computed at tree level by Wick contractions \cite{Hirano:2012vz,Young:2014lka}:
	 \begin{align}\label{eq:treeC123BPS}
	     C_{123}^{(p)}\overset{\lambda\to 0}{=}\frac{\sqrt{L_1L_2L_3}}{N}\times \begin{cases}2\delta_{p,0}
	     \qquad &\sum_{k=1}^{3}L_k:\text{ even}\\
	     \delta_{p,\frac{1}{2}}+\delta_{p,-\frac{1}{2}}\qquad &\sum_{k=1}^{3}L_k:\text{ odd}\end{cases}\period
	 \end{align}
	 On the other hand, \cite{Bastianelli:1999en,Hirano:2012vz} performed the computation at strong coupling and the result reads
	 \begin{align}
	 \begin{aligned}\label{eq:strongC123BPS}
	     C_{123}^{(p)}\,\overset{\lambda\to\infty}{=}\,&\frac{1}{N}\sqrt{\frac{h(\lambda)}{\pi}}\frac{\prod_{j=1}^{3}\sqrt{1+2L_j}\,\Gamma[1+L_j] }{\Gamma[1+\tfrac{L_1+L_2+L_3}{2}]}\prod_{\{i,j,k\}}\frac{\Gamma[1+\tfrac{L_{ij|k}}{2}]}{\Gamma [1-p+\tfrac{L_{ij|k}}{2}]\Gamma [1+p+\tfrac{L_{ij|k}}{2}]}\comma
	 \end{aligned}
	 \end{align}
	 where $\prod_{\{i,j,k\}}$ denotes a product over cyclic permutations of $\{1,2,3\}$ and $h(\lambda)$ is the so-called {\it interpolating function} \cite{Gromov:2014eha} which can be expanded at weak and strong couplings as
	 \begin{align}
	     h(\lambda)=\begin{cases}\lambda-\frac{\pi^2\lambda^3}{3}+\cdots &\lambda\to 0\\\sqrt{\frac{1}{2}\left(\lambda-\frac{1}{24}\right)}-\frac{\log 2}{\pi}+\cdots &\lambda\to\infty\period\end{cases}
	 \end{align}
	 The difference between \eqref{eq:treeC123BPS} and \eqref{eq:strongC123BPS} provides clear evidence for the coupling dependence of structure constants $C_{123}^{(p)}$.
	
	 The expression \eqref{eq:structureBPS3pt} can be readily generalized to the three-point function of two sub-determinant operators and one single-trace BPS operator, which is the main subject of this paper. The resulting expression is
	 \begin{align}\label{eq:structureBPSdet3pt}
	 \begin{aligned}
	     &\frac{\langle \mathcal{D}_M (x_1;n_1,\bar{n}_1)\mathcal{D}_M (x_2;n_2,\bar{n}_2)\mathcal{O}^{\circ}_L (x_3;n_3,\bar{n}_3)\rangle}{\mathcal{N}_{\mathcal{D}_M}\sqrt{\mathcal{N}_{\mathcal{O}_{L}^{\circ}}}}=\\
	     &\hspace{10pt}(d_{12}d_{21})^{M}\left(\frac{d_{23}d_{32}d_{31}d_{13}}{d_{12}d_{21}}\right)^{\frac{L}{2}}\sum_{p=-\frac{L}{2}}^{\frac{L}{2}}D^{(p)}_{M|L} \,\xi^{p}\comma
	     \end{aligned}
	 \end{align}
	 where $D_{M|L}^{(p)}$'s are structure constants. See \eqref{eq:BPSstructureintegral} for the tree-level result for $D_{M|L}^{(p)}$.
\paragraph{Non-BPS operators.} We now make a few comments on non-BPS operators. In the spin-chain approach \cite{Minahan:2008hf}, we start from the vacuum state, which corresponds to the $1/3$-BPS operator, and introduce excitations (magnons) in order to describe non-BPS operators. The operators constructed in this way depend on two sets of data;
\begin{enumerate}
    \item The R-symmetry polarizations ($n$ and $\tilde{n}$) of the vacuum state ${\rm tr}\left[\left((n\cdot Y)(\bar{n}\cdot \bar{Y})\right)^{L}\right]$.
    \item A set of rapidities of magnons ${\bf u}$.
\end{enumerate}
In this paper, we define the normalization of non-BPS operators $\mathcal{N}_{\mathcal{O}}$ in terms of the two-point function with a canonical choice of the R-symmetry polarizations $n_0\equiv (1,0,0,0)$ and $\bar{n}_0\equiv (0,0,0,1)$:
\begin{align}\label{eq:conformal2ptnon}
    \left< \mathcal{O}(x_1;n_0,\bar{n}_0)\left[\mathcal{O} (x_2;n_0,\bar{n}_0)\right]^{\dagger}\right>=\frac{\mathcal{N}_{\mathcal{O}}}{x_{12}^{2\Delta_{\mathcal{O}}}}\period
\end{align}
Here $\mathcal{O}^{\dagger}$ is a Hermitian-conjugate of the operator $\mathcal{O}$ and $\Delta_{\mathcal{O}}$ is the conformal dimension.

We then define the normalized three-point function by
\begin{align}\label{eq:conformal3ptnon}
    \frac{\langle \mathcal{O}_1 (x_1;n_1,\bar{n}_1)\mathcal{O}_2 (x_2;n_2,\bar{n}_2)\mathcal{O}_3 (x_3;n_3,\bar{n}_3)\rangle}{\sqrt{\mathcal{N}_{\mathcal{O}_1}\mathcal{N}_{\mathcal{O}_2}\mathcal{N}_{\mathcal{O}_3}}}=\frac{F_{123}}{x_{12}^{\Delta_{\mathcal{O}_1}+\Delta_{\mathcal{O}_2}-\Delta_{\mathcal{O}_3}}x_{23}^{\Delta_{\mathcal{O}_2}+\Delta_{\mathcal{O}_3}-\Delta_{\mathcal{O}_1}}x_{31}^{\Delta_{\mathcal{O}_3}+\Delta_{\mathcal{O}_1}-\Delta_{\mathcal{O}_2}}}\comma
\end{align}
where $F_{123}$ is a sum of all possible allowed $R$-symmetry and Lorentz invariants times the corresponding structure constants. Non-BPS operators typically have more quantum numbers than the BPS operators. Therefore the number of allowed structures for the non-BPS operators is larger than the one for the BPS operators. In addition, the structures highly depend on the Lorentz and the R-symmetry representations of the operator. These features make it difficult to write down a simple universal expression like \eqref{eq:structureBPS3pt}. In what follows, we instead focus on special kinematic configurations, called the {\it twisted-translated frame}.
\paragraph{Phase ambiguity of structure constants.} Note that the normalization of non-BPS operators as defined in \eqref{eq:conformal2ptnon} does not fix the overall phase of the operator $\mathcal{O}$ since a multiplication of a phase to $\mathcal{O}$ ($\mathcal{O}\to e^{i\phi}\mathcal{O}$) does not change $\mathcal{N}_{\mathcal{O}}$. On the other hand, the three-point functions---and therefore structure constants---do change by such a phase multiplication. This means that the overall phase of the structure constants will not be fixed by our analysis\fn{This phase ambiguity was pointed out already in \cite{Escobedo:2010xs}.} and one needs to impose further conditions in order to determine it.  In section \ref{sec:result}, we will show that there is one choice of a phase with which the result from integrability takes a simple form. However, this is more like an answer analysis and we do not have a physical explanation on why it simplifies the final expression. It would be important to come up with a field-theory argument on why this choice of a phase is preferred.
\subsection{Twisted translation}
\label{sec:2.3}
The twisted-translated frame was introduced originally in $\mathcal{N}=4$ SYM in four dimensions in \cite{Drukker:2009sf} and used also in the integrability analysis \cite{Basso:2015zoa}. A key feature of this frame is that it makes the correlation functions of $1/2$ BPS operators position-independent. A similar topological sector in a large class of $\mathcal{N}\geq 4$ SCFTs in three dimensions was studied by localization in \cite{Mezei:2017kmw,Dedushenko:2016jxl,Dedushenko:2017avn,Dedushenko:2018icp}. For the case of ABJM theory, the full localization results are not yet available\fn{Two important exceptions are ABJM theory for $k=1$ \cite{Mezei:2017kmw,Gaiotto:2020vqj} and the correlation function of an operator corresponding to the mass deformation\cite{Chester:2018aca,Binder:2019mpb,Chester:2020jay}.}, but general symmetry properties of the twisted sector were discussed in \cite{Liendo:2015cgi} and the perturbative computation was performed in \cite{Gorini:2020new}. Much like in $\mathcal{N}=4$ SYM, the twisted-translated frame in ABJM theory provides a useful setup for analyzing the non-BPS three-point functions as well. In fact, it was also discussed in an unpublished work \cite{OSPS:2016,Pereira:2017unx} which attempted to construct the hexagon formalism for ABJM theory. As we see below, the structure constant computed in this frame is a particular linear combination of the full structure constants. In principle, we can recover the full structure constants by acting the R-symmetry transformation to each operator, which in the spin-chain language corresponds to adding zero-momentum magnons.

\paragraph{BPS correlation functions.} For the correlation functions of BPS operators, the twisted-translated frame is defined by placing all the operators along a single line and aligning the R-symmetry polarizations along an $U(1)$ direction inside $SU(4)$. Written explicitly, the twisted-translated BPS single-trace operator reads
\begin{align}\label{eq:twistedsingle}
    \hat{\mathcal{O}}^{\circ}_{L}(a)\equiv {\rm tr}\left[(\mathcal{Y}(a)\bar{\mathcal{Y}}(a))^{L}\right]\comma
\end{align}
where $\mathcal{Y}$ and $\bar{\mathcal{Y}}$ are given by
\begin{align}\label{eq:twisteddet}
\begin{aligned}
    \mathcal{Y}(a)&\equiv \left(Y^{1}+\kappa a Y^{4}\right)(0,a,0)\comma\\
    \bar{\mathcal{Y}}(a)&\equiv \left(\bar{Y}_{4}-\kappa a \bar{Y}_{1}\right)(0,a,0)\period
    \end{aligned}
\end{align}
Here $\kappa$ is an arbitrary parameter with mass dimension $1$. Similarly, the sub-determinant operators in the twisted-translated frame can be defined by
\begin{align}
    \hat{\mathcal{D}}_M(a)\equiv \frac{1}{M!}\delta^{[b_1\cdots b_M]}_{[a_1\cdots a_M]}(\mathcal{Y}(a)\bar{\mathcal{Y}}(a))^{a_1}_{b_1}\cdots (\mathcal{Y}(a)\bar{\mathcal{Y}}(a))^{a_M}_{b_M}\period
\end{align}

An alternative way to describe the twisted-translated frame is to use the {\it twisted translation} generator, which is a linear combination of the translation and the R-symmetry rotation,
\begin{align}
    \mathcal{T}\equiv i P_2 +\kappa R^{4}_{1}\comma
\end{align}
where $P_2$ is the translation along the $x_2$ direction while $R^{4}_{1}$ is the R-symmetry generator which rotates $Y^{1}$ and $Y^{4}$. Using $\mathcal{T}$, we can express $\mathcal{Y}$ and $\bar{\mathcal{Y}}$ as
\begin{align}
\begin{aligned}
    \mathcal{Y}(a)=e^{\mathcal{T} a}Y^{1}(0)e^{-\mathcal{T}a}\comma\qquad
    \bar{\mathcal{Y}}(a)=e^{\mathcal{T} a}\bar{Y}_{4}(0)e^{-\mathcal{T}a}\period
    \end{aligned}
\end{align}
An important property of $\mathcal{T}$ is that it is $\mathcal{Q}$ exact, i.e.~
\begin{align}
    \mathcal{T}=\{\mathcal{Q},\bullet\}\comma
\end{align}
where $\mathcal{Q}$ is a linear combination of the supersymmetry and superconformal generators which is nilpotent and annihilates the $1/3$ BPS operator made out of $\mathcal{Y}(0)$ and $\bar{\mathcal{Y}}(0)$. Because of this property, the correlation functions of twisted-translated $1/3$ BPS operators \eqref{eq:twistedsingle} and \eqref{eq:twisteddet} become independent of the positions $a$'s and define a topological subsector. See \cite{Liendo:2015cgi,Pereira:2017unx} for more detailed explanation.

This position independence can also be seen directly from the general structure of BPS two- and three-point functions \eqref{eq:BPS2ptsingle}, \eqref{eq:BPS2ptdet}, \eqref{eq:structureBPS3pt} and \eqref{eq:structureBPSdet3pt}. For this purpose, we simply need to set
\begin{align}
    n_j= (1,0,0,\kappa a_1)\comma\quad \bar{n}_j =(-\kappa a_j,0,0,1)\comma\quad x_j=(0,a_j,0)\comma
\end{align}
in those equations. This leads to $d_{ij}=\kappa \,{\rm sgn}(a_i-a_j)$, and we get
    \begin{align}
    &\langle \hat{\mathcal{O}}_{L_1}^{\circ}(a_1)\hat{\mathcal{O}}_{L_2}^{\circ}(a_2)\rangle=\delta_{L_1,L_2}\mathcal{N}_{\mathcal{O}^{\circ}_{L_1}}(-\kappa^2)^{L_1}\comma\qquad \langle \hat{\mathcal{D}}_M(a_1)\hat{\mathcal{D}}_{M}(a_2)\rangle=\mathcal{N}_{\mathcal{D}_M}(-\kappa^2)^{M}\comma\\
    &\frac{\langle\hat{\mathcal{O}}_{L_1}^{\circ}(a_1)\hat{\mathcal{O}}_{L_2}^{\circ}(a_2)\hat{\mathcal{O}}_{L_3}^{\circ}(a_3)\rangle}{\sqrt{\mathcal{N}_{\mathcal{O}_{L_1}^{\circ}}\mathcal{N}_{\mathcal{O}_{L_2}^{\circ}}\mathcal{N}_{\mathcal{O}_{L_3}^{\circ}}}}=(-\kappa^2)^{\frac{L_1+L_2+L_3}{2}} \mathfrak{C}^{\circ}_{123}\comma\\
    &\frac{\langle \hat{\mathcal{D}}_M (a_1)\hat{\mathcal{D}}_M (a_2)\hat{\mathcal{O}}_L^{\circ} (a_3)\rangle}{\mathcal{N}_{\mathcal{D}_M}\sqrt{\mathcal{N}_{\mathcal{O}_{L}^{\circ}}}}=(-\kappa^2)^{M+\frac{L}{2}}\mathfrak{D}_{M|L}^{\circ}\period
\end{align}
Here $\mathcal{N}_{\mathcal{O}_L^{\circ}}$ and $\mathcal{N}_{\mathcal{D}_M}$ are normalizations defined in \eqref{eq:BPS2ptsingle} and \eqref{eq:BPS2ptdet} while $\mathfrak{C}_{123}^{\circ}$ and $\mathfrak{D}_{M|L}^{\circ}$ are the structure constants in the twisted-translated frame, and they are given by linear combinations of the structure constants defined in \eqref{eq:structureBPS3pt} and \eqref{eq:structureBPSdet3pt}:
\begin{align}
    \begin{aligned}
    \mathfrak{C}_{123}^{\circ}&=\sum_{p=-\ell_{123}}^{\ell_{123}}(-1)^{p}C_{123}^{(p)}\comma\qquad
    \mathfrak{D}_{M|L}^{\circ}=\sum_{p=-\frac{L}{2}}^{\frac{L}{2}}(-1)^{p} D_{M|L}^{(p)}\period
    \end{aligned}
\end{align}
\paragraph{Non-BPS operators.} To define the twisted-translated frame for non-BPS operators, we first construct a non-BPS operator by adding magnons (denoted in red below) on top of ${\rm tr}\left[(Y^{1}\bar{Y}_4)^{L}\right]$ at the origin:
\begin{align}
    \mathcal{O}(0)\equiv {\rm tr}\left[\cdots (Y^{1}\bar{Y}_4) (\red{Y^{2}}\bar{Y}_{4})(Y^{1}\red{\bar{Y}_3})(Y^{1}\bar{Y}_4)\cdots\right]+\cdots|_{x^{\mu}=0}\period
\end{align}
As we explain in more detail in the next section, these operators are in one-to-one correspondence with states in the alternating $SU(4)$ invariant spin chain,
\begin{align}
    {\rm tr}\left[\cdots (Y^{1}\bar{Y}_4) (\red{Y^{2}}\bar{Y}_{4})(Y^{1}\red{\bar{Y}_3})(Y^{1}\bar{Y}_4)\cdots\right] \quad \leftrightarrow \quad |1\,\bar{4}\,\red{2}\,\bar{4}\,1\,\red{\bar{3}}\,1\,\bar{4}\cdots \rangle\period
\end{align}
We then act the twisted translation $e^{\mathcal{T}a}$ to obtain an operator at a shifted position:
\begin{align}
\begin{aligned}
     \hat{\mathcal{O}}(a)\equiv e^{\mathcal{T}a}\mathcal{O}(0)e^{-\mathcal{T}a}\period
    \end{aligned}
\end{align}
Note that the twisted translation leaves $Y^{2,3}$ and $\bar{Y}_{2,3}$ invariant:
\begin{align}
    e^{\mathcal{T}a}\,Y^{2,3}\,e^{-\mathcal{T}a}=Y^{2,3}\comma\qquad e^{\mathcal{T}a}\,\bar{Y}_{2,3}\,e^{-\mathcal{T}a}=\bar{Y}_{2,3}\period
\end{align}
As is clear from the definition above, the twisted-translation only involves an $SU(2)$ subgroup of the full $SU(4)$ R-symmetry. As demonstrated  e.g.~in Appendix A of \cite{Kazama:2014sxa}, the $SU(2)$ Ward identity allows us to express the kinematic dependence of the three-point function in terms of the conformal dimension $\Delta$ and the $U(1)$ R-charge $J$, which assigns $+1/2$ charge to $Y^1$ and $\bar{Y}_4$ and $-1/2$ charge to $Y^4$ and $\bar{Y}_{1}$.
The result reads
\begin{align}
    \frac{\langle \hat{\mathcal{O}}_{1}(a_1)\hat{\mathcal{O}}_{2}(a_2)\hat{\mathcal{O}}_{3}(a_3)\rangle}{\sqrt{\mathcal{N}_{\mathcal{O}_1}\mathcal{N}_{\mathcal{O}_2}\mathcal{N}_{\mathcal{O}_3}}}=\frac{(-\kappa^2)^{\frac{J_1+J_2+J_3}{2}}\mathfrak{C}_{123}}{(a_1-a_2)^{\gamma_{12|3}}(a_2-a_3)^{\gamma_{23|1}}(a_3-a_1)^{\gamma_{31|2}}}\comma
\end{align}
with
\begin{align}
    \gamma_{ij|k}\equiv (\Delta_i-J_i)+(\Delta_j-J_j)-(\Delta_k-J_k)\period
\end{align}
Here $\mathcal{N}_{\mathcal{O}}$ is the normalization defined in \eqref{eq:conformal2ptnon}, and $\mathfrak{C}_{123}$ is the structure constant in the twisted-translated frame, which is $F_{123}$ in \eqref{eq:conformal3ptnon} specialized to the twisted-translated kinematics.

These expressions can be generalized to the correlation function of two sub-determinant operators and a single-trace non-BPS operator. The result reads
\begin{align}\label{eq:setupfinal}
    \frac{\langle \hat{\mathcal{D}}_M(a_1)\hat{\mathcal{D}}_M(a_2)\hat{\mathcal{O}}(a_3)\rangle}{\mathcal{N}_{\mathcal{D}_M}\sqrt{\mathcal{N}_{\mathcal{O}}}}=(-\kappa^2)^{M+\frac{J}{2}}\left(\frac{(a_1-a_2)}{(a_2-a_3)(a_3-a_1)}\right)^{\Delta-J}\mathfrak{D}_{M|\mathcal{O}}\comma
\end{align}
with $\mathfrak{D}_{M|\mathcal{O}}$ being the structure constant. The main goal of this paper is to compute $\mathfrak{D}_{M|\mathcal{O}}$ at tree level using the spin-chain description.
\section{SU(4) Invariant Alternating Spin Chain}
\label{sec:CBA}
In this section, we explain the spin-chain description of single-trace operators made out of scalar fields. After reviewing the basic facts of the $SU(4)$ invariant alternating spin chain and its relation to single-trace operators, we discuss its coordinate Bethe ansatz. In particular, we present explicit expressions for the coordinate Bethe ansatz wave functions, which we will use later to evaluate the structure constants \eqref{eq:setupfinal}. An alternative way to construct the wave function is to use the algebraic Bethe ansatz which we review in Appendix \ref{sec:NABA}.
\subsection{Hamiltonian and Bethe equations}
%%%%%%%%%%%%%%%%%%%%%%%%%%%%%%%%%%%%%%%%%%%%%%%%%%%%%
The dilatation operator in the scalar sector of ABJM theory at two-loop order is described by the $SU(4)$ invariant alternating spin chain \cite{Minahan:2008hf,Bak:2008cp}. The Hamiotnonian of the spin chain is given by
\begin{align}
\label{eq:Ham}
\mathbb{H}=\frac{\lambda^2}{2}\sum_{l=1}^{2L}\left(2-2\rP_{l,l+2}+\rP_{l,l+2}\rK_{l,l+1}+\rK_{l,l+1}\rP_{l,l+2}\right)\comma
\end{align}
where $\rP_{ab}$ and $\rK_{ab}$ are permutation and trace operators acting on the $a$-th and $b$-th sites. We denote the set of orthonormal basis of the Hilbert space at each site by $|i\rangle$, $i=1,\cdots,4$. The two operators act as
\begin{align}
\rP|i\rangle\otimes|j\rangle=|j\rangle\otimes|i\rangle,\qquad \rK|i\rangle\otimes|j\rangle=\delta_{ij}\sum_{k=1}^4|k\rangle\otimes|k\rangle\period
\end{align}
Using these definitions, it is straightforward to show that $\rP_{ab}$ and $\rK_{ab}$ obey the following relations
\begin{align}
\rP_{ab}\rP_{ab}=1,\qquad \rK_{ab}\rK_{ab}=4\rK_{ab},\qquad \rP_{ab}\rK_{bc}=\rK_{ac}\rP_{ab}=\rK_{ac}\rK_{bc}\period
\end{align}
The spin chain under consideration is an alternating spin chain. The odd and even sites sit in the $\mathbf{4}$ and $\bar{\mathbf{4}}$ representation of $SU(4)$ group and correspond to fields $Y^{A}$ and $\bar{Y}_A$ in ABJM theory respectively. In order to distinguish odd and even sites, we put bars to the odd sites, namely
\begin{align}
Y^A\mapsto |A\rangle,\qquad \bar{Y}_A\mapsto |\bar{A}\rangle\period
\end{align}
In this notation, the single-trace operator ${\rm tr}\left(Y^{A_1}\bar{Y}_{B_1}Y^{A_2}\bar{Y}_{B_2}\cdots\right)$ is mapped to the following spin-chain state;
\begin{align}
  {\rm tr}\left(Y^{A_1}\bar{Y}_{B_1}Y^{A_2}\bar{Y}_{B_2}\cdots\right)\quad \mapsto\quad |A_1\bar{B}_1A_2\bar{B}_2\cdots\rangle\period
\end{align}
One special feature of the Hamiltonian (\eqref{eq:Ham}) is that the permutation operator $\rP_{l,l+2}$ only act on the odd- or even-site spin chains while the trace operator $\rK_{l,l+1}$ only mix the two. We have
\begin{align}
\rK|A\rangle\otimes|\bar{B}\rangle=\delta_{AB}\sum_{C=1}^4|C\rangle\otimes|\bar{C}\rangle\period
\end{align}

\begin{figure}[t!]
\centering
\includegraphics[scale=0.5]{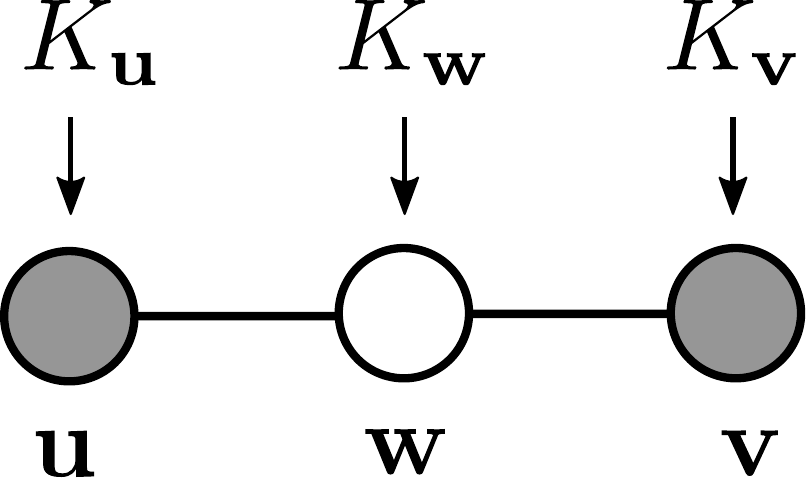}
\caption{Dynkin diagram of the SU(4) alternating spin chain. The gray circles are the momentum carrying nodes.}
\label{fig:dynkin}
\end{figure}

The Hamiltonian \eqref{eq:Ham} is known to be integrable and solvable by the Bethe ansatz. Reflecting the alternating strucrure of the spin chain, the Bethe state is described by two sets of the momentum carrying Bethe roots which we denote by ${\bf u}$ and ${\bf v}$ (see figure \ref{fig:dynkin}). In addition, there are auxiliary Bethe roots ${\bf w}$. The numbers of rapidities of $\mathbf{u},\mathbf{v},\mathbf{w}$ are denoted by $K_{\bf u},K_{\bf v},K_{\bf w}$ respectively.
The rapidities satisfy the Bethe equations
\begin{align}\label{eq:betheansatzeq}
    \begin{aligned}
        &1=e^{i\phi_{u_j}}=\left(\frac{u_j+\frac{i}{2}}{u_j-\frac{i}{2}}\right)^{L}\prod_{\substack{k=1\\k\neq j}}^{K_{\bf u}}S(u_j,u_k)\prod_{k=1}^{K_{\bf w}}\tilde{S}(u_j,w_k)\comma\\
        &1=e^{i\phi_{w_j}}=\prod_{\substack{k=1\\k\neq j}}^{K_{\bf w}}S(w_j,w_k)\prod_{k=1}^{K_{\bf u}}\tilde{S}(w_j,u_k)\prod_{k=1}^{K_{\bf v}}\tilde{S}(w_j,v_k)\comma\\
        &1=e^{i\phi_{v_j}}=\left(\frac{v_j+\frac{i}{2}}{v_j-\frac{i}{2}}\right)^{L}\prod_{\substack{k=1\\k\neq j}}^{K_{\bf v}}S(v_j,v_k)\prod_{k=1}^{K_{\bf w}}\tilde{S}(v_j,w_k)\comma
    \end{aligned}
\end{align}
where the S-matrices $S(u,v)$ and $\tilde{S}(u,v)$ are given by
\begin{align}
    S(u,v)\equiv \frac{u-v-i}{u-v+i}\comma\qquad \tilde{S}(u,v)\equiv \frac{u-v+\frac{i}{2}}{u-v-\frac{i}{2}}\period
\end{align}
The two-loop anomalous dimension of the operator is given by
\begin{align}
    \Delta -L=\lambda^2\left(\sum_{k=1}^{K_{\bf u}}\frac{1}{u_k^2+\frac{1}{4}}+\sum_{k=1}^{K_{\bf v}}\frac{1}{v_k^2+\frac{1}{4}}\right)\period
\end{align}
The $U(1)$ R-charge $J$ discussed in section \ref{sec:2.3} can be expressed in terms of $K_{{\bf u},{\bf v}}$ as
\begin{align}
    J=L-\frac{K_{\bf u}+K_{\bf v}}{2}\period
\end{align}

\paragraph{Zero-momentum condition.} As is the case with $\mathcal{N}=4$ SYM, not all solutions to the Bethe equation corresponds to a single-trace operator in ABJM theory. This is because the single-trace operator has an additional {\it cyclicity} property. In the spin-chain language, this is equivalent to the zero-momentum condition,
\begin{align}
    1=\prod_{j=1}^{K_{\bf u}}\frac{u_j+\frac{i}{2}}{u_j-\frac{i}{2}}\prod_{j=1}^{K_{\bf v}}\frac{v_j+\frac{i}{2}}{v_j-\frac{i}{2}}\period
\end{align}
\subsection{Coordinate Bethe ansatz}
The eigenvector of the spin chain \eqref{eq:Ham} can be constructed by the nested coordinate Bethe ansatz (CBA). To describe the CBA construction, we map each scalar field to a specific combination of the Bethe roots following the notation in Appendix E of \cite{Basso:2017khq}. More precisely, we first express odd- and even-sites by bullets $\bullet$ and circles $\circ$ respectively, and place Bethe roots on top of them. Then, the relation between stacks of Bethe roots and the fields in ABJM theory is given by
\begin{align}
\label{eq:Ystring}
\begin{aligned}
&Y^1\mapsto|1\rangle=|\bullet\rangle,\qquad Y^2\mapsto|2\rangle=|\overset{\textcolor{blue}{u}}{\bullet}\rangle\comma\qquad Y^3\mapsto|3\rangle=|\overset{\overset{w}{\textcolor{blue}{u}}}{\bullet}\rangle\comma\qquad Y^4\mapsto|4\rangle=|\overset{\overset{w}{\textcolor{blue}{u}\textcolor{red}{v}}}{\bullet}\rangle\comma\\
& \bar{Y}_1\mapsto|\bar{1}\rangle=|\overset{\overset{w}{\textcolor{blue}{u}\textcolor{red}{v}}}{\circ}\rangle\comma\qquad  \bar{Y}_2\mapsto|\bar{2}\rangle=|\overset{\overset{w}{\textcolor{red}{v}}}{\circ}\rangle\comma\qquad \bar{Y}_3\mapsto|\bar{3}\rangle=|\overset{\textcolor{red}{v}}{\circ}\rangle\comma\qquad
\bar{Y}_4\mapsto|\bar{4}\rangle=|\circ\rangle\period
\end{aligned}
\end{align}

We now briefly outline the procedure of constructing eigenvectors. It can be achieved in two steps. In the first step, we distribute the rapidities on different sites of the spin chain and sum over all such possibilities. In the second step, we construct the wave function for each distribution and multiply it to the corresponding distribution. Plugging in the physical solutions of the Bethe roots, we obtain an eigenvector of the Hamiltonian.

\begin{figure}[h!]
\centering
\includegraphics[scale=0.3]{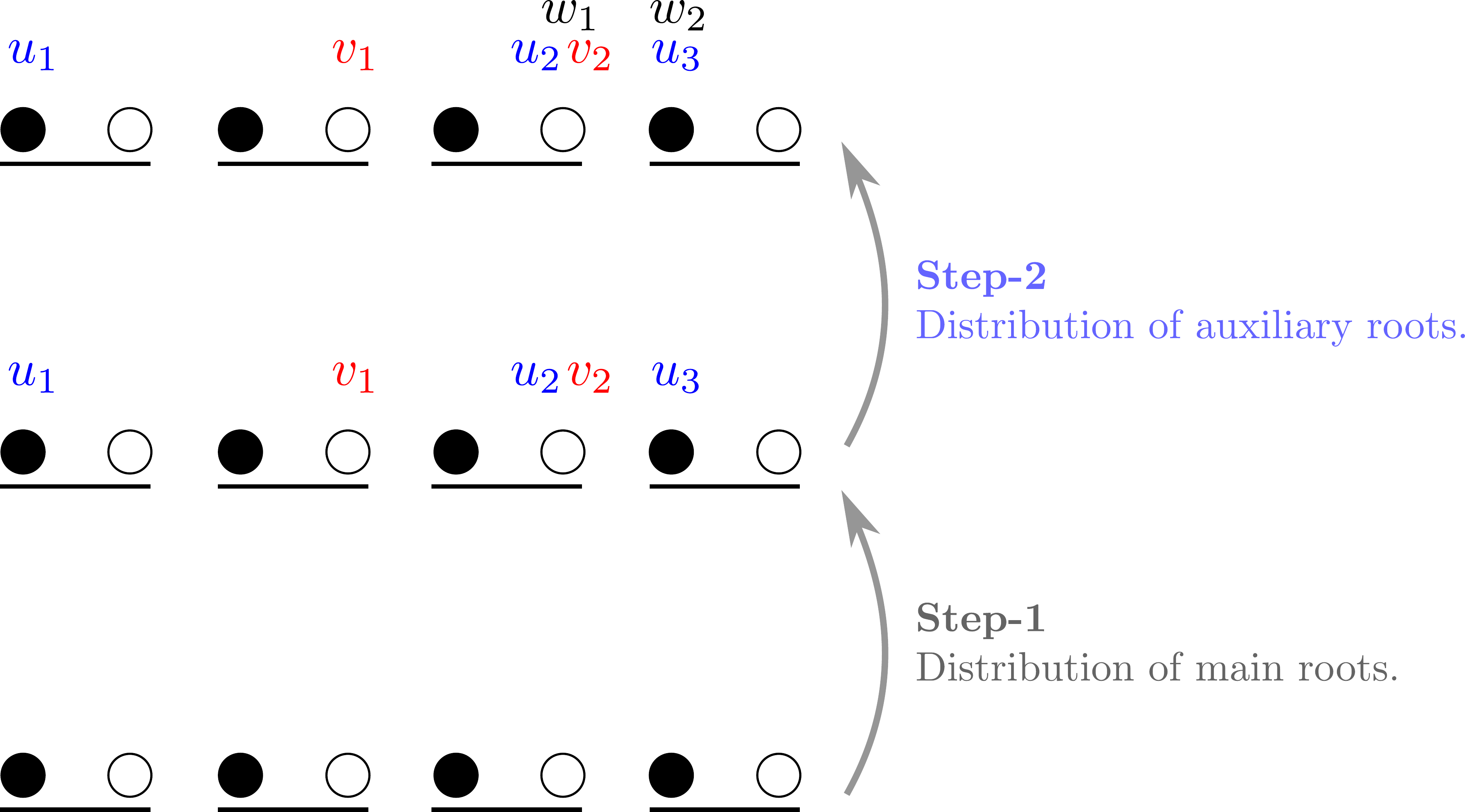}
\caption{Construction of (one of) the Bethe string configuration corresponding to the ket state $\tr(\textcolor{red}{Y^2}\bar{Y}_4)(Y^1\textcolor{blue}{\bar{Y}_3})(Y^1\textcolor{blue}{\bar{Y}_1})(\textcolor{red}{Y^3}\bar{Y}_4)
=|2,\bar{4},1,\bar{3},1,\bar{1},3,\bar{4}\rangle$.}
\label{fig:egstring}
\end{figure}

\paragraph{Distributing excitations.}
We consider the spin chain of length $2L$.  We first distribute these rapidities on different sites by the following procedure
\begin{enumerate}
\item Firstly we distribute the momentum carrying rapidities $\{u_1,\cdots,u_{K_{\bf u}}\}$ and $\{v_1,\cdots,v_{K_{\bf v}}\}$ on top of the ground state $|\bullet_1,\circ_1,\cdots\bullet_L,\circ_L\rangle$. Each odd site can support either a single $u$-type rapidity, or two rapdities one of $uv$-type. Likewise, each even site can support a single $v$-type rapidity, or two rapidities of $uv$-type.
\item Now we view the physical rapidities $\{u_1,u_2,\cdots,u_{K_{\bf u}}\}$ and $\{v_1,v_2,\cdots,v_{K_{\bf v}}\}$ as inhomogeneities of an \emph{emergent spin chain} of length $K_{\bf u}+K_{\bf v}$ and distribute the rapidities $\{w_1,w_2,\cdots,w_{K_{\bf w}}\}$ on the emergent chain. Each $w_j$ can be distributed on top of either $u_i$ or $v_i$.
\item In the previous two steps, we also generate string configurations without field theory correspondence in (\ref{eq:Ystring}). For example, $\overset{uv}{\bullet}$ does not corresponds to any of the fields $Y^A$ or $\bar{Y}_A$. We set all the states which contain such string configurations to zero.
\end{enumerate}
We give an example of the procedure described above in figure~\ref{fig:egstring}.
%We consider some special limits and check the consistency of our construction. Firstly, consider the limit where the auxiliary root is absent. Then $u$ and $v$ cannot land on the same inhomogeneity since it does not correspond to any scalar field. Then the spin chain decouple into two standard $SU(2)$ spin chain whose rapidities are $\{u_1,\cdots,u_{M}\}$ and $\{v_1,\cdots,v_{N}\}$. Secondly, we consider the limit where the $v$-roots are absent. Then the auxiliary root can only land on the $u$-roots. We therefore obtain a standard $SU(3)$ spin chain. This is indeed consistent with the analysis of the spin chain.

In what follows, we denote the stack of Bethe roots at site $n$ by ${\sf s}_n$. The possible configurations are
\begin{align}
\mathsf{s}_{2n-1}=\{\bullet_n,\quad\overset{\textcolor{blue}{u_k}}{\bullet_n},\quad\overset{\overset{w_a}{\textcolor{blue}{u_k}}}{\bullet_n},\quad
\overset{\overset{w_a}{\textcolor{blue}{u_k}\textcolor{red}{v_j}}}{\bullet_n}\}\comma\qquad
\mathsf{s}_{2n}=\{\circ_n,\quad\overset{\textcolor{red}{v_k}}{\circ_n},\quad\overset{\overset{w_a}{\textcolor{red}{v_k}}}{\circ_n},\quad
\overset{\overset{w_a}{\textcolor{blue}{u_k}\textcolor{red}{v_j}}}{\circ_n}\}\period
\end{align}
A collection of all ${\sf s}_n$'s for a given distribution will be denoted by $\vec{\sf s}$
\begin{align}
    \vec{\sf s}\equiv \{{\sf s}_1,{\sf s}_2,\ldots, {\sf s}_{2L-1},{\sf s}_{2L}\}\period
\end{align}

\paragraph{The wave functions: ordered configurations.}The next step is to write down a wave function for each distribution of rapidities. We first discuss the distribution in which the rapidities are in the canonical order (namely in the same order as ${\bf u}$, ${\bf v}$ and ${\bf w}$). We call such distributions {\it ordered distributions}. The wave function for a ordered distribution of Bethe roots $\vec{\sf s}$ is given by a product of ``single-site wave functions" $\Phi ({\sf s})$;
\begin{align}
\Psi_{\vec{\mathsf{s}}}\big(\textcolor{blue}{\mathbf{u}},\textcolor{red}{\mathbf{v}},\mathbf{w}\big)=\prod_{n=1}^{2L}\Phi(\mathsf{s}_n)\period
\end{align}

The single-site wave functions with at most a single rapidity are
\begin{align}
\label{eq:phiF1}
\Phi(\bullet_n)=&\,\Phi(\circ_n)=1\comma\qquad
\Phi(\overset{\textcolor{blue}{u_k}}{\bullet_n})=\,\left(\frac{u_k+\tfrac{i}{2}}{u_k-\tfrac{i}{2}}\right)^n\comma\qquad
\Phi(\overset{\textcolor{red}{v_k}}{\circ_n})=\,\left(\frac{v_k+\tfrac{i}{2}}{v_k-\tfrac{i}{2}}\right)^n\period
\end{align}
It is slightly more complicated to write a single-site wave function when there is $w$ because the expression depends on the Bethe roots on other sites; or more precisely on the momentum carrying rapidities to the left of the site. For instance, $\Phi ({\sf s})$'s corresponding to $\{\overset{\overset{w}{\textcolor{blue}{u}}}{\bullet_n},\overset{\overset{w}{\textcolor{red}{v}}}{\circ_n}\}$ are given by
\begin{align}
\label{eq:phiF2}
&\,\Phi(\overset{\overset{w_a}{\textcolor{blue}{u_k}}}{\bullet_n})
=\left(\frac{u_k+\tfrac{i}{2}}{u_k-\tfrac{i}{2}}\right)^n\times\psi(w_a|\bm{z}^{<})\times \frac{-1}{w_a-u_k-\tfrac{i}{2}}\comma\\\nonumber
&\,\Phi(\overset{\overset{w_a}{\textcolor{red}{v_k}}}{\circ_n})=\left(\frac{v_k+\tfrac{i}{2}}{v_k-\tfrac{i}{2}}\right)^n\times\psi(w_a|\bm{z}^{<})
\times \frac{1}{w_a-v_k-\tfrac{i}{2}}\comma
\end{align}
where $\bm{z}^{<}$ denotes all the momentum carrying rapidities that are to the left of the site which supports $w$. The explicit form is given by
\begin{align}
\psi(w|\bm{z}^{<})=\prod_{j}\frac{w-z_j^{<}+\tfrac{i}{2}}{w-z_j^{<}-\tfrac{i}{2}}.
\end{align}
Similarly, the single-site wave functions for the configurations $\{\overset{\overset{w}{\textcolor{blue}{u}\textcolor{red}{v}}}{\bullet_n},\overset{\overset{w}{\textcolor{blue}{u}\textcolor{red}{v}}}{\circ_n}\}$ are given by
\begin{align}
\begin{aligned}
\label{eq:phiF3}
&\,\Phi(\overset{\overset{w_a}{\textcolor{blue}{u_k}\textcolor{red}{v_j}}}{\bullet_n})
=\left(\frac{u_k+\tfrac{i}{2}}{u_k-\tfrac{i}{2}}\right)^n\left(\frac{v_j+\tfrac{i}{2}}{v_j-\tfrac{i}{2}}\right)^{n}\times\psi(w_a|\bm{z}^{<})
\times\frac{-(v_j-\tfrac{i}{2})}{(w_a-u_k-\tfrac{i}{2})(w_a-v_j-\tfrac{i}{2})}\comma\\
&\Phi(\overset{\overset{w_a}{\textcolor{blue}{u_k}\textcolor{red}{v_j}}}{\circ_n})
=\left(\frac{u_k+\tfrac{i}{2}}{u_k-\tfrac{i}{2}}\right)^n\left(\frac{v_k+\tfrac{i}{2}}{v_k-\tfrac{i}{2}}\right)^{n}\times\psi(w_a|\bm{z}^{<})
\times\frac{+(u_k+\tfrac{i}{2})}{(w_a-u_k-\tfrac{i}{2})(w_a-v_j-\tfrac{i}{2})}\period
\end{aligned}
\end{align}

\paragraph{The wave functions: general configurations.}More general configurations of the rapidities can be obtained from the ordered configurations by performing permutations of the Bethe roots. For such configuraions, the wave function consists of two factors. The first factor is a product of single-site wave functions $\Phi$,
\begin{align}
    \prod_{n=1}^{2L}\Phi ({\sf s}_n)\period
\end{align}
The second factor is a product of S-matrices, which are needed to bring the rapidities in the distribution into the canonical order. For instance, for the distribution in which $u_2$ is to the left of $u_1$, we need to multiply
\begin{align}
    S(u_1,u_2)=\frac{u_1-u_2-i}{u_1-u_2+i}\period
\end{align}
We multiply similar factors ($S(v_j,v_k)$ and $S(w_j,w_k)$) also for $v$'s and $w$'s. Therefore, the wave function for a general distribution is given by
\begin{align}
    \Psi_{\vec{\mathsf{s}}}\big(\textcolor{blue}{\mathbf{u}},\textcolor{red}{\mathbf{v}},\mathbf{w}\big)=\mathbb{S}\times\prod_{n=1}^{2L}\Phi(\mathsf{s}_n)\comma
\end{align}
where $\mathbb{S}$ is a product of S-matrices described above.

The Bethe eigenstate is given by a summation of all possible distributions; namely
\begin{align}
    |\Psi_{\textcolor{blue}{\mathbf{u}},\textcolor{red}{\mathbf{v}},\mathbf{w}}\rangle=\sum_{\vec{\sf s}\in \substack{\text{all possible}\\\text{distributions}}}\Psi_{\vec{\mathsf{s}}}\big(\textcolor{blue}{\mathbf{u}},\textcolor{red}{\mathbf{v}},\mathbf{w}\big)|\vec{\sf s}\rangle\period
\end{align}

\paragraph{Examples.}
As explicit examples of the procedure described above, we give three simple states for $L=2$. The vacuum state is given by
\begin{align}
|\Omega\rangle=|1,\bar{4},1,\bar{4}\rangle\period
\end{align}
First, we consider the state with $K_{\bf u}=1$, $K_{\bf v}=1$, $K_{\bf w}=0$.
The state is given by a sum of four terms, and below we list them along with the wave function for each distribution:
\begin{align}\nonumber
\begin{aligned}
&|2,\bar{3},1,\bar{4}\rangle\equiv|\overset{\textcolor{blue}{u}}{\bullet}_1,\overset{\textcolor{red}{v}}{\circ}_1,\bullet_2,\circ_2\rangle\quad
\left(\frac{u+\tfrac{i}{2}}{u-\tfrac{i}{2}}\right)\left(\frac{v+\tfrac{i}{2}}{v-\tfrac{i}{2}}\right)\comma\\
&|2,\bar{4},1,\bar{3}\rangle\equiv|\overset{\textcolor{blue}{u}}{\bullet}_1,\circ_1,\bullet_2,\overset{\textcolor{red}{v}}{\circ}_2\rangle\quad
\left(\frac{u+\tfrac{i}{2}}{u-\tfrac{i}{2}}\right)\left(\frac{v+\tfrac{i}{2}}{v-\tfrac{i}{2}}\right)^2\comma\\
&|1,\bar{3},2,\bar{4}\rangle\equiv|\bullet_1,\overset{\textcolor{red}{v}}{\circ}_1,\overset{\textcolor{blue}{u}}{\bullet}_2,\circ_2\rangle\quad
\left(\frac{u+\tfrac{i}{2}}{u-\tfrac{i}{2}}\right)^2\left(\frac{v+\tfrac{i}{2}}{v-\tfrac{i}{2}}\right)\comma\\
&|1,\bar{4},2,\bar{3}\rangle\equiv|\bullet_1,\circ_1,\overset{\textcolor{blue}{u}}{\bullet}_2,\overset{\textcolor{red}{v}}{\circ}_2\rangle\quad
\left(\frac{u+\tfrac{i}{2}}{u-\tfrac{i}{2}}\right)^2\left(\frac{v+\tfrac{i}{2}}{v-\tfrac{i}{2}}\right)^2\period
\end{aligned}
\end{align}

Second, we consider the state with $K_{\bf u}=2$, $K_{\bf v}=K_{\bf w}=0$. It consists of the following two terms
\begin{align}\nonumber
    \begin{aligned}
        &|2,\bar{4},2,\bar{4}\rangle_1\equiv |\overset{\textcolor{blue}{u_1}}{\bullet}_1,\circ_1,\overset{\textcolor{blue}{u_2}}{\bullet}_2,\circ_2\rangle\qquad \left(\frac{u_1+\frac{i}{2}}{u_1-\frac{i}{2}}\right)\left(\frac{u_2+\frac{i}{2}}{u_2-\frac{i}{2}}\right)^2\comma\\
        &|2,\bar{4},2,\bar{4}\rangle_2\equiv |\overset{\textcolor{blue}{u_2}}{\bullet}_1,\circ_1,\overset{\textcolor{blue}{u_1}}{\bullet}_2,\circ_2\rangle\qquad \frac{u_1-u_2-i}{u_1-u_2+i}\left(\frac{u_2+\frac{i}{2}}{u_2-\frac{i}{2}}\right)\left(\frac{u_1+\frac{i}{2}}{u_1-\frac{i}{2}}\right)^2\period
    \end{aligned}
\end{align}
The first term is an ordered distribution while the second term is not. This is why the second term comes with an extra factor of the S-matrix $S(u_1,u_2)=(u_1-u_2-i)/(u_1-u_2+i)$.

As the third example, we consider the state with $K_{\bf u}=1$, $K_{\bf v}=1$, $K_{\bf w}=1$.
In this case, there are 12 terms, which we list below
\begin{align}\nonumber
\begin{aligned}
&|2,\bar{2},1,\bar{4}\rangle\equiv |\overset{\textcolor{blue}{u}}{\bullet}_1,\overset{\overset{w}{\textcolor{red}{v}}}{\circ}_1,\bullet_2,\circ_2\rangle\qquad
\left(\frac{u+\tfrac{i}{2}}{u-\tfrac{i}{2}}\right)\left(\frac{v+\tfrac{i}{2}}{v-\tfrac{i}{2}}\right)\frac{w-u+\tfrac{i}{2}}{w-u-\tfrac{i}{2}}
\frac{1}{w-v-\tfrac{i}{2}}\comma\\
&|2,\bar{4},1,\bar{2}\rangle\equiv |\overset{\textcolor{blue}{u}}{\bullet}_1,\circ_1,\bullet_2,\overset{\overset{w}{\textcolor{red}{v}}}{\circ}_2\rangle\qquad
\left(\frac{u+\tfrac{i}{2}}{u-\tfrac{i}{2}}\right)\left(\frac{v+\tfrac{i}{2}}{v-\tfrac{i}{2}}\right)^2\frac{w-u+\tfrac{i}{2}}{w-u-\tfrac{i}{2}}
\frac{1}{w-v-\tfrac{i}{2}}\comma\\
&|1,\bar{2},2,\bar{4}\rangle\equiv|\bullet_1,\overset{\overset{w}{\textcolor{red}{v}}}{\circ}_1,\overset{\textcolor{blue}{u}}{\bullet}_2,\circ_2\rangle\qquad
\left(\frac{u+\tfrac{i}{2}}{u-\tfrac{i}{2}}\right)^2\left(\frac{v+\tfrac{i}{2}}{v-\tfrac{i}{2}}\right)\frac{1}{w-v-\tfrac{i}{2}}\comma\\
&|1,\bar{4},2,\bar{2}\rangle\equiv|\bullet_1,\circ_1,\overset{\textcolor{blue}{u}}{\bullet}_2,\overset{\overset{w}{\textcolor{red}{v}}}{\circ}_2\rangle
\qquad
\left(\frac{u+\tfrac{i}{2}}{u-\tfrac{i}{2}}\right)^2\left(\frac{v+\tfrac{i}{2}}{v-\tfrac{i}{2}}\right)^2\frac{w-u+\tfrac{i}{2}}{w-u-\tfrac{i}{2}}
\frac{1}{w-v-\tfrac{i}{2}}\comma\\
&|3,\bar{3},1,\bar{4}\rangle\equiv|\overset{\overset{w}{\textcolor{blue}{u}}}{\bullet}_1,\overset{\textcolor{red}{v}}{\circ}_1,\bullet_2,\circ_2\rangle\qquad
\left(\frac{u+\tfrac{i}{2}}{u-\tfrac{i}{2}}\right)\left(\frac{v+\tfrac{i}{2}}{v-\tfrac{i}{2}}\right)\frac{-1}{w-u-\tfrac{i}{2}}\comma\\
&|3,\bar{4},1,\bar{3}\rangle\equiv|\overset{\overset{w}{\textcolor{blue}{u}}}{\bullet}_1,\circ_1,\bullet_2,\overset{\textcolor{red}{v}}{\circ}_2\rangle\qquad
\left(\frac{u+\tfrac{i}{2}}{u-\tfrac{i}{2}}\right)\left(\frac{v+\tfrac{i}{2}}{v-\tfrac{i}{2}}\right)^2\frac{-1}{w-u-\tfrac{i}{2}}\comma\\
&|1,\bar{4},3,\bar{3}\rangle\equiv|\bullet_1,\circ_1,\overset{\overset{w}{\textcolor{blue}{u}}}{\bullet}_2,\overset{\textcolor{red}{v}}{\circ}_2\rangle\qquad
\left(\frac{u+\tfrac{i}{2}}{u-\tfrac{i}{2}}\right)^2\left(\frac{v+\tfrac{i}{2}}{v-\tfrac{i}{2}}\right)^2\frac{-1}{w-u-\tfrac{i}{2}}\comma\\
&|1,\bar{3},3,\bar{4}\rangle\equiv|\bullet_1,\overset{\textcolor{red}{v}}{\circ}_1,\overset{\overset{w}{\textcolor{blue}{u}}}{\bullet}_2,\circ_2\rangle\qquad
\left(\frac{u+\tfrac{i}{2}}{u-\tfrac{i}{2}}\right)^2\left(\frac{v+\tfrac{i}{2}}{v-\tfrac{i}{2}}\right)\frac{w-v+\tfrac{i}{2}}{w-v-\tfrac{i}{2}}
\frac{-1}{w-u-\tfrac{i}{2}}\comma\\
&|4,\bar{4},1,\bar{4}\rangle\equiv|\overset{\overset{w}{\textcolor{blue}{u}\textcolor{red}{v}}}{\bullet}_1,\circ_1,\bullet_2,\circ_2\rangle\qquad
\left(\frac{u+\tfrac{i}{2}}{u-\tfrac{i}{2}}\right)\left(\frac{v+\tfrac{i}{2}}{v-\tfrac{i}{2}}\right)
\frac{-(v-\tfrac{i}{2})}{(w-v-\tfrac{i}{2})(w-u-\tfrac{i}{2})}\comma\\
&|1,\bar{1},1,\bar{4}\rangle\equiv|\bullet_1,\overset{\overset{w}{\textcolor{blue}{u}\textcolor{red}{v}}}{\circ}_1,\bullet_2,\circ_2\rangle\qquad
\left(\frac{u+\tfrac{i}{2}}{u-\tfrac{i}{2}}\right)\left(\frac{v+\tfrac{i}{2}}{v-\tfrac{i}{2}}\right)
\frac{+(u+\tfrac{i}{2})}{(w-v-\tfrac{i}{2})(w-u-\tfrac{i}{2})}\comma\\
&|1,\bar{4},4,\bar{1}\rangle\equiv|\bullet_1,\circ_1,\overset{\overset{w}{\textcolor{blue}{u}\textcolor{red}{v}}}{\bullet}_2,\circ_2\rangle\qquad
\left(\frac{u+\tfrac{i}{2}}{u-\tfrac{i}{2}}\right)^2\left(\frac{v+\tfrac{i}{2}}{v-\tfrac{i}{2}}\right)^2
\frac{-(v-\tfrac{i}{2})}{(w-v-\tfrac{i}{2})(w-u-\tfrac{i}{2})}\comma\\
&|1,\bar{4},1,\bar{1}\rangle\equiv|\bullet_1,\circ_1,\bullet_2,\overset{\overset{w}{\textcolor{blue}{u}\textcolor{red}{v}}}{\circ}_2\rangle\qquad
\left(\frac{u+\tfrac{i}{2}}{u-\tfrac{i}{2}}\right)^2\left(\frac{v+\tfrac{i}{2}}{v-\tfrac{i}{2}}\right)^2
\frac{+(u+\tfrac{i}{2})}{(w-v-\tfrac{i}{2})(w-u-\tfrac{i}{2})}\period
\end{aligned}
\end{align}

%%%%%%%%%%%%%%%%%%%%%%%%%%%%%%%%%%%%%%%%%%%%%%%%%%%%%%%%%%%%%
	%%%%%%%%%%%%%%%%%%%%%%%%%%%%%%%%%%%%%%%%%%%%%%%%%%%%%%%%%%%%%
\section{Matrix Product States from Giant Gravitons}
\label{sec:MPS}
To compute the structure constant \eqref{eq:setupfinal} using integrability, we first need to map it to a quantity in the spin chain. For $\mathcal{N}=4$ SYM, this was done\fn{The analysis was recently generalized to the four-point function of determinant operators in $\mathcal{N}=4$ SYM \cite{Vescovi:2021fjf}.} in \cite{Jiang:2019xdz,Jiang:2019zig,Chen:2019gsb}, which showed that the structure constant \eqref{eq:setupfinal} corresponds to an overlap between a matrix product state (MPS) and a Bethe eigenstate. This analysis was later generalized to ABJ(M) theory in \cite{Chen:2019kgc}. In this section, we review and extend the results in \cite{Chen:2019kgc}. Two main new results are
\begin{enumerate}
\item We compute tree-level structure constants of two non-maximal giant gravitons and a single-trace BPS operator.
\item We derive an explicit expression for MPS of two non-maximal giant gravitons in the twisted-translated frame which can be readily used in later sections.
\end{enumerate}

%%%%%%%%%%%%%%%%%%%%%%%%%%%%%%%%%%%%%%%%%%%%%%%%%%%%%%%%%%%%%
\subsection{Derivation of matrix product states}
Although the main goal of this paper is to compute the structure constant in the twisted-translated frame \eqref{eq:setupfinal}, in this subsection we consider a slightly more general setup; the correlation function of $m$ giant gravitons and a single-trace operator in a general kinematics. More precisely we consider the following correlation function
\begin{align}
G_{m}=\left<\left(\prod_{j=1}^{m}\mathcal{G}_j\right)\mathcal{O}(y)\right>\comma
\end{align}
where $\mathcal{G}_j$ is a generating function for giant gravitons (cf.~\eqref{eq:defgenerating}):
\begin{align}
\mathcal{G}_j\equiv \det \pmatrix{cc}{{\bf 1}&-t_j (\bar{n}_j\cdot \bar{Y})\\ t_j (n_j\cdot Y)&{\bf 1}}(x_j)\period
\end{align}
To extract the correlation functions of individual giant gravitons, we simply need to perform integrals of $t_j$'s,
\begin{align}\label{eq:projection}
  \left<\left(\prod_{j=1}^{m}\mathcal{D}_{M_j}\right)\mathcal{O}(y)\right>=  \left(\prod_{j=1}^{m}\oint \frac{dt_j}{t_j^{1+2M_j}}\right)G_m \period
\end{align}
Since giant gravitons carry large R-charges $M_j\sim \mathcal{O}(N)$, most of these integrals in the large $N$ limit can be computed by the saddle-point approximation. We will later see how this works in practice. For now we focus on computing the correlation functions of generating functions.

 Let us now compute $G_m$ at tree level. This is given by the following path integral
\begin{align}
G_m=\frac{1}{Z_Y}\int DY^I\,D\bar{Y}_I\left(\prod_{j=1}^m\mathcal{G}_j\right)\mathcal{O}\,\exp\left[-\frac{k}{4\pi}\int\rd^3x\,\tr(\partial_{\mu}\bar{Y}_I\,\partial^{\mu}Y^I) \right]\comma
\end{align}
with
\begin{align}
{Z_Y}=\int DY^I\,D\bar{Y}_I\exp\left[-\frac{k}{4\pi}\int\rd^3x\,\tr(\partial_{\mu}\bar{Y}_I\,\partial^{\mu}Y^I) \right]\comma
\end{align}
The generating functions can be written in terms of path integral of fermions
\begin{align}
\begin{aligned}
\mathcal{G}_j=\int d\eta_jd\bar{\eta}_jd\chi_jd\bar{\chi}_j\exp \left[(\bar{\chi}_j\chi_j)+(\bar{\eta}_j\eta_j)+t_j\bar{\chi}_j(n_j\cdot Y)\eta_j-t_j\bar{\eta}_j(\bar{n}_j\cdot \bar{Y})\chi_j\right]\period
\end{aligned}
\end{align}
Here the fermions $\chi_i$ and $\eta_i$ are in the fundamental representation of the gauge group $U(N)_k$ and $U(N)_{-k}$ respectively while $\bar{\chi}_i,\bar{\eta}_i$ are in the antifundamental representation. The indices are contracted as
\begin{align}
\begin{aligned}
\bar{\chi}(n\!\cdot\! Y)\eta=\,\bar{\chi}^a(n\!\cdot\! Y)_a^{\phantom{a}\bar{b}}\eta_{\bar{b}},\qquad
\bar{\eta}(\bar{n}\!\cdot\!\bar{Y})\chi=\,\bar{\eta}^{\bar{a}}\,(\bar{n}\!\cdot\! \bar{Y})_{\bar{a}}^{\phantom{a}b}\,\chi_b\period
\end{aligned}
\end{align}
We can then rewrite the correlation function $G_m$ as
\begin{align}
\label{eq:Gmpathintegral1}
G_m=\frac{1}{Z_Y}\int DY^I\,D\bar{Y}_I\prod_{i=1}^m d\bar{\chi}_i d\eta_i d\bar{\eta}_id\chi_i\,\mathcal{O}\,\exp\left(-S_{\chi,\eta}\right)\comma
\end{align}
where
\begin{align}
\begin{aligned}
&S_{\chi,\eta}=-\sum_{i=1}^m \left((\bar{\eta}_i \eta_i)+(\bar{\chi}_i \chi_i)\right)+\frac{k}{4\pi}\int\rd^3x\Big[\tr(\partial_{\mu}\bar{Y}_I\partial^{\mu}Y^I)
\\
&\qquad-\frac{4\pi}{k}\sum_{i=1}^m t_i\bar{\chi}_i(n_i\!\cdot\!Y)\eta_i\,\delta^3(x-x_i)+\frac{4\pi}{k}\sum_{i=1}^mt_i\bar{\eta}_i(\bar{n}_i\!\cdot\!\bar{Y})\chi_i\,\delta^3(x-x_i) \Big]\period
\end{aligned}
\end{align}
Since this action is quadratic in $Y^I$, $\bar{Y}_I$, we can integrate them out. In practice, this is equivalent to solving the equations of motion for these fields,
\begin{align}
\begin{aligned}
\label{eq:eomYYb}
\square (Y^I)_a^{\phantom{a}\bar{b}}(x)=&\,-\frac{4\pi}{k}\sum_{i=1}^m t_i \bar{n}_i^I\, {\chi}_{i,a}\bar{\eta}_i^{\bar{b}} \delta^3(x-x_i),\\
\square (Y^{\dagger}_I)_{\bar{a}}^{\phantom{a}b}(x)=&\,\frac{4\pi}{k}\sum_{i=1}^m t_i{n}_{i,I}\,\eta_{i,\bar{a}}\bar{\chi}_i^b \delta^3(x-x_i)\period
\end{aligned}
\end{align}
These equations can be solved by Green's function. Using the fact that
\begin{align}
\square \frac{1}{|x-y|}=-4\pi\,\delta^3(x-y)\comma
\end{align}
the solution to (\ref{eq:eomYYb}) is given by
\begin{align}
\label{eq:solemoYYb}
Y^I(x)=S^I(x)\equiv \frac{1}{k}\sum_{i=1}^m\frac{t_i\bar{n}_i^I\,\chi_i\bar{\eta}_i }{|x-x_i|}\comma\qquad
\bar{Y}_I(x)=\bar{S}_I(x)\equiv-\frac{1}{k}\sum_{i=1}^m\frac{t_i{n}_{i,I}\,\eta_i\bar{\chi}_i }{|x-x_i|}\period
\end{align}
Plugging the solution (\ref{eq:solemoYYb}) back to (\ref{eq:Gmpathintegral1}), we obtain
\begin{align}
G_m=\int \prod_{i=1}^m d\eta_i d\bar{\eta}_i d\chi_i d\bar{\chi}_i\,\mathcal{O}^S\,\exp\left(-\widetilde{S}_{\chi,\eta}\right)\comma
\end{align}
where $\mathcal{O}^S$ is obtained by replacing all $Y^I$, $\bar{Y}_I$ by their classical solution $S^I$, $\bar{S}_I$ in the single trace operator. The effective action $\widetilde{S}_{\chi,\eta}$ is given by
\begin{align}
\widetilde{S}_{\chi,\eta}=-\sum_i \left((\bar{\eta}_i \eta_i)+(\bar{\chi}_i \chi_i)\right)-\frac{\lambda}{ N}\sum_{i,j}t_it_j\underbrace{\frac{n_i\!\cdot\!\bar{n}_j}{|x_{ij}|}}_{=d_{ij}}(\bar{\chi}_i {\chi}_j)(\bar{\eta}_j\eta_i)\period
\end{align}

For the next step, we integrate in the``bilocal" fields $\rho_{ij}$ and $\bar{\rho}_{ij}$
\begin{align}
\label{eq:Gmpathintegral2}
G_m=\frac{1}{Z_{\rho,\bar{\rho}}}\int d\rho d\bar{\rho}d\eta d\bar{\eta} d\chi d\bar{\chi} \,\mathcal{O}^S\,
\exp\Big[-\frac{N}{\lambda}\sum_{\substack{i,j\\i\ne j}}\rho_{ij}\bar{\rho}_{ji}-\tilde{S}_{\chi,\eta}\Big]\comma
\end{align}
where
\begin{align}
Z_{\rho,\bar{\rho}}=\int d\rho d\bar{\rho} \exp\Big[-\frac{ N}{\lambda}\sum_{\substack{i,j\\i\ne j}}\rho_{ij}\bar{\rho}_{ji}\Big]\period
\end{align}
Shifting the bilocal fields $\rho,\bar{\rho}$ by
\begin{align}
\rho_{ij}\mapsto \, \rho_{ij}-\frac{\lambda}{ N}\sqrt{t_it_j\,d_{ij}}(\bar{\chi}_i {\chi}_j),\qquad
\bar{\rho}_{ji}\mapsto \,\bar{\rho}_{ji}-\frac{\lambda}{ N}\sqrt{t_it_j\,d_{ij}}(\bar{\eta}_j\eta_i)\comma
\end{align}
the action of (\ref{eq:Gmpathintegral2}) can be written as
\begin{align}
\begin{aligned}
S_{\rho,\chi,\eta}=&\frac{ N}{\lambda}\sum_{\substack{i,j\\i\ne j}}\rho_{ij}\bar{\rho}_{ji}-\sum_{i,j}\sqrt{t_it_j\,d_{ij}}\big(\rho_{ij}(\bar{\eta}_j\eta_i)+(\bar{\chi}_i {\chi}_j)\bar{\rho}_{ji} \big)\\
&-\sum_i \left((\bar{\eta}_i \eta_i)+(\bar{\chi}_i \chi_i)\right)\period
\end{aligned}
\end{align}

Finally we integrate out the fermions and obtain the expression
\begin{align}
\label{eq:GmSeff}
G_m=\frac{1}{Z_{\rho,\bar{\rho}}}\int d\rho d\bar{\rho}\,\langle\mathcal{O}^S\rangle_{\chi,\eta}\,\exp\big(NS_{\text{eff}}\big)\period
\end{align}
Let us explain the definitions of each quantity in the formula:
First the effective action is given by
\begin{align}\label{eq:Seff}
S_{\text{eff}}=-\frac{1}{\lambda}\text{tr}_m(\rho\bar{\rho})+\text{tr}_{m}\log\Sigma_{\eta}+\text{tr}_{m}\log\Sigma_{\chi}\comma
\end{align}
where ${\rm tr}_k$ means a trace of matrices of size $k$. Second $\Sigma_{\eta,\chi}$ are $m\times m$ matrices defined by
\begin{align}
    \Sigma_{\eta}\equiv {\bf 1}_m+\hat{\rho}\comma\qquad \Sigma_{\chi}\equiv {\bf 1}_m+\hat{\bar{\rho}}\comma
\end{align}
where ${\bf 1}_m$ is the identity matrix of size $m$ and $\hat{\rho}$ and $\hat{\bar{\rho}}$ are defined by
\begin{align}
\hat{\rho}_{ij}=\sqrt{t_id_{ij}t_j}\rho_{ij},\qquad \hat{\bar{\rho}}_{ij}=\sqrt{t_jd_{ji}t_i}\bar{\rho}_{ij}\period
\end{align}
Third the expectation value $\langle\mathcal{O}^S\rangle_{\chi,\eta}$ is given by
\begin{align}\label{eq:fermiexpectation}
\langle\mathcal{O}^S\rangle_{\chi,\eta}=\frac{\int d\eta d\bar{\eta}d\chi d\bar{\chi} \,\mathcal{O}^S\exp(\bar{\eta}\Sigma_{\eta}^{T}\eta+\bar{\chi}\Sigma_{\chi}^{T}\chi)}
{\int d\eta d\bar{\eta}d\chi d\bar{\chi} \,\exp(\bar{\eta}\Sigma_{\eta}^{T}\eta+\bar{\chi}\Sigma_{\chi}^{T}\chi)}\comma
\end{align}
where $\Sigma_{\eta,\chi}^{T}$ means a transposition of the matrix $\Sigma_{\eta,\chi}$.

Let us now explain how to evaluate \eqref{eq:fermiexpectation} in practice using the following single-trace operator as an example:
\begin{align}
\mathcal{O}(y)=\tr(Y^{A_1}\bar{Y}_{B_1}\ldots Y^{A_L}\bar{Y}_{B_L})(y)\period
\end{align}
We first replace the scalar fields in the trace by their classical solutions (\ref{eq:solemoYYb})
\begin{align}
\mathcal{O}(y)\mapsto\mathcal{O}^S(y)=\frac{(-1)^{L}}{k^{2L}}\sum_{\{i,j\}}\prod_{s=1}^L\frac{t_{i_s}t_{j_s}\bar{n}_{i_s}^{A_s}\,{n}_{j_s,B_s}}{|y-x_{i_s}||y-x_{j_s}|}
\prod_{s=1}^L\,\chi_{i_s}\bar{\eta}_{i_s}\eta_{j_s}\bar{\chi}_{j_s}\comma
\end{align}
where $\sum_{\{i,j\}}$ denotes a summation of $i_s$'s and $j_s$'s from 1 to $m$. Then we take the expectation value
\begin{align}
\label{eq:expectationOS}
\langle\mathcal{O}^S\rangle_{\chi,\eta}=\frac{(-1)^{L}}{k^{2L}}\sum_{\{i,j\}}\prod_{s=1}^L\frac{t_{i_s}t_{j_s}\bar{n}_{i_s}^{A_s}\,{n}_{j_s,B_s}}{|y-x_{i_s}||y-x_{j_s}|}
\times\big< \chi_{i_1}\bar{\eta}_{i_1}\eta_{j_1}\bar{\chi}_{j_1}\chi_{i_2}\bar{\eta}_{i_2}\eta_{j_2}\bar{\chi}_{j_2}\cdots\big>_{\chi,\eta}\period
\end{align}
The expectation value of $\chi$'s and $\eta$'s can be computed by the Wick contractions, which are given by
\begin{align}
\label{eq:propfermion}
\langle\bar{\eta}_i^{a}\,\eta_{j,b}\rangle=\delta^{a}_{b}\left(\Sigma_{\eta}^{-1}\right)_{ij}\comma\qquad \langle\bar{\chi}_i^{a}\,\chi_{j,b}\rangle=\delta^{a}_{b}\left(\Sigma_{\chi}^{-1}\right)_{ij}\comma
\end{align}
where $a$ and $b$ are gauge indices while $i$ and $j$ run from $1$ to $m$ (= the number of giant gravitons).
\paragraph{Large $N$ limit.} To proceed, we consider the large $N$ limit. There are two simplifications in the large $N$ limit. Firstly, the integrations over $\rho$ and $\bar{\rho}$ are  dominated by the saddle points of   the action $S_{\text{eff}}$ in (\ref{eq:GmSeff}). We denote the saddle point solutions by $\rho^*$ and $\bar{\rho}^*$ and the propagators (\ref{eq:propfermion}) can be evaluated at the saddle point. Secondly, the Wick contraction in (\ref{eq:expectationOS}) is dominated by the nearest neighboring contraction (see \cite{Jiang:2019xdz} for more detailed explanation). For instance, if we neglect the prefactors in \eqref{eq:expectationOS}, the Wick contraction gives
\begin{align}
\label{eq:WickContraction}
\begin{aligned}
\big< \chi_{i_1}\bar{\eta}_{i_1}\eta_{j_1}\bar{\chi}_{j_1}\chi_{i_2}\bar{\eta}_{i_2}\eta_{j_2}\bar{\chi}_{j_2}\cdots\big>
\overset{N\to\infty}{=}&-\langle\bar{\eta}_{i_1}\eta_{j_1}\rangle\langle \bar{\chi}_{j_1}\chi_{i_2}\rangle\ldots\langle\bar{\eta}_{i_L}\eta_{j_L}\rangle
\langle \bar{\chi}_{j_L}\chi_{i_1}\rangle\\
=&-N^{2L}\text{tr}_m\big[\Sigma_{\eta}^{-1}\Sigma_{\chi}^{-1}\cdots \Sigma_{\eta}^{-1}\Sigma_{\chi}^{-1}\big]\period
\end{aligned}
\end{align}
Reinstating the prefactors, (\ref{eq:expectationOS}) can be written as
\begin{align}
\langle\mathcal{O}^S\rangle_{\chi,\eta}=-\text{tr}_m\big[T^{A_1}\bar{T}_{B_1}\cdots T^{A_L}\bar{T}_{B_L}\big]\comma
\end{align}
where
\begin{align}
T^A=&\,-\text{diag}\left(\frac{\lambda t_1 \bar{n}_1^{A}}{|y-x_1|},\frac{\lambda t_2\bar{n}_2^{A}}{|y-x_2|}\cdots,\frac{\lambda t_m\bar{n}_m^A}{|y-x_m|}\right)\cdot\Sigma_{\eta}^{-1}\comma\\ \nonumber
\bar{T}_B=&\,\text{diag}\left(\frac{\lambda t_1{n}_{1,B}}{|y-x_1|},\frac{\lambda t_2{n}_{2,B}}{|y-x_2|}\cdots,\frac{\lambda t_m{n}_{m,B}}{|y-x_m|}\right)\cdot\Sigma_{\chi}^{-1}\comma
\end{align}
are $m\times m$ matrices.
Alternatively, we can express this as an overlap between a spin-chain state $|A_1\bar{B}_1\cdots A_L \bar{B}_L\rangle$ and a matrix product state $|{\rm MPS}\rangle$ defined by
\begin{align}
    \langle\mathcal{O}^S\rangle_{\chi,\eta}=\langle {\rm MPS}|A_1\bar{B}_1\cdots A_L \bar{B}_L\rangle\comma
\end{align}
with
\begin{align}\label{eq:MPSdef}
    \langle {\rm MPS}|\equiv-\sum_{\{A_s,B_s\}}\langle A_1\bar{B}_1\cdots A_L \bar{B}_L|\text{tr}_m\big[T^{A_1}\bar{T}_{B_1}\cdots T^{A_L}\bar{T}_{B_L}\big]\period
\end{align}
In deriving these expressions, we implicitly assumed that there is no flat direction around the saddle point. However, this is not always the case as we see below for $m=2$. See section \ref{subsec:twopnt} for discussions on this point.

Physically the saddle-point value of $e^{NS_{\rm eff}}$ gives the $m$-point function of giant gravitons while the expectation value $\langle \mathcal{O}^{S}\rangle_{\chi,\eta}$ gives a ratio of the $(m+1)$-point function and the $m$-point function,
\begin{align}\label{eq:ratiofermiexp}
    \langle \mathcal{O}^{S}\rangle_{\chi,\eta} =\frac{\left<\left(\prod_{j=1}^{m}\mathcal{D}_{M_j}\right)\mathcal{O}(y)\right>}{\left<\prod_{j=1}^{m}\mathcal{D}_{M_j}\right>}\period
\end{align}
\subsection{Two-point function: saddle point and MPS} \label{subsec:twopnt}

To find the solution of saddle point equations, we focus on the case of interest with two giant gravitons, namely we take $m=2$. We furthermore set the charges of the two giant gravitons to be identical $M_1=M_2=M$ and express the ratio $M/N$ by
\begin{align}
    \omega \equiv \frac{M}{N}\period
\end{align}
Note that $\omega$ takes values from $0$ to $1$ and the case with $\omega=1$ corresponds to the maximal giant graviton.

Setting $m=2$ in \eqref{eq:Seff}, we find that the effective action is given by
\begin{align}
S_{\text{eff}}=-\frac{1}{\lambda}(\rho_{12}\bar{\rho}_{21}+\rho_{21}\bar{\rho}_{12})+\log \left[(1-t_1t_2\sqrt{d_{12}d_{21}}\rho_{12}\rho_{21})(1-t_1t_2\sqrt{d_{12}d_{21}}\bar{\rho}_{12}\bar{\rho}_{21})\right]\period
\end{align}
Taking variation with respect to $\rho$'s and $\bar{\rho}$'s and imposing $\frac{\delta S_{\text{eff}}}{\delta\rho_{i,j}}=\frac{\delta S_{\text{eff}}}{\delta\bar{\rho}_{i,j}}=0$, we obtain
\begin{align} \label{eq:rhosaddles} \rho_{12}^{\ast}\bar{\rho}^{\ast}_{21}=\rho_{21}^{\ast}\bar{\rho}^{\ast}_{12}=\lambda\frac{t_1 t_2\sqrt{d_{12}d_{21}}\rho^{\ast}_{12}\rho^{\ast}_{21}}{t_1 t_2\sqrt{d_{12}d_{21}}\rho^{\ast}_{12}\rho^{\ast}_{21}-1}=\lambda\frac{t_1 t_2\sqrt{d_{12}d_{21}}\bar{\rho}^{\ast}_{12}\bar{\rho}^{\ast}_{21}}{t_1 t_2\sqrt{d_{12}d_{21}}\bar{\rho}^{\ast}_{12}\bar{\rho}^{\ast}_{21}-1}\period
\end{align}
Solving these equations, we find\footnote{In fact there is another solution to the saddle point equations,
\begin{align}
 \rho^{\ast}_{ij}&=-\bar{\rho}_{ij}^{\ast}\comma\qquad \rho_{12}^{\ast}\rho_{21}^{\ast}=\bar{\rho}^{\ast}_{12}\bar{\rho}^{\ast}_{21}=-\lambda+\frac{1}{t_1 t_2\sqrt{d_{12}d_{21}}}\period
\end{align}
However it  will eventually give the same saddle point action and the same MPS (after the integration over $\theta$ which will be defined soon), and therefore do not modify our result for the ratio of correlators. Then we will focus only on the first solution in the rest of the paper.}

\begin{align}
    \rho^{\ast}_{ij}=\bar{\rho}_{ij}^{\ast}\comma\qquad \rho_{12}^{\ast}\rho_{21}^{\ast}=\bar{\rho}^{\ast}_{12}\bar{\rho}^{\ast}_{21}=\lambda+\frac{1}{t_1 t_2\sqrt{d_{12}d_{21}}}\period
\end{align}
The effective action at this saddle point is given by
\begin{align}
S_{\text{eff}}^*=-2-\frac{2}{\lambda t_1 t_2\sqrt{d_{12}d_{21}}}+\log \left[\lambda^2t_1^2t_2^2d_{12}d_{21}\right] \period
\end{align}

To compute the correlation function of giant gravitons with fixed charges, we need to perform integrals of $t$'s \eqref{eq:projection}. Since $M$ is $\mathcal{O}(N)$, some of those integrals can be evaluated using the saddle-point approximation. Solving the saddle-point equations
\begin{align}
    2\omega\frac{\del \log t_{1,2}}{\del t_{1,2}}-\frac{\del S_{\rm eff}^{\ast}}{\del t_{1,2}}=0\comma
\end{align}
we get
\begin{align}\label{eq:tsaddles}
    t_1^{\ast}t^{\ast}_{2}=\frac{1}{\lambda (\omega-1)\sqrt{d_{12}d_{21}}}\period
\end{align}
As is clear from this expression, the saddle-point equation only determines the product $t_{1}t_2$. Therefore, we still need to perform an integral of the ratio
\begin{align}\label{eq:theta}
    e^{i\theta}\equiv\sqrt{\frac{t_1}{t_2}}\period
\end{align}
as we see below.

\paragraph{MPS for two giant gravitons.}
Let us place the single trace operator at $y=x_3$,
\begin{align}
\mathcal{O}(y)=\tr(Y^{A_1}\bar{Y}_{B_1}\ldots Y^{A_L}\bar{Y}_{B_L})(x_3)\comma
\end{align}
 and compute the three-point function of two giant gravitons and $\mathcal{O}$.

Evaluating the matrices $T^A$ and $\bar{T}_B$ at the saddle point \eqref{eq:rhosaddles} and \eqref{eq:tsaddles}, we get
\begin{align}\label{eq:Kijdef}
    \begin{aligned}
    T^{A}&=-\lambda\,{\rm diag}\left(\tfrac{t_1^{\ast}\bar{n}_1^{A}}{|x_{13}|},\tfrac{t_2^{\ast}\bar{n}_2^{A}}{|x_{23}|}\right)\cdot \pmatrix{cc}{1-\omega&-i\rho^{\ast}_{12}\sqrt{\frac{1-\omega}{\lambda}}\left(\frac{d_{12}}{d_{21}}\right)^{\frac{1}{4}}\\-i\rho_{21}^{\ast}\sqrt{\frac{1-\omega}{\lambda}}\left(\frac{d_{21}}{d_{12}}\right)^{\frac{1}{4}}&1-\omega}\comma\\
    \bar{T}_{B}&=\lambda\,{\rm diag}\left(\tfrac{t_1^{\ast}n_{1,B}}{|x_{13}|},\tfrac{t_2^{\ast}n_{2,B}}{|x_{23}|}\right)\cdot \pmatrix{cc}{1-\omega&-i\rho^{\ast}_{12}\sqrt{\frac{1-\omega}{\lambda}}\left(\frac{d_{21}}{d_{12}}\right)^{\frac{1}{4}}\\-i\rho_{21}^{\ast}\sqrt{\frac{1-\omega}{\lambda}}\left(\frac{d_{12}}{d_{21}}\right)^{\frac{1}{4}}&1-\omega}\period
    \end{aligned}
\end{align}
To proceed, we make use of the fact that the MPS \eqref{eq:MPSdef} is invariant under the transformation $T^{A}\mapsto U T^{A}U^{-1}$ and $\bar{T}_B\mapsto U\bar{T}_B U^{-1}$. By choosing an appropriate $U$,  we can replace $\rho_{12}^{\ast}$ and $\rho_{21}^{\ast}$ in \eqref{eq:Kijdef} with $\sqrt{\rho_{12}^{\ast}\rho_{21}^{\ast}}$. As a result, we obtain the following matrices
\begin{align}\label{eq:matricesMPSrep}
    \begin{aligned}
        T^{A}_{\theta}&=\frac{\sqrt{\lambda}}{(d_{12}d_{21})^{\frac{1}{4}}}\pmatrix{cc}{-\frac{\bar{n}_1^{A}e^{i\theta}}{|x_{13}|}\sqrt{1-\omega}&i\frac{\bar{n}_1^{A}e^{i\theta}}{|x_{13}|}\sqrt{\omega}\left(\frac{d_{12}}{d_{21}}\right)^{\frac{1}{4}}\\i\frac{\bar{n}_2^{A}e^{-i\theta}}{|x_{23}|}\sqrt{\omega}\left(\frac{d_{21}}{d_{12}}\right)^{\frac{1}{4}}&-\frac{\bar{n}_2^{A}e^{-i\theta}}{|x_{23}|}\sqrt{1-\omega}}\comma\\
        \bar{T}_{B,\theta}&=\frac{\sqrt{\lambda}}{(d_{12}d_{21})^{\frac{1}{4}}}\pmatrix{cc}{\frac{n_{1,B}e^{i\theta}}{|x_{13}|}\sqrt{1-\omega}&-i\frac{n_{1,B}e^{i\theta}}{|x_{13}|}\sqrt{\omega}\left(\frac{d_{21}}{d_{12}}\right)^{\frac{1}{4}}\\-i\frac{n_{2,B}e^{-i\theta}}{|x_{23}|}\sqrt{\omega}\left(\frac{d_{12}}{d_{21}}\right)^{\frac{1}{4}}&\frac{n_{2,B}e^{-i\theta}}{|x_{23}|}\sqrt{1-\omega}}\period
    \end{aligned}
\end{align}
Here $e^{i\theta}$ is the ratio of $t$'s defined in \eqref{eq:theta}. As is clear from this expression, the MPS depens on $\theta$ which parametrizes the flat direction around the saddle point. Therefore, the correct result would be given by an integral over this direction;
\begin{align}\label{eq:MPSthetaintegral}
    \langle\mathcal{O}^S\rangle_{\chi,\eta}=\int_{0}^{2\pi} \frac{d\theta}{2\pi}\langle {\rm MPS}_{\theta}|A_1\bar{B}_1\cdots A_L \bar{B}_L\rangle\comma
\end{align}
with
\begin{align}\label{eq:MPSthetarep}
    \langle {\rm MPS}_{\theta}|\equiv-\sum_{\{A_s,B_s\}}\langle A_1\bar{B}_1\cdots A_L \bar{B}_L|\text{tr}_m\big[T^{A_1}_{\theta}\bar{T}_{B_1,\theta}\cdots T^{A_L}_{\theta}\bar{T}_{B_L,\theta}\big]\period
\end{align}

\subsection{Structure constants of BPS single-trace operators}
Let us now use the results above to derive an expression for the structure constant of the single-trace operator.

\paragraph{Integral representation.}We first consider the BPS single-trace operator, given by
\begin{align}
\mathcal{O}^{\circ}_{L}(x_3)=\tr[(n_3\!\cdot\!Y\,\bar{n}_3\!\cdot\!\bar{Y})^L](x_3)\period
\end{align}
Using the results above, we find
\begin{align}\label{eq:BPSlengthLresult}
    \frac{\langle\mathcal{D}_M\mathcal{D}_M\mathcal{O}^{\circ}_{L}(x_3)\rangle}{\langle\mathcal{D}_M\mathcal{D}_M\rangle}=\int^{2\pi}_{0}\frac{d\theta}{2\pi}{\rm tr}_{2}\left[(\mathcal{T}_\theta \bar{\mathcal{T}}_{\theta})^{L}\right]\comma
\end{align}
with
\begin{align}
    \begin{aligned}
        \mathcal{T}_{\theta}&\equiv \sum_{A}n_{3,A} T_{\theta}^{A}=\frac{\sqrt{\lambda}}{(d_{12}d_{21})^{\frac{1}{4}}}\pmatrix{cc}{-d_{31}e^{i\theta}\sqrt{1-\omega}&id_{31}e^{i\theta}\sqrt{\omega}\left(\frac{d_{12}}{d_{21}}\right)^{\frac{1}{4}}\\id_{32}e^{-i\theta}\sqrt{\omega}\left(\frac{d_{21}}{d_{12}}\right)^{\frac{1}{4}}&-d_{32}e^{-i\theta}\sqrt{1-\omega}}\comma\\
        \bar{\mathcal{T}}_{\theta}&\equiv \sum_{B}\bar{n}_{3}^{B} \bar{T}_{\theta,B}=\frac{\sqrt{\lambda}}{(d_{12}d_{21})^{\frac{1}{4}}}\pmatrix{cc}{d_{13}e^{i\theta}\sqrt{1-\omega}&-id_{13}e^{i\theta}\sqrt{\omega}\left(\frac{d_{21}}{d_{12}}\right)^{\frac{1}{4}}\\-id_{23}e^{-i\theta}\sqrt{\omega}\left(\frac{d_{12}}{d_{21}}\right)^{\frac{1}{4}}&d_{23}e^{-i\theta}\sqrt{1-\omega}}\period
    \end{aligned}
\end{align}
To evaluate \eqref{eq:BPSlengthLresult} it is convenient to consider a {\it generating function of structure constants},
\begin{align}\label{eq:forapB1}
    \sum_{L=0}^{\infty}s^{L}\frac{\langle\mathcal{D}_M\mathcal{D}_M\mathcal{O}^{\circ}_{L}(x_3)\rangle}{\langle\mathcal{D}_M\mathcal{D}_M\rangle}=-\int^{2\pi}_{0}\frac{d\theta}{2\pi}{\rm tr}_2\left[\frac{1}{1-s \mathcal{T}_{\theta}\bar{\mathcal{T}}_{\theta}}\right]\period
\end{align}
To proceed, we evaluate the right hand side by diagonalizing the matrices $\mathcal{T}_{\theta}\bar{\mathcal{T}}_{\theta}$, perform the integral, and read off the coefficient in front of $s^{L}$.\footnote{Here it is important to keep $s$ being  small to make the  power series convergent in \eqref{eq:forapB1}. We will give the details of the computations  in Appendix \ref{sec:derBPS}.} The result reads
\begin{align}
\begin{aligned}\label{eq:forapB2}
    &\frac{\langle\mathcal{D}_M\mathcal{D}_M\mathcal{O}^{\circ}_{L}(x_3)\rangle}{\langle\mathcal{D}_M\mathcal{D}_M\rangle}=-\lambda^{L}\left(\frac{d_{23}d_{32}d_{31}d_{13}}{d_{12}d_{21}}\right)^{\frac{L}{2}}\xi^{-\frac{L}{2}}\oint\frac{ds}{2\pi i s^{1+L}}\frac{1-\xi s^2 }{\sqrt{\mathcal{P}(s)}}\comma
    \end{aligned}
\end{align}
where $\mathcal{P}(s)$ is given by
\begin{align}
\begin{aligned}
    \mathcal{P}(s)\equiv& 1-2 \omega (1+\xi)s+\left[-2\xi+8\xi \omega+\omega^2(1-\xi)^2\right]s^2
    -2 \omega\xi (1+\xi)s^3+\xi^2 s^4\period
    \end{aligned}
\end{align}
Dividing this by the normalization factor $\mathcal{N}_{\mathcal{O}_{L}^{\circ}}$ and stripping off the kinematic factors, we arrive at the following expression for the structure constants;
\begin{align}\label{eq:BPSstructureintegral}
    \sum_{p=-\frac{L}{2}}^{\frac{L}{2}}\xi^{p}\,D_{M|L}^{(p)}=-\frac{\xi^{-\frac{L}{2}}}{\sqrt{L}}\oint\frac{ds}{2\pi i s^{1+L}}\frac{1-\xi s^2 }{\sqrt{\mathcal{P}(s)}}\period
\end{align}

For a given $L$, the integral \eqref{eq:BPSstructureintegral} can be readily evaluated. For instance, for small values of $L$, we obtain
\begin{align}
    \begin{aligned}
     L=1:&\quad\sum_{p=-\frac{1}{2}}^{\frac{1}{2}}\xi^{p}\,D_{M|1}^{(p)}=-\omega\left(\xi^{-\frac{1}{2}}+\xi^{\frac{1}{2}}\right)\comma\\
     L=2:&\quad\sum_{p=-1}^{1}\xi^{p}\,D_{M|2}^{(p)}=-\frac{\omega^2}{\sqrt{2}} (\xi^{-1}+4(1-\omega^{-1})+\xi)\comma\\
     L=3:&\quad\sum_{p=-\frac{3}{2}}^{\frac{3}{2}}\xi^{p}\,D_{M|3}^{(p)}=-\frac{\omega^{3}}{\sqrt{3}}\left[\xi^{-\frac{3}{2}}+\frac{3(1-\omega)(1-3\omega)}{\omega^2}\left(\xi^{-\frac{1}{2}}+\xi^{\frac{1}{2}}\right)+\xi^{\frac{3}{2}}\right]\period
    \end{aligned}
\end{align}
\paragraph{Closed-form expressions for special cases.}It is however difficult to write down a closed-form expression for general $L$, $\xi$ and $\omega$. The exceptions are for the maximal giant graviton ($\omega=1$) and for the twisted-translated kinematics ($\xi=-1$). The former is given by
\begin{align}
    \sum_{p=-\frac{L}{2}}^{\frac{L}{2}}\xi^{p}\,D_{N|L}^{(p)}=-\frac{1}{\sqrt{L}}\left(\xi^{-\frac{L}{2}}+\xi^{\frac{L}{2}}\right)\comma
\end{align}
which is equivalent to
\begin{align}
    D_{N|L}^{(p)}=-\frac{1}{\sqrt{L}}\left(\delta_{p,\frac{L}{2}}+\delta_{p,-\frac{L}{2}}\right)\period
\end{align}

On the other hand, in the twisted-translated kinematics $(\xi=-1)$, we have
\begin{align}
\begin{aligned}
    &\mathfrak{D}_{M|L}=\sum_{p=-\frac{L}{2}}^{\frac{L}{2}}(-1)^{p}\,D_{M|L}^{(p)}=\\
    &\begin{cases}-\frac{(-1)^{\frac{L}{2}}}{\sqrt{L}}\left[P_{\frac{L}{2}}\left(-1+4\omega-2\omega^2\right)+P_{\frac{L}{2}-1}\left(-1+4\omega-2\omega^2\right)\right]\qquad &L:\text{ even}\\
    0\qquad &L:\text{ odd}
    \end{cases}\comma
    \end{aligned}
\end{align}
where $P_{\frac{L}{2}}$ is the Legendre polynomial.
\subsection{Structure constants of non-BPS single-trace operators} We next study the structure constant of the non-BPS single-trace operator.
As mentioned in section \ref{sec:setup}, in this paper we focus on the twisted-translated frame, which amounts to setting
\begin{align}
    n_{1,2}=(1,0,0,\kappa a_{1,2})\comma\qquad \bar{n}_{1,2}=(-\kappa a_{1,2},0,0,1)\comma \qquad x_{1,2}=(0,a_{1,2},0)\period
\end{align}

\paragraph{Result for the structure constant.}To further simplify the analysis, we consider a symmetric configuration in which the single-trace operator is at the origin and the giant gravitons are placed symmetrically around it; namely we set $a_{1}=-a_2=+1$ and $a_3=0$ in \eqref{eq:setupfinal}. Then the matrices that enter in the MPS representation, \eqref{eq:matricesMPSrep}, simplify to the following,
\begin{align}
\begin{aligned}
T^{1}_{\theta}&=\sqrt{\kappa\lambda} e^{-i\frac{\pi}{4}}\pmatrix{cc}{e^{i\theta} \sqrt{1-\omega}&-ie^{i(\theta-\frac{\pi}{4})}\sqrt{\omega}\\ie^{-i(\theta-\frac{\pi}{4})}\sqrt{\omega}&-e^{-i\theta} \sqrt{1-\omega}}\comma\\ \bar{T}_{4,\theta}&=\sqrt{\kappa\lambda} e^{-i\frac{\pi}{4}}\pmatrix{cc}{e^{i\theta} \sqrt{1-\omega}&-ie^{i(\theta+\frac{\pi}{4})}\sqrt{\omega}\\ie^{-i(\theta+\frac{\pi}{4})}\sqrt{\omega}&-e^{-i\theta} \sqrt{1-\omega}}\comma\\
T^{4}_{\theta}&=\sqrt{\frac{\lambda}{\kappa}} e^{-i\frac{\pi}{4}}\pmatrix{cc}{-e^{i\theta} \sqrt{1-\omega}&ie^{i(\theta-\frac{\pi}{4})}\sqrt{\omega}\\ie^{-i(\theta-\frac{\pi}{4})}\sqrt{\omega}&-e^{-i\theta} \sqrt{1-\omega}}\comma\\
\bar{T}_{1,\theta}&=\sqrt{\frac{\lambda}{\kappa}} e^{-i\frac{\pi}{4}}\pmatrix{cc}{e^{i\theta} \sqrt{1-\omega}&-ie^{i(\theta+\frac{\pi}{4})}\sqrt{\omega}\\-ie^{-i(\theta+\frac{\pi}{4})}\sqrt{\omega}&e^{-i\theta} \sqrt{1-\omega}}\comma
\end{aligned}
\end{align}
and all the other $T^{A}_{\theta}$'s and $\bar{T}_{B,\theta}$ are zero.

To compute the structure constant $\mathfrak{D}_{M|\mathcal{O}}$, we then compute the matrix trace in \eqref{eq:MPSthetarep}, perform the $\theta$ integral in \eqref{eq:MPSthetaintegral} and divide the result by the normalization $\mathcal{N}_{\mathcal{O}}$. As discussed in \cite{Escobedo:2010xs}, $\mathcal{N}_{\mathcal{O}}$ can be expressed in terms of the norm of the spin-chain state $\langle \mathcal{O}|\mathcal{O}\rangle$ as follows
\begin{align}
    \mathcal{N}_{\mathcal{O}}=L\lambda^{2L}\langle\mathcal{O}|\mathcal{O} \rangle\comma
\end{align}
where the prefactor $L$ is the number of different Wick contractions related by cyclic permutations (see \cite{Escobedo:2010xs} for details) and $\lambda^{2L}$ comes from the normalization of propagators. Finally, factoring out the kinematic factor in \eqref{eq:setupfinal}, we arrive at the expression,
\begin{align}\label{eq:finalstructureconst}
    \mathfrak{D}_{M|\mathcal{O}}=\frac{(-1)^{J+1}}{2^{\Delta-J}\sqrt{L\langle \mathcal{O}|\mathcal{O}\rangle}}\int_{0}^{2\pi} \frac{d\theta}{2\pi}\langle B_{\theta}|\mathcal{O}\rangle\comma
\end{align}
where the matrix product state $\langle B_{\theta}|$ is defined by
\begin{align}
    \langle B_{\theta}|\equiv \sum_{A_s,B_s=1,4}\langle A_1\bar{B}_1\cdots A_{L}\bar{B}_L|{\rm tr}_2\left[{\sf t}^{A_1}\bar{\sf t}_{B_1}\cdots {\sf t}^{A_L}\bar{\sf t}_{B_L}\right]\comma
\end{align}
with\fn{Recall that $\omega$ is related to the charge of the giant graviton $M$ by $\omega\equiv M/N$.}
\begin{align}\label{eq:tmatrices}
\begin{aligned}
{\sf t}^{1}&=\pmatrix{cc}{e^{i\theta} \sqrt{1-\omega}&-i\sqrt{\omega}\\i\sqrt{\omega}&-e^{-i\theta} \sqrt{1-\omega}}\comma\qquad \bar{\sf t}_{4}= \pmatrix{cc}{e^{i\theta} \sqrt{1-\omega}&\sqrt{\omega}\\\sqrt{\omega}&-e^{-i\theta} \sqrt{1-\omega}}\comma\\
{\sf t}^{4}&=\pmatrix{cc}{-e^{i\theta} \sqrt{1-\omega}&i\sqrt{\omega}\\i\sqrt{\omega}&-e^{-i\theta} \sqrt{1-\omega}}\comma\qquad
\bar{\sf t}_{1}=\pmatrix{cc}{e^{i\theta} \sqrt{1-\omega}&\sqrt{\omega}\\-\sqrt{\omega}&e^{-i\theta} \sqrt{1-\omega}}\period
\end{aligned}
\end{align}

\paragraph{Simplification for the maximal giant graviton.} For the maximal giant gravitons $\omega=1$, the matrices in \eqref{eq:tmatrices} all become off-diagonal, and the products of ${\sf t}$ and $\bar{\sf t}$ take the following simple form;
\begin{align}
    \begin{aligned}
        &{\sf t}^1\bar{\sf t}_4=\pmatrix{cc}{-i&0\\0&i}\comma\qquad &&{\sf t}^1\bar{\sf t}_1=\pmatrix{cc}{i&0\\0&i}\comma\\
        &{\sf t}^4\bar{\sf t}_4=\pmatrix{cc}{i&0\\0&i}\comma\qquad &&{\sf t}^4\bar{\sf t}_1=\pmatrix{cc}{-i&0\\0&i}\period
    \end{aligned}
\end{align}
Since these matrices do not depend on $\theta$, the integral of $\theta$ can be trivially performed. As a result, we can replace \eqref{eq:finalstructureconst} with the following expression
\begin{align}\label{eq:maximalmps}
    \mathfrak{D}_{N|\mathcal{O}}=-\frac{(-i)^{J}}{2^{\Delta-J}} \frac{\langle \mathcal{B}|\mathcal{O}\rangle}{\sqrt{L\langle \mathcal{O}|\mathcal{O}\rangle}}\comma
\end{align}
where the state $\langle \mathcal{B}|$ is defined by
\begin{align}
    \langle \mathcal{B}|\equiv \sum_{A_s,B_s=1,4}\langle A_1\bar{B}_1\cdots A_L\bar{B}_L|\left(1+(-1)^{J}\right)\period
\end{align}
Here $J$ is the $U(1)$ R-charge, which counts
\begin{align}
    J=L-(\text{number of $4$ on odd sites})-(\text{number of $\bar{1}$ on even sites})\period
\end{align}
We can also check that the result \eqref{eq:maximalmps} matches the one computed from a different approach explained in Appendix \ref{sec:PCGG}.

In the rest of this paper, we use \eqref{eq:finalstructureconst} and \eqref{eq:maximalmps} to evaluate the structure constants of non-BPS single-trace operators. For this purpose, we need to compute two quantities in the spin chain; the overlap $\langle B|\mathcal{O}\rangle$ and the norm $\langle \mathcal{O}|\mathcal{O}\rangle$. In the next section, we present conjectures for these observables based on the coordinate Bethe ansatz we developed in section \ref{sec:CBA}.

%%%%%%%%%%%%%%%%%%%%%%%%%%%%%%%%%%%%%%%%%%%%%%%%%%%%%%%

%%%%%%%%%%%%%%%%%%%%%%%%%%%%%%%%%%%%%%%%%%%%%%%%%%%%%

\section{Main Results}
\label{sec:result}
In this section, we use the coordinate Bethe ansatz in section \ref{sec:CBA} and evaluate the overlap $\langle B|\mathcal{O}\rangle$ and the norm $\langle \mathcal{O}|\mathcal{O}\rangle$. For reader's convenience, we first present a summary of results and later explain the details. The numerical data used for checking our formula, such as the solutions to the Bethe equations and the results for the overlap, is summarized in Appendix \ref{sec:numerics}.

In what follows, we label a Bethe state in terms of its Bethe roots; namely we express $|\mathcal{O}\rangle$ as $|{\bf u},{\bf w}, {\bf v}\rangle$ where ${\bf u}$, ${\bf w}$ and ${\bf v}$ are sets of Bethe roots for each Dynkin node (see figure \ref{fig:dynkin}).
\subsection{Summary of results}
\paragraph{Selection rules for the overlap.}
As a result of the numerical experiments, we found a set of selection rules in order for the overlap to be non-zero. The selection rules apply both to the maximal giant graviton $\langle \mathcal{B}|\mathcal{O}\rangle$ and the non-maximal giant gravitons $\int \frac{d\theta}{2\pi}\langle B_{\theta}|\mathcal{O}\rangle$.

The first and the most obvious selection rule is
\begin{enumerate}
    \item[0.] The numbers of Bethe roots for each node need to satisfy
    \begin{align}
        K_{\bf u}=K_{\bf v}=K_{\bf w}\period
    \end{align}
\end{enumerate}
This simply follows from the fact that $\langle B_{\theta}|$ only contains fields $Y^{1}$, $Y^{4}$, $\bar{Y}_1$ and $\bar{Y}_4$: In order for $|{\bf u}, {\bf w}, {\bf v}\rangle$ to contain kets involving $Y^{1}$, $Y^{4}$, $\bar{Y}_1$, $\bar{Y}_4$ only, we need to set $K_{\bf u}=K_{\bf v}=K_{\bf w}$. Note that this is just a consequence of the global symmetry, not the integrability of the matrix product state.

In addition to this constraint,  the overlap also obeys the following ``parity conditions":
\begin{enumerate}
    \item The $U(1)$ R-charge of the state, $J= L-\frac{K_{\bf u}+K_{\bf v}}{2}$, must be even.
    \item The rapidities of the right node must be $(-1)$ times the rapidities of the left node:
    \begin{align}
        {\bf v}=-{\bf u}\period
    \end{align}
    \item The rapidities of the middle node should be parity-symmetric, namely
    \begin{align}
        \begin{aligned}
            {\bf w}=\begin{cases}(w_1,-w_1,w_2,-w_2,\ldots)\qquad &K_{\bf w}:{\rm even}\comma\\
            (w_1,-w_1,w_2,-w_2,\ldots,0)\qquad &K_{\bf w}:{\rm odd}\period
            \end{cases}
        \end{aligned}
    \end{align}
\end{enumerate}
These selection rules are almost identical to the selection rules found for the structure constant of two giant gravitons and a single-trace operator in the $SO(6)$ sector of $\mathcal{N}=4$ SYM \cite{Jiang:2019xdz,Jiang:2019zig}. The only difference is the roles of the rapidities: In ABJM, the left and the right nodes are momentum carrying nodes and the middle node is auxiliary while in $\mathcal{N}=4$ SYM, the middle node is momentum carrying and the left and right nodes are auxiliary.

As discussed in \cite{Jiang:2019xdz,Piroli:2017sei,Pozsgay:2019,Ghoshal:1993tm}, the existence of these selection rules indicate that the matrix product states $\langle \mathcal{B}|$ and $\int \frac{d\theta}{2\pi}\langle B_{\theta}|$ are integrable boundary states. In particular, we want to emphasize that the non-maximal giant gravitons also satisfy the selection rule, implying that they correspond to integrable boundary states as well.

\paragraph{Determinant formula for the maximal giant graviton.} For the maximal giant graviton $(\omega=1)$, we found a closed-form expression for the structure constant. Before writing down the result, let us first clarify our convention and ordering of the rapidities satisfying the selection rules;
\begin{align}\label{eq:rapconvention}
    \begin{aligned}
        {\bf u}&=(u_1,u_2,\ldots, u_{K_{\bf u}})\comma\\
        {\bf v}&=-{\bf u}=(-u_1,-u_2,\ldots, -u_{K_{\bf u}})\comma\\
        {\bf w}&=\begin{cases}(w_1,-w_1,w_2,-w_2,\ldots, w_{\frac{K_{\bf w}}{2}},-w_{\frac{K_{\bf w}}{2}})\qquad &K_{\bf w}:\text{ even}\\
        (w_1,-w_1,w_2,-w_2,\ldots, w_{\lceil\frac{ K_{\bf w}}{2}\rceil },-w_{\lceil \frac{K_{\bf w}}{2}\rceil},0)\qquad &K_{\bf w}:\text{ odd}\end{cases}\comma
    \end{aligned}
\end{align}
where $\lceil \frac{M}{2}\rceil$ means the largest integer not larger than $\frac{M}{2}$.

The result for the normalized overlap reads
\begin{align}\label{eq:mainresult1}
    \frac{\langle \mathcal{B}|{\bf u},{\bf w},-{\bf u}\rangle}{\sqrt{\langle {\bf u},{\bf w},-{\bf u}|{\bf u},{\bf w},-{\bf u}\rangle}}=2\sqrt{\prod_{j=1}^{K_{\bf u}}\left(u_j^2+\tfrac{1}{4}\right)\prod_{k=1}^{\lceil\frac{K_{\bf w}}{2}\rceil}\frac{1}{w_k^2(w_k^2+\frac{1}{4})}\frac{\det G_{+}}{\det G_{-}}}\period
\end{align}
Here $\det G_{\pm}$ are the Gaudin-like determinants which we will explain in the next subsection.
The result leads to the following expression for the structure constant:
\begin{align}\label{eq:mainresult2}
    \mathfrak{D}_{N|\mathcal{O}}=\frac{i^{J}+(-i)^{J}}{2^{\Delta-J}}\sqrt{\prod_{j=1}^{K_{\bf u}}\left(u_j^2+\tfrac{1}{4}\right)\prod_{k=1}^{\lceil\frac{K_{\bf w}}{2}\rceil}\frac{1}{w_k^2(w_k^2+\frac{1}{4})}\frac{\det G_{+}}{\det G_{-}}}\period
\end{align}
In \eqref{eq:mainresult1} and \eqref{eq:mainresult2}, we neglected an overall phase factor which is ambiguous (see the discussion in section \ref{subsec:structure}).
These are the main results of this paper. Below we explain the details of the formula and the derivation.

For the non-maximal giant gravitons, we have not been able to find a closed-form expression analogous to \eqref{eq:mainresult2}. The results for a sample of Bethe states are summarized in Appendix \ref{sec:numerics}. We leave it for future investigations to find a determinant formula for the non-maximal giant gravitons.

\subsection{Norms and Gaudin determinant}
\paragraph{Norms for the coordinate Bethe states.}
The norms of the Bethe states are often given by the so-called Gaudin determinant, which are given by the logarithmic derivative of the Bethe equations. In the case at hand, the relevant Gaudin matrix is given by
\begin{align}
G=\left(
\begin{array}{ccc}
\partial_{u_i} \phi_{u_j} & \partial_{u_i}\phi_{w_j}  &\partial_{u_i} \phi_{v_j} \\
\partial_{w_i} \phi_{u_j} & \partial_{w_i} \phi_{w_j} & \partial_{w_i} \phi_{v_j}\\
\partial_{v_i} \phi_{u_j} & \partial_{v_i}\phi_{w_j}  &\partial_{v_i} \phi_{v_j} \\
\end{array}
\right)\period
\end{align}
where $\phi$'s was defined through  the  Bethe ansatz equations \eqref{eq:betheansatzeq}.

We conjecture that the norms of the Bethe states in the $SU(4)$-invariant alternating spin chain are given by
\begin{align}\label{eq:normconjecture}
    \begin{aligned}
\langle{\bf u}, {\bf w}, {\bf v} |{\bf u}, {\bf w}, {\bf v} \rangle=&\left(\prod_{i<j }\frac{S(u_i,u_j)}{S(u_i^{\ast},u_j^{\ast})}\right)^{\frac{1}{2}}\left(\prod_{i<j }\frac{S(v_i,v_j)}{S(v_i^{\ast},v_j^{\ast})}\right)^{\frac{1}{2}}\left(\prod_{i<j }\frac{S(w_i,w_j)}{S(w_i^{\ast},w_j^{\ast})}\right)^{\frac{1}{2}} \\
&\times \left(\prod_{j}\frac{1}{\del_{u}p(u_j)}\right)\left(\prod_{k}\frac{1}{\del_{v}p(v_k)}\right) \det G\comma
\end{aligned}
\end{align}
where the states are normalized using the coordinate Bethe ansatz described in section \ref{sec:CBA} and $\langle {\bf u},{\bf w}, {\bf v}|\equiv \left[|{\bf u}, {\bf w}, {\bf v}\rangle\right]^{\dagger}$. $S(u,v)$ is the  S-matrix
\beq
S(u,v)\equiv \frac{u-v-i}{u-v+i}\comma
\eeq
while $p(u)$ is the momentum of a magnon with rapidity $u$
\beq
p(u)\equiv \frac{1}{i}\log \frac{u+\frac{i}{2}}{u-\frac{i}{2}}\quad \Rightarrow \quad \del_{u}p(u)=\frac{-1}{u^2+\frac{1}{4}}\period
\eeq

The conjecture \eqref{eq:normconjecture} passes several nontrivial tests: It correctly reproduces the results for the $SU(2)\times SU(2)$ sector (see for instance \cite{Gromov:2012uv}) and it is satisfied by all the Bethe states listed in Appendix \ref{sec:numerics}.
\paragraph{Simplifications for the parity-symmetric states.} For the states satisfying the parity conditions $|{\bf u}, {\bf w}, -{\bf u}\rangle$, the formula for the norm simplifies further.

First the prefactor in \eqref{eq:normconjecture} simplifies to
\begin{align}
    \langle{\bf u}, {\bf w}, -{\bf u} |{\bf u}, {\bf w}, -{\bf u} \rangle=\prod_{k=1}^{\lceil \frac{K_{\bf w}}{2}\rceil}\frac{(w_k-\frac{i}{2})(w_k^{\ast}+\frac{i}{2})}{(w_k+\frac{i}{2})(w_k^{\ast}-\frac{i}{2})}\prod_{j=1}^{K_{\bf u}}\left(u_j^2+\frac{1}{4}\right)^2\, \det G\period
\end{align}
Note that we are using the convention and the ordering given in \eqref{eq:rapconvention}.

Second the Gaudin determinant for the parity-symmetric states factorizes into a determinants of submatrices. This factorization was already observed in $\mathcal{N}=4$ SYM, and we will show below that a similar factorization holds also for ABJM. Since the derivations for the even $K_{\bf w}$ cases is similar to the one for the odd $K_{\bf w}$ cases, we first discuss the odd $K_{\bf w}$ cases.

For this purpose, we divide the rapidities $\textbf{w}=(w_1, -w_1, w_2, -w_2,
\cdots, 0)$ into three parts, $\textbf{w}^{(+)}=(w_1, w_2, \cdots)$ denotes the ``postive" part of Bethe roots, $\textbf{w}^{(-)}=(-w_1, -w_2, \cdots)$ denotes the ``negative" part, and $\textbf{w}^{(0)}=(0)$. As the first step, We reorder the rows and columns of Gaudin determinant to express it as
\begin{align}
\label{Guadin det2}
\det G=\det \left(
\begin{array}{ccccc}
U_u & U_+  & U_0 & U_v& U_- \\
W_{+u} & W_{++} & W_{+0} & W_{+v} & W_{+-}\\
W_{0u} & W_{0+} & W_{00} & W_{0v} & W_{0-}\\
V_u & V_+  & V_0 & V_v& V_- \\
W_{-u} & W_{-+} & W_{-0} & W_{-v} & W_{--}\\
\end{array}
\right)\comma
\end{align}
with
\begin{align}
\begin{aligned}
&[U_u]_{ij}\equiv \partial_{u_i}\phi_{u_j}\comma \quad \left[U_\pm\right]_{ij}\equiv \partial_{u_i}\phi_{{w\pm}_j}\comma\quad [U_0]_{ij}\equiv \partial_{u_i}\phi_{w^{(0)}}\comma\quad
 [U_v]_{ij}\equiv\partial_{u_i}\phi_{v_j}\comma  \\  &[W_{\pm u}]_{ij}\equiv \partial_{w^{(\pm)}_i}\phi_{u_j}\comma\quad [W_{\pm\pm}]_{ij}\equiv \partial_{w^{(\pm)}_i}\phi_{w^{(\pm)}_j}\comma\quad [W_{\pm0}]_{ij}\equiv \partial_{w^{(\pm)}_i}\phi_{w^{(0)}}\comma\\
 &[W_{\pm\mp}]_{ij}\equiv \partial_{w^{(\pm)}_i}\phi_{w^{(\mp)}_j}\comma  \quad [W_{\pm v}]_{ij}\equiv \partial_{w^{(\pm)}_i}\phi_{v_j}\comma\quad [W_{0u}]_{ij}\equiv \partial_{w^{(0)}}\phi_{u_j}\comma\\
 &[W_{0\pm}]_{ij}\equiv \partial_{w^{(0)}}\phi_{w^{(\pm)}_j}\comma\quad [W_{00}]_{ij}\equiv \partial_{w^{(0)}}\phi_{w^{(0)}}\comma\quad [W_{0v}]_{ij}\equiv \partial_{w0}\phi_{v_j}\comma\\
 &
[V_u]_{ij}\equiv \partial_{v_i}\phi_{u_j}\comma \quad \left[V_\pm\right]_{ij}\equiv \partial_{v_i}\phi_{{w^{(\pm)}}_j}\comma\quad [V_0]_{ij}\equiv \partial_{v_i}\phi_{w^{(0)}} \comma\quad  [V_v]_{ij}\equiv\partial_{v_i}\phi_{v_j}\period
\end{aligned}
\end{align}
To proceed, we use the following relations that are valid when the selection rules are satisfied,
\begin{align}
\nonumber
&U_u=V_v\comma \quad U_v=V_u\comma\quad W_{\pm u}=W_{\mp v}\comma\quad  U_{\pm}=V_{\mp}\comma\quad U_0=V_0\comma\\
&W_{+\pm}=W_{-\pm}\comma\quad W_{0u}=W_{0v}\comma\quad W_{0+}=W_{0-}\comma\quad W_{+0}=W_{-0}\period
\end{align}

Then we can rewrite the determinant by adding the rows and subtracting the columns.
\begin{align}
\label{oddGaudindet}
\det G&=\det \left(
\begin{array}{ccccc}
U_u & U_+  & U_0 & U_v& U_- \\
W_{+u} & W_{++} & W_{+0} & W_{+v} & W_{+-}\\
W_{0u} & W_{0+} & W_{00} & W_{0v} & W_{0+}\\
U_v & U_-  & U_0 & U_u& U_+\\
W_{+v} & W_{+-} & W_{+0} & W_{+u} & W_{++}\\
\end{array}
\right) \\
&=\det\left(
\begin{array}{ccccc}
U_u+U_v & U_++U_-  & 2U_0 & 0& 0 \\
W_{+u}+W_{+v} & W_{++}+W_{+-} & 2 W_{+0} & 0 & 0\\
W_{0u} & W_{0+} & W_{00} & 0 & 0\\
U_v & U_-  & U_0 & U_u-U_v& U_+-U_-\\
W_{+v} & W_{+-} & W_{+0} & W_{+u}-W_{+v} & W_{++}-W_{+-}\\
\end{array}
\right)\period
\end{align}
We thus obtain the factorization formula for the Gaudin determinant of the parity-symmetric states $\det G=\det G_{+}\det G_{-}$ with
\begin{align}
\begin{aligned}
    \text{Odd }K_{\bf w}:\qquad &G_{+}=\pmatrix{ccc}{U_u+U_v & U_++U_-  & 2U_0\\W_{+u}+W_{+v} & W_{++}+W_{+-} & 2 W_{+0} \\W_{0u} & W_{0+} & W_{00}}\comma\\
   &G_{-}=\pmatrix{cc}{U_u-U_v& U_+-U_-\\W_{+u}-W_{+v} & W_{++}-W_{+-}}\period
   \end{aligned}
\end{align}

For the even $K_{\bf w}$ cases, we can repeat the derivation above simply by omitting the row and the colum involving ${\bf w}_0$. As a result we obtain the following factorization formula $\det G=\det G_{+}\det G_{-}$;
\begin{align}\label{eq:normparitysymmetric}
\begin{aligned}
    \text{Even }K_{\bf w}:\qquad &G_{+}=\pmatrix{cc}{U_u+U_v & U_++U_-\\W_{+u}+W_{+v} & W_{++}+W_{+-}}\comma\\
    &G_{-}=\pmatrix{cc}{U_u-U_v& U_+-U_-\\W_{+u}-W_{+v} & W_{++}-W_{+-}}\period
\end{aligned}
\end{align}
The same submatrices $G_{\pm}$ appear in the expression for the overlap \eqref{eq:mainresult1}.

To summarize, the norm of the parity-symmetric state is given by
\begin{align}
    \langle {\bf u},{\bf w},-{\bf u}|{\bf u},{\bf w},-{\bf u}\rangle=\prod_{k=1}^{\lceil \frac{K_{\bf w}}{2}\rceil}\frac{(w_k-\frac{i}{2})(w_k^{\ast}+\frac{i}{2})}{(w_k+\frac{i}{2})(w_k^{\ast}-\frac{i}{2})}\prod_{j=1}^{K_{\bf u}}\left(u_j^2+\frac{1}{4}\right)^2\, \det G_{+}\det G_{-}\period
\end{align}

\subsection{Some details on the overlap}
We now explain the details of how we arrive at the formula \eqref{eq:mainresult1}. For this purpose, let us first present our conjecture for the overlap $\langle \mathcal{B}|{\bf u},{\bf w},-{\bf u}\rangle$ itself:
\begin{align}
\langle\mathcal{B}|\textbf{u}, \textbf{w}, -\textbf{u}\rangle=2 (-1)^L \prod_{j=1}^M \left(u_j^2+\tfrac{1}{4}\right)\left(u_j+\frac{i}{2}\right)\prod_{k=1}^{\lceil\frac{M}{2}\rceil}\frac{1}{w_k (w_k+\frac{i}2)}\,\det G_+\period
\end{align}
We stress that the state $|\textbf{u}, \textbf{w}, -\textbf{u}\rangle$ was constructed using the nested coordinate Bethe ansatz, with the order of the magnon rapidities given by \eqref{eq:rapconvention}.

Combining this with the result for the norm of the parity-symmetric state \eqref{eq:normparitysymmetric}, we obtain
\begin{align}
&\frac{\langle\mathcal{B}|\textbf{u}, \textbf{w}, -\textbf{u}\rangle}{\sqrt{
\langle{\bf u}, {\bf w}, -{\bf u} |{\bf u}, {\bf w}, -{\bf u} \rangle}}\nonumber\\
&=2(-1)^{L}\prod_{j=1}^M \left(u_j+\frac{i}{2}\right)\prod_{k=1}^{\lceil\frac{M}{2}\rceil}\frac{1}{w_k (w_k+\frac{i}{2})}\left[\prod_{k=1}^{\lceil\frac{M}{2}\rceil}\frac{(w_k+\frac{i}{2})(w_k^{\ast}-\frac{i}{2})}{(w_k-\frac{i}{2})(w_k^{\ast}+\frac{i}{2})}
\right]^{\frac{1}{4}}
\sqrt{\frac{{\rm det}G_+}{{\rm det}G_-}}\period
\end{align}

To arrive at the formula \eqref{eq:mainresult1}, we then utilize the phase ambiguity discussed in section \ref{subsec:structure}. For instance, multiplying the following phase\footnote{
Here we used the fact that ${\bf u}$ as a set must be invariant under the complex conjugation (otherwise, the energy and the higher conserved charges generically will take complex values). Similar results for XXX and XXZ spin chains were proved in \cite{vladimirov1986proof}.}
\begin{align}
\begin{aligned}
e^{i\varphi}&\equiv \left( \prod_{k=1}^{\lceil\frac{M}{2}\rceil}\frac{(w_k+\frac{i}{2})(w_k^*+\frac{i}{2})}{(w_k-\frac{i}{2})(w_k^*-\frac{i}{2})}\right)^{\frac14}
\prod_{j=1}^M \left(\frac{u_j-\frac{i}2}{u_j^*+\frac{i}2}\right)^{\frac12}\\
&= \left( \prod_{k=1}^{\lceil\frac{M}{2}\rceil}\frac{(w_k+\frac{i}{2})(w_k^*+\frac{i}{2})}{(w_k-\frac{i}{2})(w_k^*-\frac{i}{2})}\right)^{\frac14}
\prod_{j=1}^M \left(\frac{u_j-\frac{i}2}{u_j+\frac{i}2}\right)^{\frac12}\comma
\end{aligned}
\end{align}
we get
\begin{align}
\frac{e^{i\varphi}\langle \mathcal{B}|{\bf u},{\bf w},-{\bf u}\rangle}{\sqrt{\langle {\bf u},{\bf w},-{\bf u}|{\bf u},{\bf w},-{\bf u}\rangle}}=2\sqrt{\prod_{j=1}^{K_{\bf u}}\left(u_j^2+\tfrac{1}{4}\right)\prod_{k=1}^{\lceil\frac{K_{\bf w}}{2}\rceil}\frac{1}{w_k^2(w_k^2+\frac{1}{4})}\frac{\det G_{+}}{\det G_{-}}}\comma
\end{align}
which coincides with \eqref{eq:mainresult1}.

%%%%%%%%%%%%%%%%%%%%%%%%%%%%%%%%%%%%%%%%%%%%%%%%%%%%%%%%%%%%%

%%%%%%%%%%%%%%%%%%%%%%%%%%%%%%%%%%%%%%%%%%%%%%%%%%%%%%%%%%%%%
\section{Conclusion}
\label{sec:conclusion}
In this paper, we studied the tree-level structure constants of a single-trace non-BPS operator and two sub-determinant operators (also known as giant gravitons) in ABJM theory in the planar limit. Much like in $\mathcal{N}=4$ SYM, these structure constants can be computed by overlaps between a matrix product state and a Bethe eigenstate. To evaluate them explicitly, we developed the coordinate Bethe ansatz for the alternating SU(4) spin chain. As a result of the computation, we found that the overlap obeys a selection rule similar to the ones found in defect one-point functions in $\mathcal{N}=4$ SYM. We also found a closed-form expression for the overlaps for the maximal giant gravitons.

This paper is the first installment of our studies of the structure constants of giant gravitons in ABJM theory. In the second paper \cite{secondpaper}, we will analyze these quantities at strong coupling using a holographic description. The results of this paper and the next will be used in the third paper \cite{thirdpaper} to test and verify the nonperturbative approach based on the integrable bootstrap.

There are several future directions worth pursuing. First it would be desirable to prove the determinant formula for the overlap conjectured in this paper. One possible strategy is to map it to a partition function of a lattice model (a vertex model) as was done for $\mathcal{N}=4$ SYM \cite{Foda:2015nfk}. Another possible strategy is to use the algebraic Bethe ansatz and Separation of Variables\fn{The relation between overlaps of integrable boundary states and Separation of Variables was studied also in recent papers \cite{Caetano:2020dyp,Cavaglia:2021mft} from different perspectives.} following the recent work \cite{Gombor:2021uxz}. Second the coordinate Bethe ansatz developed in this paper will be useful for computing other quantities in ABJM theory, most notably the three-point functions of single-trace operators\fn{See \cite{Bissi:2012ff} for results in the SU(2)$\times$SU(2) sector.}. Such computations will provide valuable data for developing the hexagon formalism for ABJM theory, which is yet to be established. Works in this direction are in progress. Third it is important to perform the computation at a loop level in order to see if the structure found in this paper persists, and to have more data to test the bootstrap approach discussed in the third paper. Finally the results in this paper suggest that non-maximal giant gravitons in ABJM theory might also correspond to integrable boundary states. It would be interesting to explore this further\fn{It would also be useful to revisit the analysis in \cite{Chen:2019kgc} in order to understand the difference between $\mathcal{N}=4$ SYM and ABJM. This is important since, even for $\mathcal{N}=4$ SYM, there are contradicting claims on the (non-)integrability of the non-maximal giant gravitons in the literature \cite{Berenstein:2006qk,Ciavarella:2010tp} (see also \cite{Koch:2016qkp,deMelloKoch:2018tlb}).}, e.g.~by  analyzing the boundary condition of the string worldsheet at strong coupling \cite{Linardopoulos:2021rfq,Dekel:2011ja} or by computing the Hamiltonian of the open spin chain attached to the non-maximal giant gravitons (see \cite{Chen:2018sbp,Bai:2019soy,Chen:2019igg} for the results on the maximal giant graviton). In addition, it would be desirable to derive a closed-form expression for the overlap for the non-maximal giant gravitons.  Another interesting question is to generalize the analysis in this paper to the dual giant gravitons and see whether they lead to integrable boundary conditions.

%%%%%%%%%%%%%%%%%%%%%%%%%%%%%%%%%%%%%%%%%%%%%%%%%%%%%%%%%%%%%
\subsection*{Acknowledgement}
We are  very thankful to Yang Zhang for kindly providing us
computation resource. We would also like to thank  Junpeng Cao and De-Liang Zhong for very helpful discussions.
The work of PY and JBW  is  supported in part  by the National Natural Science Foundation of China, Grant No.~11975164, 11935009, 12047502, 11947301, and  Natural Science Foundation of Tianjin under Grant No.~20JCYBJC00910. The work of SK was supported in part by DOE grant number DE-SC0009988.

\appendix
%%%%%%%%%%%%%%%%%%%%%%%%%%%%%%%%%%%%%%%%%%%%%%%%%%%%%%%%%%%%%
\section{Partially Contracted Giant Graviton}
\label{sec:PCGG}

%%%%%%%%%%%%%%%%%%%%%%%%%%%%%%%%%%%%%%%%%%%%%%%%%%%%%%%
In this appendix, we study the structure constants of two giant graviton and one single-trace operator of length $2L$ using the partially-contracted-giant-graviton (PCGG) approach \cite{Jiang:2019xdz} in order to cross-check the results in the main text. For simplicity, we only discuss the maximal giant gravitons but the analysis can be readiliy extended to non-maximal giant gravitons.

\paragraph{Two-point function.}The main focus of this appendix is the three-point function
\begin{align}\label{eq:tostudy}
G_2=\langle\mathcal{D}_1(x_1)\mathcal{D}_2(x_2)\mathcal{O}(x_3)\rangle\comma
\end{align}
where
\begin{align}
\label{eq:Dfactorization}
\mathcal{D}_i(x_i)=\det\left((n_i\!\cdot\! Y)(\bar{n}_i\!\cdot\!\bar{Y})\right)=\det(n_i\!\cdot\! Y)\det(\bar{n}_i\!\cdot\!\bar{Y})\period
\end{align}
However, it is useful to first analyze the two-point function
\begin{align}
\label{eq:DfactorizationCor}
\langle\mathcal{D}_1(x_1)\mathcal{D}_2(x_2)\rangle=\langle \det(n_1\!\cdot\! Y)\det(\bar{n}_2\!\cdot\!\bar{Y})\rangle
\langle \det(n_2\!\cdot\! Y)\det(\bar{n}_1\!\cdot\!\bar{Y})\rangle\period
\end{align}
At the leading order, we can compute it by Wick contraction. The propagator is given by
\begin{align}
\langle (n_1\!\cdot\! Y(x))_a^{\phantom{a}\bar{b}}\,(\bar{n}_2\!\cdot\! \bar{Y}(y))_{\bar{c}}^{\phantom{c}d}\rangle=\frac{1}{k}\frac{n_1\!\cdot\!\bar{n}_2}{|x-y|}\delta_a^d\,\delta_{\bar{c}}^{\bar{b}}\period
\end{align}
The determinant can be written as
\begin{align}
\det X=\frac{1}{N!}\epsilon_{a_1\cdots a_N}\epsilon^{b_1\cdots b_N}X_{b_1}^{a_1}\cdots X_{b_N}^{a_N}\period
\end{align}
Therefore, the two-point function of a single determinant is given by
\begin{align}
\langle \det(n_1\!\cdot\! Y)(x_1)\det(\bar{n}_2\!\cdot\!\bar{Y})(x_2)\rangle=&\,\frac{1}{k^N(N!)^2}\epsilon_{a_1\cdots a_N}\epsilon^{b_1\cdots b_N}
\epsilon_{c_1\cdots c_N}\epsilon^{d_1\cdots d_N}\\\nonumber
&\,\times \frac{(n_1\!\cdot\!\bar{n}_2)^N}{|x_{12}|^N}\delta_{a_1}^{d_1}\cdots\delta_{a_N}^{d_N}\delta_{b_1}^{c_1}\cdots\delta_{b_N}^{c_N}\cdot N!\\\nonumber
=&\,\frac{(n_1\!\cdot\!\bar{n}_2)^N N!}{k^N|x_{12}|^N}=\frac{ d_{12}^N}{k^N} N!\comma
\end{align}
where we have used $\epsilon_{a_1\cdots a_N}\epsilon^{a_1\cdots a_N}=N!$. Then, we get the following result for the two-point function of giant gravitons
\begin{align}\label{eq:dettwopnt}
\langle\mathcal{D}_1(x_1)\mathcal{D}_2(x_2)\rangle=\frac{1}{k^{2N}}(d_{12}d_{21})^N(N!)^2\period
\end{align}
\paragraph{Partially contracted giant graviton.} To compute the three-point function involving a single-trace operator of length $2L$ \eqref{eq:tostudy}, we first perform the Wick contractions between the two giant gravitons \emph{partially} so that it leaves precisely $2L$ uncontracted scalar fields. We then perform the Wick contraction between these scalar fields with the single trace operator.
We see that, in ABJM theory, the giant graviton factorizes into the product of two determinants (\ref{eq:Dfactorization}). In the large $N$ limit, the partially contracted giant graviton at the leading order is given by the sum of the following two contributions:
\begin{itemize}
\item Contract completely $\det(n_1 \cdot Y)$ and $\det(\bar{n}_2 \cdot \bar{Y})$ and take the leading term of the partially contracted $\det(n_2 \cdot Y)$ and $\det(\bar{n}_1 \cdot \bar{Y})$.
\item Contract completely $\det(n_2 \cdot Y)$ and $\det(\bar{n}_1 \cdot \bar{Y})$ and take the leading term of the partially contracted $\det(n_1 \cdot Y)$ and $\det(\bar{n}_2 \cdot \bar{Y})$.
\end{itemize}
The leading term of the partially contracted determinant operators have been worked out in \cite{Jiang:2019xdz} in the study of giant gravitons in $\mathcal{N}=4$ SYM, and we can simply recycle the result. For instance, the PCGG obtained from the partial contraction of $\det(n_2 \cdot Y)$ and $\det(\bar{n}_1 \cdot \bar{Y})$ is a non-local single trace operator
\begin{align}
(N-L)! \frac{( d_{21})^{N-L}}{k^{N-L}}\frac{(-1)^{L+1}}{L}\tr\left[\left((n_2\cdot Y)(x_2)(\bar{n}_1\cdot \bar{Y})(x_1)\right)^L\right]\period
\end{align}
A similar expression can be derived for the partial contraction of $\det(n_1 \cdot Y)$ and $\det(\bar{n}_2 \cdot \bar{Y})$. Combining the two contributions, we find that the full PCGG is given by
\begin{align}
\begin{aligned}
&G_L(x_1,x_2)=\,N!(N-L)!\frac{( d_{21}d_{12})^{N}}{k^{2N-L}}\frac{(-1)^{L+1}}{L}\\
&\times\left(\frac{\tr\left[\left((n_2\cdot Y)(x_2)(\bar{n}_1\cdot \bar{Y})(x_1)\right)^L\right]}{(d_{21})^L}+
\frac{\tr\left[\left((n_1\cdot Y)(x_1)(\bar{n}_2\cdot \bar{Y})(x_2)\right)^L\right]}{( d_{12})^L}\right)\period
\end{aligned}
\end{align}
In order to take the large $N$ limit, it is useful to divide it by the two-point function of giant gravitons \eqref{eq:dettwopnt}. We then get
\begin{align}\label{eq:pcggexpression}
\begin{aligned}
    &\frac{G_{L}(x_1,x_2)}{\langle \mathcal{D}_1(x_1)\mathcal{D}_2 (x_2)\rangle}\overset{N\to \infty}{=}\frac{(-1)^{L+1}}{L}\left(\frac{1}{\lambda}\right)^{L}\\
    &\times \left(\frac{\tr\left[\left((n_2\cdot Y)(x_2)(\bar{n}_1\cdot \bar{Y})(x_1)\right)^L\right]}{( d_{21})^L}+
\frac{\tr\left[\left((n_1\cdot Y)(x_1)(\bar{n}_2\cdot \bar{Y})(x_2)\right)^L\right]}{( d_{12})^L}\right)\period
\end{aligned}
\end{align}
\paragraph{Twisted-translated frame and MPS.}
To compare this with the results in the main text, we consider the twisted-translated kinematics and set $a_1=-a_2=1$ and $a_3=0$. Computing the Wick contractions between the PCGG \eqref{eq:pcggexpression} and the single-trace operator and dividing the result by the normalization $\mathcal{N}_{\mathcal{O}}$, we find that the structure constant is given by
\begin{align}
    \mathfrak{D}_{N|\mathcal{O}}=-\frac{(-i)^{J}}{2^{\Delta-J}} \frac{\langle \mathcal{B}|\mathcal{O}\rangle}{\sqrt{L\langle \mathcal{O}|\mathcal{O}\rangle}}\comma
\end{align}
which precisely matches with the result in the main text \eqref{eq:maximalmps}.

%%%%%%%%%%%%%%%%%%%%%%%%%%%
\section{Derivation of BPS Three-Point Funcitons}\label{sec:derBPS}
Here we explain how to derive the expression for the three-point function of two non-maximal giant gravitons and a single-trace BPS operator given in \eqref{eq:forapB2}.

The starting point of the analysis is the generating function of structure constants \eqref{eq:forapB1}. In order to extract the structure constant for a single-trace operator with a fixed charge, we perform the integration over $s$; namely
\begin{align}
    \frac{\langle \mathcal{D}_M\mathcal{D}_M\mathcal{O}_L^{\circ}\rangle}{\langle \mathcal{D}_M\mathcal{D}_M\rangle}=-\oint_{|s|=\epsilon\ll 1}\frac{ds}{2\pi i s^{1+L}}\oint_{|z|=1}\frac{dz}{2\pi i z}{\rm tr}_2\left[\frac{1}{1-s\mathcal{T}_{\theta}\bar{\mathcal{T}}_{\theta}}\right]\period
\end{align}
Here we replaced the integration variable $\theta$ with $z\equiv e^{i\theta}$. Note also that the integration of $s$ is performed in a region near the origin $|s|\ll 1$ since the generating function $1/(1-s\mathcal{T}_{\theta}\bar{\mathcal{T}}_{\theta})$ is expected to have a finite radius of convergence when expanded in $s$ and therefore one has to take $|s|$ to be sufficiently small in order to use the formula.

As the next step, we diagonalize the matrix $1/(1-s\mathcal{T}_{\theta}\bar{\mathcal{T}}_{\theta})$. We also change the integration variables as 
\begin{align}
\begin{aligned}
    \tilde{s}&=\lambda \frac{d_{13}d_{32}}{d_{12}} s\comma\qquad \tilde{z}=\left(\frac{d_{13}d_{31}}{d_{23}d_{32}}\right)^{\frac{1}{4}}z\comma
    \end{aligned}
\end{align}
in order to simplify the expression. As a result we get
\begin{align}
    \begin{aligned}
    \frac{\langle \mathcal{D}_M\mathcal{D}_M\mathcal{O}_L^{\circ}\rangle}{\langle \mathcal{D}_M\mathcal{D}_M\rangle}=&-\left(\lambda \frac{d_{13}d_{32}}{d_{12}}\right)^{L}\oint_{|\tilde{s}|=\epsilon\ll 1}\frac{d\tilde{s}}{2\pi i \tilde{s}^{1+L}}\oint_{|z|=1}\frac{d\tilde{z}}{2\pi i \tilde{z}} \\
    &\times \left[1+\frac{\tilde{z}^2 \left(1-\xi  \tilde{s}^2\right)}{\sqrt{\xi } \tilde{s} (\omega -1)(1+
   \tilde{z}^4)+\tilde{z}^2 ((\xi +1) \tilde{s} \omega -2)}\right]^{-1}\period
    \end{aligned}
\end{align}
We then perform the integration of $\tilde{z}$ by closing the contour and picking up contributions from poles inside the contour. When $|\tilde{s}|\ll 1$, we find that the integrand has three poles inside the contour $|z|=1$:
\begin{align}
    \begin{aligned}
    \tilde{z}=0\comma\quad \pm \frac{1}{\xi^{1/4}}\sqrt{\frac{1+\tilde{s}^2\xi-(1+\xi)\tilde{s}\omega-\sqrt{(1+\tilde{s}^2\xi-\tilde{s}(1+\xi)\omega)^2-4\tilde{s}^2\xi(\omega-1)^2}}{2\tilde{s} (\omega-1)}}\period
    \end{aligned}
\end{align}
Computing the residues from these poles (and redefining $\tilde{s}$ as $s$), we obtain \eqref{eq:forapB2}.

\section{Data and Numerics}
\label{sec:numerics}
%%%%%%%%%%%%%%%%%%%%%%%%%%%%%%%%%%%%%%%%%%%%%%%%%%%%%%%%%%%%%
In this appendix, we explain our method to numerically solve the Bethe equation and provide some data for the overlap with the non-maximal giant graviton.
\subsection{Numerical solutions to the Bethe equations}

\paragraph{Solving the Bethe equation.}
To solve the Bethe equations \eqref{eq:betheansatzeq} numerically, we follow the following strategy:
    First we divide the auxiliary roots ${\bf w}$ into two subsets ${\bf w}={\bf w}^{(1)}\cup {\bf w}^{(2)}$, introduce a parameter $\epsilon$ to modify the Bethe equations into the following:
    \begin{align}\label{eq:modifiedbethe}
    \begin{aligned}
\left(\frac{u_j+\frac{i}{2}}{u_j-\frac{i}{2}}\right)^L=&\prod_{\substack{k=1\\k\neq j}}^{K_{\bf u}}\dfrac{u_j-u_k+i}{u_j-u_k-i}\prod_{k=1}^{K_{\bf w}^{(1)}}\dfrac{u_j-w_k^{(1)}-\dfrac{i}{2}}{u_j-w_k^{(1)}+\dfrac{i}{2}}\prod_{k=1}^{K_{\bf w}^{(2)}}\dfrac{u_j-w_k^{(2)}-\dfrac{i}{2}\epsilon}{u_j-w_k^{(2)}+\dfrac{i}{2}\epsilon}\comma\\
\left(\dfrac{v_j+\frac{i}{2}}{v_j-\frac{i}{2}}\right)^L=&\prod_{\substack{k=1\\k\neq j}}^{K_{\bf v}}\dfrac{v_j-v_k+i}{v_j-v_k-i}\prod_{k=1}^{K_{\bf w}^{(1)}}\dfrac{v_j-w_k^{(1)}-\dfrac{i}{2}}{v_j-w_k^{(1)}+\dfrac{i}{2}}\prod_{k=1}^{K_{\bf w}^{(2)}}\dfrac{v_j-w_k^{(2)}-\dfrac{i}{2}\epsilon}{v_j-w_k^{(2)}+\dfrac{i}{2}\epsilon}\comma\\ 1=\prod_{k=1}^{K_{\bf u}}\dfrac{w_j^{(1)}-u_k-\dfrac{i}{2}}{w_j^{(1)}-u_k+\dfrac{i}{2}}&\prod_{k=1}^{K_{\bf v}}\dfrac{w_j^{(1)}-v_k-\dfrac{i}{2}\epsilon}{w_j^{(1)}-v_k+\dfrac{i}{2}\epsilon}\prod_{\substack{k=1\\k\neq j}}^{K_{\bf w}^{(1)}}\dfrac{w_j^{(1)}-w_k^{(1)}+i}{w_j^{(1)}-w_k^{(1)}-i}\prod_{k=1}^{K_{\bf w}^{(2)}}\dfrac{w_j^{(1)}-w_k^{(2)}+i\epsilon}{w_j^{(1)}-w_k^{(2)}-i\epsilon}\\ 1=\prod_{k=1}^{M}\dfrac{w_j^{(2)}-u_k-\dfrac{i}{2}\epsilon}{w_j^{(2)}-u_k+\dfrac{i}{2}\epsilon}&\prod_{k=1}^{N}\dfrac{w_j^{(2)}-v_k-\dfrac{i}{2}}{w_j^{(2)}-v_k+\dfrac{i}{2}}\prod_{\substack{k=1\\k\neq j}}^{K_{\bf w}^{(2)}}\dfrac{w_j^{(2)}-w_k^{(2)}+i}{w_j^{(2)}-w_k^{(2)}-i}\prod_{k=1}^{K_{\bf w}^{(1)}}\dfrac{w_j^{(2)}-w_k^{(1)}+i\epsilon}{w_j^{(2)}-w_k^{(1)}-i\epsilon}\period
\end{aligned}
\end{align}
The modified Bethe equations \eqref{eq:modifiedbethe} go back to the original equations when $\epsilon=1$ while they become two decoupled $SU(3)$ Bethe equations when $\epsilon=0$.

Second we solve the $SU(3)$ Bethe equations using the rational Q-system\fn{However, the Q system there only works for the case when $L-M\ge M-N\ge N$, where $L$ is the length of the $SU(3)$ spin chain and $M, N$ are number of two types of magnons. Sometimes this constraint is too strong to obtain the $SU(4)$ solutions. In some cases, we can fix this problem using the bosonic duality \cite{Gromov:2007ky}.} by Marboe and Volin \cite{Marboe:2016yyn}. We then use these solutions as ``seed" solutions for $\epsilon=0$ and use {\bf FindRoots} in {\it Mathematica} to generate solutions for $\epsilon=1$.

As a result, we obtained the solutions to the original Bethe equations \eqref{eq:betheansatzeq}. In what follows we only consider the solutions with $K_{\bf u}=K_{\bf v}=K_{\bf w}$ that are relevant for the computation of the overlap. Note also that the solutions listed below are by no means exhaustive. It would be desirable to improve the algorithm for solving the Bethe equations and perform more systematic checks. In particular, the generalization of Marboe-Volin algorithm for the alternating spin chain is an important open problem.
\paragraph{Solutions to the Bethe equations.} Let us now list the solutions we obtained. Here the Bethe roots denoted in red satisfy both the selection rules and the zero-momentum condition while the Bethe roots denoted in blue only satisfy the zero-momentum condition. All the other Bethe roots do not satisfy the zero-momentum condition and therefore do not correspond to single-trace operators in ABJM theory. For later convenience, we also numbered each solution in red, as {\tt [n]}.
\newpage
{\centering
	\begin{longtable}{p{0.3cm}|p{0.6cm}|p{13cm}}
		\hline
		\hline
		
		$L$& $K_{\bf w}$ & $[{\bf u}, {\bf v}, {\bf w}]$ \\
		\hline
		1 &1& \textcolor{red}{$[ \{0\}, \{0\},\{ 0\}]$}   {\tt [1]} \\
		\hline
		2 &1& \textcolor{blue}{$ [\{\tfrac{1}{2 \sqrt{3}}\},\{ -\tfrac{1}{2 \sqrt{3}}\},\{0\}] $}\\
\cline{3-3}
& & \textcolor{black}{$[\{\tfrac{1}{2}\},\{\tfrac{1}{2}\},\{\tfrac{1}{2}\} ]$}\\
\cline{3-3}
& & \textcolor{black}{$[\{-\tfrac{1}{2}\},\{-\tfrac{1}{2}\},\{-\tfrac{1}{2}\}] $}\\
		
		\hline
		3 &1& \textcolor{red}{$[\{ 0\},\{ 0\},\{ 0\}]$} {\tt [2]}\\
         \cline{3-3}
				&&\textcolor{red}{$ [\{\tfrac{1}{2}\},\{-\tfrac{1}{2}\},\{0\}]$}   {\tt [3]}\\
		\cline{3-3}
	&&	$ [\{ -\tfrac{\sqrt{3}}{2}\},\{-\tfrac{\sqrt{3}}{2}\},\{-\tfrac{\sqrt{3}}{2} \}$ \\
		\hline
		4&1&  $[\{\tfrac{\sqrt{7}-2}{6}\},\{-\tfrac{\sqrt{7}+2}{6}\},\{ -\tfrac{1}{3}\}$\\
		\hline
		5 &1& \textcolor{red}{$[\{0\},\{0\},\{0\}]$}  {\tt [4]}\\
        \cline{3-3} && \textcolor{red}{$[\{-\tfrac{1}{\sqrt{12}}\},\{\tfrac{1}{\sqrt{12}}\},\{0\}]$} {\tt [5]}\\
		\hline
		6 &1&  		\textcolor{blue}{$[\{-0.3987366944412021\}, \{0.3987366944412021\}, \{0\}]$}\\
\cline{3-3}
&&$ [\{0.13397459621556135\}, \{ 0.13397459621556138\}, \{ 0.13397459621556135\}]$\\
\cline{3-3}
		&&		$[\{0.45440762733328904\}, \{0.15990657179334503\}, \{0.307157099563317\}]$\\
		\cline{3-3}
&&	$[\{\tfrac{1}{2}\},\{\tfrac{1}{2}\},\{\tfrac{1}{2}\}] $\\
		\hline
		2 &2& \textcolor{red}{$[\{\sqrt{\tfrac{3}{20}},-\sqrt{\tfrac{3}{20}}\},\{-\sqrt{\tfrac{3}{20}},\sqrt{\tfrac{3}{20}}\},\{\tfrac{1}{\sqrt{5}},-\tfrac{1}{\sqrt{5}}\}]$}  {\tt [6]}\\
		\hline
		3 &2& 	\textcolor{blue}{$[\{ -0.34554024732023516,  0.34554024732023516\},$}\\
		&&\textcolor{blue}{$\{
			0.34554024732023516,  -0.34554024732023516\},$}\\ &&\textcolor{blue}{ $\{ 0.51108108452939387i,  -0.51108108452939387 i\}]$}\\
\cline{3-3}
		&&
		\textcolor{blue}{$[\{ -0.51320279936202029, 0.062678427970585504\}, $}
		
		\textcolor{blue}{$\{ 0.51320279936202029, -0.062678427970585504\},$}
		
		\textcolor{blue}{$\{ -0.41623324836102756,  0.41623324836102756\}]$}\\
\cline{3-3}
		&&
		$[\{-0.67910322332093357, 0.025611245506665032\},$
		
		$\{ 0.53448799950517913, -0.74702142547534923\}$,
		
		$\{ -0.76136871888869386,  0.32835601699647453\}]$\\
		\hline
		4&2&\textcolor{red}{$[\{ -0.56944513222254722, 0.56944513222254722\},$}

		\textcolor{red}{$ \{0.56944513222254722,  -0.56944513222254722\},$}
		
		\textcolor{red}{$\{ 0.55472575089337952 i,	 -0.55472575089337952 i\}]$}   {\tt [7]}\\
		\cline{3-3}
	&&	\textcolor{red}{$[\{ -0.66931302374657110,  -0.10647062449199778\}, $}
		
		\textcolor{red}{$\{ 0.66931302374657110, 0.10647062449199778\},$}
		
		\textcolor{red}{$\{ -0.48978976193047651,  0.48978976193047651\}]$}  {\tt [8]}\\

\cline{3-3}
&&$[\{ -0.028299146374196785, 	 0.55999384176787242\}, 	$
		
		$\{-0.028299146374196785, 0.55999384176787242\},$
		
		$\{ 0.26584734769683781 - 0.50624190637875312 i,$
		
		$0.26584734769683781 + 0.50624190637875312 i\}]$\\

		\hline
		6&2& 		\textcolor{red}{$[\{ 0.34675919853122809,  0.99366751548216189\},$}	
		
		\textcolor{red}{$ \{-0.34675919853122809, -0.99366751548216189\},$}
		
		\textcolor{red}{$ \{0.71060441235307684, 	 -0.71060441235307684\}]$}  {\tt [9]}\\\cline{3-3}
&&
$[\{-1.0773108398074073,  0.49798319908651183\},$

$ \{0.089846608200097142, 0.42002515253515016\}$,

$\{ -0.59531374844468910,  0.56058580845186502\}]$\\
\hline
		6&2&
		$[\{0.40946507692652007,  0.10041044458589016\},$

		$\{ -0.39457524195942078,
		0.15869120668880070\},$
		
		$\{ 0.44210986670775702, -0.30511412358686194\}],$\\
		\cline{3-3}
	 &&
      $[\{0.11567004621958770,  0.44116926182762157\},$

      $\{ -0.40026535781351243,0.48672949230748789\},$

      $\{ 0.57166303815947364,  -0.25001131688888128\}]$\\
		\cline{3-3}
	 &&	$[\{ 0.37830228050349018, 	 1.0458410010838975\}, $
	
	 $\{ -0.055390813251937810, -0.35320869561294462\},$
	
	 $\{ 0.78252761450158240, -0.27475572814032976\}]$\\
	\cline{3-3}
	 &&	
		$[\{0.36681212663617484,  1.0519697940662345\},$
		
		$\{ 0.20899528319467480, -1.0040495234017774\}$
		
		$\{ 0.82203932258846606,  -0.51017548234081269\}]$\\
	\cline{3-3}
	 &&
		$[\{0.13241485029334295,  0.45759251757083131\}$,
		
		$\{ -0.098732170839668265, 0.48070339312713823\}$,
		
		$\{ 0.59462910995790549,  -0.10863981488208338\}]$\\
		\hline
		3&3&\textcolor{red}{$[\{ -0.61842989257770833, \, 0\,, 0.61842989257770833\}$}
		
		\textcolor{red}{$ \{ 0.61842989257770833, \, 0\, ,  -0.61842989257770833\},$}
		
		\textcolor{red}{$\{0.71410132990930250,  -0.71410132990930250, 	 0\}] 	$}  {\tt [10]}\\
		\hline
		
		5&3& 	\textcolor{red}{$[\{ -0.90018200552947429,  0.031648482693564728,  0.37680381182936445\}$}
		
		\textcolor{red}{$\{ 0.90018200552947429,  -0.031648482693564728,  -0.37680381182936445\},$}
		
		\textcolor{red}{$\{ -0.75676650863660755,  0.75676650863660755,  0\}]$}  {\tt [11]}\\
		\hline
		6&4&	\textcolor{red}{$[\{ \pm 0.53714639382219377 \pm 	0.51401134667644172 i\}$}

		\textcolor{red}{$\{ \mp 0.53714639382219377 \mp 0.51401134667644172 i\},$}

		\textcolor{red}{$\{ \pm 1.6004024229193260 i,  \pm 0.50622703524559230 i\}]$}  {\tt [12]}\\
\cline{3-3}
		
		&&\textcolor{red}{$[\{ -0.098635382965062615, 	0.14753207194168752,$}
		
		\textcolor{red}{$-1.0179905746843994, 0.47877616601418219\},$}

		\textcolor{red}{$\{ 0.098635382965062615,  -0.14753207194168752,$}
		
		\textcolor{red}{$1.0179905746843994,  -0.47877616601418219\}$,}
		
		\textcolor{red}{$ \{-0.89439365253962075, 0.89439365253962075,$}
		
		\textcolor{red}{$0.22001470275637868,	-0.22001470275637868\}]$}  {\tt [13]} \\
		\hline
		\hline
	\end{longtable}
\par}

\subsection{Overlaps for non-maximal giant gravitons}
Here we summarize the results for overlaps for non-maximal giant gravitons $\omega\neq 1$. Below we list the results for the ratio between the structure constant of the maximal giant graviton $\mathfrak{D}_{N|\mathcal{O}}$ and the structure constant of the non-maximal giant graviton $\mathfrak{D}_{M|\mathcal{O}}$; $r\equiv \mathfrak{D}_{M|\mathcal{O}}/\mathfrak{D}_{N|\mathcal{O}}$.
\begin{align}
    \begin{aligned}
    {\tt [1]}\qquad r=&\omega\comma\\
    {\tt [2]}\qquad r=&-2 \omega ^3+4 \omega ^2-\omega\comma\\
    {\tt [3]}\qquad r=&-2 \omega ^3+4 \omega ^2-\omega\comma\\
    {\tt [4]}\qquad r=&6 \omega ^5-24 \omega ^4+30 \omega ^3-12 \omega ^2+\omega\comma\\
    {\tt [5]}\qquad r=&6 \omega ^5-24 \omega ^4+30 \omega ^3-12 \omega ^2+\omega\comma\\
    {\tt [6]}\qquad r=&2 \omega ^2-\omega\comma\\
    {\tt [7]}\qquad r=&-4 \omega ^4+8.80557589 \omega ^3-3.77669700 \omega ^2-0.028878886 \omega\comma\\
    {\tt [8]}\qquad r=&-4 \omega ^4+8.21090774 \omega ^3-3.39983829 \omega ^2+0.188930549 \omega\comma\\
    {\tt [9]}\qquad r=&12 \omega ^6-47.8851733 \omega ^5+62.8683251 \omega ^4-30.0180998 \omega ^3\\
    &+3.96919074
   \omega ^2+0.065757247 \omega\comma\\
    {\tt [10]}\qquad r=&-2 \omega ^3+4 \omega ^2-\omega\comma\\
    {\tt [11]}\qquad r=&-8 \omega ^5+20.17658812 \omega ^4-12.92497528 \omega ^3\\&+1.74975464 \omega ^2-0.001367480
   \omega\comma\\
    {\tt [12]}\qquad r=&-16\omega^6+42.978665\omega^5-33.996976\omega^4+8.5586366\omega^3\\
    &-0.54003958\omega^2-0.00028639193\omega\comma\\
    {\tt [13]}\qquad r=&-16\omega^6+49.281471\omega^5-51.190705\omega^4+23.285677\omega^3\\
    &-5.0615656\omega^2+0.68512348\omega\period
    \end{aligned}
\end{align}
Here ${\tt [n]}$'s refer to the numbers in the table above.
\section{Nested Algebraic Bethe Ansatz}\label{sec:NABA}

In this appendix, we review the nested algebraic Bethe ansatz approach  for the $SU(4)$ alternating spin chain \cite{Minahan:2008hf,Bak:2008cp}. We begin with the following set of four $R-$matrices,
\begin{align}
R_{ab}&=u+\rP_{ab}\comma\\
R_{\bar{a}\bar{b}}&=u+\rP_{\bar{a}\bar{b}}\comma\\
R_{a\bar{b}}&=-(u+2)+\rK_{a\bar{b}}\comma\\
R_{\bar{a}b}&=-(u+2)+\rK_{\bar{a}b}\comma
\end{align}
where $a$ ($\bar{a}$) denotes the spin in the fundamental (anti-fundamental) representation of $SU(4)$.
These $R-$matrices satisfy the following eight Yang-Baxter equations,
\begin{align}
R_{\textbf{a}\textbf{b}}(u-v) R_{\textbf{a}\textbf{c}}(u)R_{\textbf{b}\textbf{c}}(v)=
R_{\textbf{b}\textbf{c}}(v)R_{\textbf{a}\textbf{c}}(u)R_{\textbf{a}\textbf{b}}(u-v)\comma
\end{align}
where $\textbf{a}$ can take value of $a$ or $\bar{a}$ independently.

We now define two monodromy matrices using two different auxiliary spaces in $\textbf{4}$ and $\bar{\textbf{4}}$ representations,
\begin{align}
T_a(u)=R_{a1}(u)R_{a\bar{1}}(u)\cdots R_{aL}(u)R_{a\bar{L}}(u)\comma\\
T_{\bar{a}}(u)=R_{\bar{a}1}(u)R_{\bar{a}\bar{1}}(u)\cdots R_{\bar{a}L}(u)R_{\bar{a}\bar{L}}(u)\period
\end{align}
The corresponding   transfer matrices
\begin{align}
\tau(u)={\rm Tr}_a T_a(u), \,\, \bar{\tau}(u)={\rm Tr}_{\bar{a}}T_{\bar{a}}(u)\comma
\end{align}
commute with each other
\begin{align}\label{eq:tau}
[\tau(u), \tau(v)]=0, \,\, [\bar{\tau}(u), \bar{\tau}(v)]=0, \,\, [\tau(u), \bar{\tau}(v)]=0\comma
\end{align}
due to the above Yang-Baxter equations.
We can generate two Hamiltonians from the transfer matrices,
\begin{align}
H_{odd}=(\tau(0))^{-1}\frac{d}{du}\tau(u)\bigg|_{u=0}, \, \,  H_{even}=(\bar{\tau}(0))^{-1}\frac{d}{du}\bar{\tau} (u)\bigg|_{u=0}\period
\end{align}
The true Hamiltonian is given by $H=H_{odd}+H_{even}$ which is the same as planar two-loop anomalous dimension matrix in the scalar sector of ABJM theory, up to  rescaling and shifting by a constant.

Due to (\ref{eq:tau}), there exist $u-$indepedent common eigenstates of $\tau(u)$ and $\bar{\tau}(u)$. We now construct eigenstates of $\tau(u)$
using nested algebraic Bethe ansatz \cite{Kulish:1983rd}. For a nice review, we refer to \cite{Slavnov:2019hdn}. We first write the monodromy matrix $T_a(u)$ as $4\times 4$
matrix whose elements are operators acting on the Hilbert space of the spin chain $\mathcal{H}\cong ({\mathbf{C}}^4)^{\otimes 2L}$,
\begin{align}
\nonumber
T_a(u)=\left(
\begin{array}{cccc}
A(u) & B_1(u)  & B_2(u) & B_3(u) \\
C_1(u) & D_{22}(u) &D_{23}(u) &D_{24}(u)\\
C_2(u) & D_{32}(u) & D_{33}(u) &D_{34}(u)\\
C_3(u) & D_{42}(u) & D_{43}(u) & D_{44}(u) \\
\end{array}
\right)\period
\end{align}

The  RTT relation
\begin{align}
R_{ab}(u-v)T_a(u)T_b(v)=T_b(v)T_a(u)R_{ab}(u-v),
\end{align}
from Yang-Bexter equations, leads to the following commutation relations,
\begin{align}
A(u)B_i(v)&=\frac{u-v-1}{u-v}B_i(v)A(u){\color{red}+\frac{1}{u-v}B_i(u)A(v)}\comma\\
D_{ij}(u)B_k(v)&=\frac{1}{u-v}B_{k^\prime}(v)D_{ij^\prime}(u)\left(R^{SU(3)}(u-v)\right)^{k^\prime j^\prime}_{kj}{\color{red}-\frac1{u-v}B_j(u)D_{ik}(v)}\comma
\end{align}
where $i, j, k, \cdots$ take values in $2, 3, 4$ and
\begin{align}
\left(R^{SU(3)}(u-v)\right)^{k^\prime j^\prime}_{kj}=(u-v)\delta^{k^\prime}_k\delta^{j^\prime}_j+\delta^{k^\prime}_j\delta^{j^\prime}_k
\end{align}
is an $SU(3)$ R-matrix. And the red part in the above formulas is the original of the unwanted terms in the following.

Now we consider the subspace $\mathcal{H}_1$ of $\mathcal {H}$ spanned by the state
\begin{align}  |1, \bar{i}_1, 1, \bar{i}_2, \cdots, 1, \bar{i}_L\rangle \end{align}
with $\bar{i}_k=2, 3, 4$.
Any state $|1\rangle$ in this subspace satisfy the following important property,
\begin{align}
A(u)|1\rangle&=(-u-2)^L(u+1)^L |1\rangle\comma\\
C_i(u)|1\rangle&=0\comma\\
D_{ij}(u)|1\rangle&\in {\mathcal{H}}_1\period
\end{align}

Consider the following state in ${\mathcal H}$,
\begin{align}
|\psi\rangle=B_{i_1}(\mu_1)\cdots B_{i_M}(\mu_M)X^{i_1\cdots i_M}|1\rangle,
\end{align}
with $|1\rangle \in \mathcal{H}_1$ and $X^{i_1\cdots i_M}$ to-be-determined coefficients.
The acts of $\tau(u)=A(u)+D_{jj}(u)$ on $|\psi\rangle$ is given by
\begin{align}
A(u)|\psi\rangle&=(-u-2)^L (u+1)^L \prod_{i=1}^M \frac{u-\mu_i-1}{u-\mu_i}|\psi\rangle+ \text{\color{red} unwanted terms}\comma\\
D_{jj}(u)|\psi\rangle&=\left(\prod_{i=1}^M\frac1{u-\mu_i}\right)B_{k_1}(\mu_1)\cdots B_{k_M}(\mu_M){\color{blue} D_{j l_M}(u)
R^{SU(3)}(u-\mu_M)^{k_M l_M}_{i_M l_{M-1}}}\\ \nonumber
&
{\color{blue}R^{SU(3)}(u-\mu_{M-1})^{k_{M-1} l_{M-1}}_{i_{M-1} l_{M-2}}\cdots R^{SU(3)}(u-\mu_2)^{k_2 l_2}_{i_2 l_1}R^{SU(3)}(u-\mu_{1})^{k_{1} l_{1}}_{i_{1} j} }\\
 \nonumber
&X^{i_1\cdots i_M}|1\rangle+\text{\color{red} unwanted terms}\period
\end{align}

Now we need to compute the eigenstate of
\begin{align}
&{\color{blue}D_{j l_M}(u)
R^{SU(3)}(u-\mu_M)^{k_M l_M}_{i_M l_{M-1}}R^{SU(3)}(u-\mu_{M-1})^{k_{M-1} l_{M-1}}_{i_{M-1} l_{M-2}}\cdots R^{SU(3)}(u-\mu_2)^{k_2 l_2}_{i_2 l_1}}\\ \nonumber
&{\color{blue}R^{SU(3)}(u-\mu_{1})^{k_{1} l_{1}}_{i_{1} j} }
\end{align}
acting on  $({\mathbf{C}}^3)^{\otimes M}\otimes {\cal H}_L$. It is not hard to see that the  the problem is reduced to finding eigenvalue of transfer matrix on  an integrable $SU(3)$ spin chain with length $M+2L$.  On this chain,  $\mu_1, \cdots, \mu_M $ play the roles of inhomogeneity parameters. One can reduce further to an $SU(2)$ spin chain and finally diagonalize the transfer matrix.  Collecting all of the wanted terms, we get
\begin{align}
\Lambda(u)&=(-u-2)^L (u+1)^L \prod_{i=1}^{K_{\textbf{u}}} \dfrac{u-\mu_i-1}{u-\mu_i}\\ \nonumber
&+(-u)^L (u+1)^L \prod_{j=1}^{K_{\textbf{v}}}\dfrac{u-\nu_j+1}{u-\nu_j}\\ \nonumber
&+(-u)^L (u+2)^L \prod_{i=1}^{K_{\textbf{u}}}\dfrac{u-\mu_i+1}{u-\mu_i}\prod_{k=1}^{K_{\textbf{w}}}\dfrac{u-\lambda_k-1}{u-\lambda_k}\\ \nonumber
&+(-u)^L (u+2)^L \prod_{j=1}^{K_{\textbf{v}}}\dfrac{u-\nu_j-1}{u-\nu_j}\prod_{k=1}^{K_{\textbf{w}}}\dfrac{u-\lambda_k+1}{u-\lambda_k}\period
\end{align}
Demanding the vanishing of the residue of the spurious pole at $u=\mu_i, \nu_j, \lambda_k$ leads to Bethe equations.
After performing the substitution $\mu_i={\rm i}u_i-1/2, \nu_j={\rm i}v_j-3/2, \lambda_k={\rm i}w_k-1$, these Bethe equations becomes exact
the Bethe ansatz equations \eqref{eq:betheansatzeq} in subsection \ref{sec:CBA}. When these Bethe equations are satisfied, the above $\Lambda(u)$ is the eigenvalue of $\tau(u)$,
and the corresponding eigenstate can be constructed using the above procedure. We have checked, for small $L$, that such states are also eigenstates of $\bar{\tau}(u)$. It should be very valuable to prove this for the general case.

The Bethe states from algebraic Bethe ansatz and the ones from coordinate Bethe ansatz are related by
\begin{align}
|\textbf{u}, \textbf{v}, \textbf{w}\rangle^{al}\propto \prod_{i<j}\dfrac{u_i-u_j+{\rm i}}{u_i-u_j}\prod_{i<j}\dfrac{v_i-v_j+{\rm i}}{v_i-v_j}\prod_{i<j}\dfrac{w_i-w_j+{\rm i}}{w_i-w_j}|\textbf{u}, \textbf{v}, \textbf{w}\rangle^{co}\comma
\end{align}
with  a  proportional factor which is invariant under the permutations of rapidities of the same type.

%\bibliographystyle{JHEP}
%\bibliography{yunfeng}

\providecommand{\href}[2]{#2}\begingroup\raggedright\endgroup

%\bibliographystyle{JHEP}
%\bibliography{MPSRef}

\begin{thebibliography}{100}

\bibitem{Minahan:2002ve}
J.~Minahan and K.~Zarembo, \emph{{The Bethe ansatz for N=4 superYang-Mills}},
  \href{https://doi.org/10.1088/1126-6708/2003/03/013}{\emph{JHEP} {\bfseries
  03} (2003) 013} [\href{https://arxiv.org/abs/hep-th/0212208}{{\ttfamily
  hep-th/0212208}}].

\bibitem{Gromov:2009tv}
N.~Gromov, V.~Kazakov and P.~Vieira, \emph{{Exact Spectrum of Anomalous
  Dimensions of Planar N=4 Supersymmetric Yang-Mills Theory}},
  \href{https://doi.org/10.1103/PhysRevLett.103.131601}{\emph{Phys. Rev. Lett.}
  {\bfseries 103} (2009) 131601}
  [\href{https://arxiv.org/abs/0901.3753}{{\ttfamily 0901.3753}}].

\bibitem{Bombardelli:2009ns}
D.~Bombardelli, D.~Fioravanti and R.~Tateo, \emph{{Thermodynamic Bethe Ansatz
  for planar AdS/CFT: A Proposal}},
  \href{https://doi.org/10.1088/1751-8113/42/37/375401}{\emph{J. Phys. A}
  {\bfseries 42} (2009) 375401}
  [\href{https://arxiv.org/abs/0902.3930}{{\ttfamily 0902.3930}}].

\bibitem{Arutyunov:2009ur}
G.~Arutyunov and S.~Frolov, \emph{{Thermodynamic Bethe Ansatz for the AdS(5) x
  S(5) Mirror Model}},
  \href{https://doi.org/10.1088/1126-6708/2009/05/068}{\emph{JHEP} {\bfseries
  05} (2009) 068} [\href{https://arxiv.org/abs/0903.0141}{{\ttfamily
  0903.0141}}].

\bibitem{Gromov:2013pga}
N.~Gromov, V.~Kazakov, S.~Leurent and D.~Volin, \emph{{Quantum Spectral Curve
  for Planar $\mathcal{N} = 4$ Super-Yang-Mills Theory}},
  \href{https://doi.org/10.1103/PhysRevLett.112.011602}{\emph{Phys. Rev. Lett.}
  {\bfseries 112} (2014) 011602}
  [\href{https://arxiv.org/abs/1305.1939}{{\ttfamily 1305.1939}}].

\bibitem{Gromov:2014caa}
N.~Gromov, V.~Kazakov, S.~Leurent and D.~Volin, \emph{{Quantum spectral curve
  for arbitrary state/operator in AdS$_{5}$/CFT$_{4}$}},
  \href{https://doi.org/10.1007/JHEP09(2015)187}{\emph{JHEP} {\bfseries 09}
  (2015) 187} [\href{https://arxiv.org/abs/1405.4857}{{\ttfamily 1405.4857}}].

\bibitem{Basso:2015zoa}
B.~Basso, S.~Komatsu and P.~Vieira, \emph{{Structure Constants and Integrable
  Bootstrap in Planar N=4 SYM Theory}},
  \href{https://arxiv.org/abs/1505.06745}{{\ttfamily 1505.06745}}.

\bibitem{Fleury:2016ykk}
T.~Fleury and S.~Komatsu, \emph{{Hexagonalization of Correlation Functions}},
  \href{https://doi.org/10.1007/JHEP01(2017)130}{\emph{JHEP} {\bfseries 01}
  (2017) 130} [\href{https://arxiv.org/abs/1611.05577}{{\ttfamily
  1611.05577}}].

\bibitem{Basso:2015eqa}
B.~Basso, V.~Goncalves, S.~Komatsu and P.~Vieira, \emph{{Gluing Hexagons at
  Three Loops}},
  \href{https://doi.org/10.1016/j.nuclphysb.2016.04.020}{\emph{Nucl. Phys. B}
  {\bfseries 907} (2016) 695}
  [\href{https://arxiv.org/abs/1510.01683}{{\ttfamily 1510.01683}}].

\bibitem{Basso:2017khq}
B.~Basso, F.~Coronado, S.~Komatsu, H.~T. Lam, P.~Vieira and D.-l. Zhong,
  \emph{{Asymptotic Four Point Functions}},
  \href{https://doi.org/10.1007/JHEP07(2019)082}{\emph{JHEP} {\bfseries 07}
  (2019) 082} [\href{https://arxiv.org/abs/1701.04462}{{\ttfamily
  1701.04462}}].

\bibitem{Basso:2017muf}
B.~Basso, V.~Goncalves and S.~Komatsu, \emph{{Structure constants at wrapping
  order}}, \href{https://doi.org/10.1007/JHEP05(2017)124}{\emph{JHEP}
  {\bfseries 05} (2017) 124}
  [\href{https://arxiv.org/abs/1702.02154}{{\ttfamily 1702.02154}}].

\bibitem{Eden:2015ija}
B.~Eden and A.~Sfondrini, \emph{{Three-point functions in ${\cal N}=4$ SYM: the
  hexagon proposal at three loops}},
  \href{https://doi.org/10.1007/JHEP02(2016)165}{\emph{JHEP} {\bfseries 02}
  (2016) 165} [\href{https://arxiv.org/abs/1510.01242}{{\ttfamily
  1510.01242}}].

\bibitem{Fleury:2017eph}
T.~Fleury and S.~Komatsu, \emph{{Hexagonalization of Correlation Functions II:
  Two-Particle Contributions}},
  \href{https://doi.org/10.1007/JHEP02(2018)177}{\emph{JHEP} {\bfseries 02}
  (2018) 177} [\href{https://arxiv.org/abs/1711.05327}{{\ttfamily
  1711.05327}}].

\bibitem{Eden:2016xvg}
B.~Eden and A.~Sfondrini, \emph{{Tessellating cushions: four-point functions in
  $\mathcal{N} $ = 4 SYM}},
  \href{https://doi.org/10.1007/JHEP10(2017)098}{\emph{JHEP} {\bfseries 10}
  (2017) 098} [\href{https://arxiv.org/abs/1611.05436}{{\ttfamily
  1611.05436}}].

\bibitem{Coronado:2018ypq}
F.~Coronado, \emph{{Perturbative four-point functions in planar $ \mathcal{N}=4
  $ SYM from hexagonalization}},
  \href{https://doi.org/10.1007/JHEP01(2019)056}{\emph{JHEP} {\bfseries 01}
  (2019) 056} [\href{https://arxiv.org/abs/1811.00467}{{\ttfamily
  1811.00467}}].

\bibitem{Belitsky:2019fan}
A.~Belitsky and G.~Korchemsky, \emph{{Exact null octagon}},
  \href{https://doi.org/10.1007/JHEP05(2020)070}{\emph{JHEP} {\bfseries 05}
  (2020) 070} [\href{https://arxiv.org/abs/1907.13131}{{\ttfamily
  1907.13131}}].

\bibitem{Belitsky:2020qrm}
A.~Belitsky and G.~Korchemsky, \emph{{Octagon at finite coupling}},
  \href{https://doi.org/10.1007/JHEP07(2020)219}{\emph{JHEP} {\bfseries 07}
  (2020) 219} [\href{https://arxiv.org/abs/2003.01121}{{\ttfamily
  2003.01121}}].

\bibitem{Kostov:2019stn}
I.~Kostov, V.~B. Petkova and D.~Serban, \emph{{Determinant Formula for the
  Octagon Form Factor in $N$=4 Supersymmetric Yang-Mills Theory}},
  \href{https://doi.org/10.1103/PhysRevLett.122.231601}{\emph{Phys. Rev. Lett.}
  {\bfseries 122} (2019) 231601}
  [\href{https://arxiv.org/abs/1903.05038}{{\ttfamily 1903.05038}}].

\bibitem{Kostov:2019auq}
I.~Kostov, V.~B. Petkova and D.~Serban, \emph{{The Octagon as a Determinant}},
  \href{https://doi.org/10.1007/JHEP11(2019)178}{\emph{JHEP} {\bfseries 11}
  (2019) 178} [\href{https://arxiv.org/abs/1905.11467}{{\ttfamily
  1905.11467}}].

\bibitem{Fleury:2020ykw}
T.~Fleury and V.~Goncalves, \emph{{Decagon at Two Loops}},
  \href{https://doi.org/10.1007/JHEP07(2020)030}{\emph{JHEP} {\bfseries 07}
  (2020) 030} [\href{https://arxiv.org/abs/2004.10867}{{\ttfamily
  2004.10867}}].

\bibitem{deLeeuw:2019qvz}
M.~De~Leeuw, B.~Eden, D.~Le~Plat, T.~Meier and A.~Sfondrini,
  \emph{{Multi-particle finite-volume effects for hexagon tessellations}},
  \href{https://doi.org/10.1007/JHEP09(2020)039}{\emph{JHEP} {\bfseries 09}
  (2020) 039} [\href{https://arxiv.org/abs/1912.12231}{{\ttfamily
  1912.12231}}].

\bibitem{Bargheer:2019kxb}
T.~Bargheer, F.~Coronado and P.~Vieira, \emph{{Octagons I: Combinatorics and
  Non-Planar Resummations}},
  \href{https://doi.org/10.1007/JHEP08(2019)162}{\emph{JHEP} {\bfseries 08}
  (2019) 162} [\href{https://arxiv.org/abs/1904.00965}{{\ttfamily
  1904.00965}}].

\bibitem{Bargheer:2019exp}
T.~Bargheer, F.~Coronado and P.~Vieira, \emph{{Octagons II: Strong Coupling}},
  \href{https://arxiv.org/abs/1909.04077}{{\ttfamily 1909.04077}}.

\bibitem{McLoughlin:2020siu}
T.~McLoughlin, R.~Pereira and A.~Spiering, \emph{{One-loop non-planar anomalous
  dimensions in super Yang-Mills theory}},
  \href{https://doi.org/10.1007/JHEP10(2020)124}{\emph{JHEP} {\bfseries 10}
  (2020) 124} [\href{https://arxiv.org/abs/2005.14254}{{\ttfamily
  2005.14254}}].

\bibitem{Bargheer:2017nne}
T.~Bargheer, J.~Caetano, T.~Fleury, S.~Komatsu and P.~Vieira, \emph{{Handling
  Handles: Nonplanar Integrability in $\mathcal{N}=4$ Supersymmetric Yang-Mills
  Theory}}, \href{https://doi.org/10.1103/PhysRevLett.121.231602}{\emph{Phys.
  Rev. Lett.} {\bfseries 121} (2018) 231602}
  [\href{https://arxiv.org/abs/1711.05326}{{\ttfamily 1711.05326}}].

\bibitem{Eden:2017ozn}
B.~Eden, Y.~Jiang, D.~le~Plat and A.~Sfondrini, \emph{{Colour-dressed hexagon
  tessellations for correlation functions and non-planar corrections}},
  \href{https://doi.org/10.1007/JHEP02(2018)170}{\emph{JHEP} {\bfseries 02}
  (2018) 170} [\href{https://arxiv.org/abs/1710.10212}{{\ttfamily
  1710.10212}}].

\bibitem{Ben-Israel:2018ckc}
R.~Ben-Israel, A.~G. Tumanov and A.~Sever, \emph{{Scattering amplitudes
  \textemdash{} Wilson loops duality for the first non-planar correction}},
  \href{https://doi.org/10.1007/JHEP08(2018)122}{\emph{JHEP} {\bfseries 08}
  (2018) 122} [\href{https://arxiv.org/abs/1802.09395}{{\ttfamily
  1802.09395}}].

\bibitem{Bargheer:2018jvq}
T.~Bargheer, J.~Caetano, T.~Fleury, S.~Komatsu and P.~Vieira, \emph{{Handling
  handles. Part II. Stratification and data analysis}},
  \href{https://doi.org/10.1007/JHEP11(2018)095}{\emph{JHEP} {\bfseries 11}
  (2018) 095} [\href{https://arxiv.org/abs/1809.09145}{{\ttfamily
  1809.09145}}].

\bibitem{Jiang:2019xdz}
Y.~Jiang, S.~Komatsu and E.~Vescovi, \emph{{Structure constants in $
  \mathcal{N} $ = 4 SYM at finite coupling as worldsheet g-function}},
  \href{https://doi.org/10.1007/JHEP07(2020)037}{\emph{JHEP} {\bfseries 07}
  (2020) 037} [\href{https://arxiv.org/abs/1906.07733}{{\ttfamily
  1906.07733}}].

\bibitem{Jiang:2019zig}
Y.~Jiang, S.~Komatsu and E.~Vescovi, \emph{{Exact Three-Point Functions of
  Determinant Operators in Planar $N=4$ Supersymmetric Yang-Mills Theory}},
  \href{https://doi.org/10.1103/PhysRevLett.123.191601}{\emph{Phys. Rev. Lett.}
  {\bfseries 123} (2019) 191601}
  [\href{https://arxiv.org/abs/1907.11242}{{\ttfamily 1907.11242}}].

\bibitem{secondpaper}
P.~Yang, Y.~Jiang, S.~Komatsu and J.-B. Wu, \emph{{D-branes and Orbit
  Average}},  \href{https://arxiv.org/abs/2103.16580}{{\ttfamily 2103.16580}}.

\bibitem{thirdpaper}
Y.~Jiang, S.~Komatsu, J.-B. Wu and P.~Yang, \emph{{Structure Constants in ABJM
  and Integrable Bootstrap}},
  \href{https://arxiv.org/abs/21xx.xxxxx}{{\ttfamily 21xx.xxxxx}}.

\bibitem{Aharony:2008ug}
O.~Aharony, O.~Bergman, D.~L. Jafferis and J.~Maldacena, \emph{{N=6
  superconformal Chern-Simons-matter theories, M2-branes and their gravity
  duals}}, \href{https://doi.org/10.1088/1126-6708/2008/10/091}{\emph{JHEP}
  {\bfseries 10} (2008) 091} [\href{https://arxiv.org/abs/0806.1218}{{\ttfamily
  0806.1218}}].

\bibitem{Minahan:2008hf}
J.~Minahan and K.~Zarembo, \emph{{The Bethe ansatz for superconformal
  Chern-Simons}},
  \href{https://doi.org/10.1088/1126-6708/2008/09/040}{\emph{JHEP} {\bfseries
  09} (2008) 040} [\href{https://arxiv.org/abs/0806.3951}{{\ttfamily
  0806.3951}}].

\bibitem{Gaiotto:2008cg}
D.~Gaiotto, S.~Giombi and X.~Yin, \emph{{Spin Chains in N=6 Superconformal
  Chern-Simons-Matter Theory}},
  \href{https://doi.org/10.1088/1126-6708/2009/04/066}{\emph{JHEP} {\bfseries
  04} (2009) 066} [\href{https://arxiv.org/abs/0806.4589}{{\ttfamily
  0806.4589}}].

\bibitem{Nishioka:2008gz}
T.~Nishioka and T.~Takayanagi, \emph{{On Type IIA Penrose Limit and N=6
  Chern-Simons Theories}},
  \href{https://doi.org/10.1088/1126-6708/2008/08/001}{\emph{JHEP} {\bfseries
  08} (2008) 001} [\href{https://arxiv.org/abs/0806.3391}{{\ttfamily
  0806.3391}}].

\bibitem{Bak:2008cp}
D.~Bak and S.-J. Rey, \emph{{Integrable Spin Chain in Superconformal
  Chern-Simons Theory}},
  \href{https://doi.org/10.1088/1126-6708/2008/10/053}{\emph{JHEP} {\bfseries
  10} (2008) 053} [\href{https://arxiv.org/abs/0807.2063}{{\ttfamily
  0807.2063}}].

\bibitem{Cavaglia:2014exa}
A.~Cavagli\`a, D.~Fioravanti, N.~Gromov and R.~Tateo, \emph{{Quantum Spectral
  Curve of the $\mathcal N=$ 6 Supersymmetric Chern-Simons Theory}},
  \href{https://doi.org/10.1103/PhysRevLett.113.021601}{\emph{Phys. Rev. Lett.}
  {\bfseries 113} (2014) 021601}
  [\href{https://arxiv.org/abs/1403.1859}{{\ttfamily 1403.1859}}].

\bibitem{Bombardelli:2017vhk}
D.~Bombardelli, A.~Cavagli\`a, D.~Fioravanti, N.~Gromov and R.~Tateo,
  \emph{{The full Quantum Spectral Curve for $AdS_4/CFT_3$}},
  \href{https://doi.org/10.1007/JHEP09(2017)140}{\emph{JHEP} {\bfseries 09}
  (2017) 140} [\href{https://arxiv.org/abs/1701.00473}{{\ttfamily
  1701.00473}}].

\bibitem{Brockmann}
M.~{Brockmann}, J.~{De Nardis}, B.~{Wouters} and J.~S. {Caux}, \emph{{A
  Gaudin-like determinant for overlaps of N{\'e}el and XXZ Bethe states}},
  \href{https://doi.org/10.1088/1751-8113/47/14/145003}{\emph{Journal of
  Physics A Mathematical General} {\bfseries 47} (2014) 145003}
  [\href{https://arxiv.org/abs/1401.2877}{{\ttfamily 1401.2877}}].

\bibitem{Pozsgay:2014}
B.~{Pozsgay}, \emph{{Overlaps between eigenstates of the XXZ spin-1/2 chain and
  a class of simple product states}},
  \href{https://doi.org/10.1088/1742-5468/2014/06/P06011}{\emph{Journal of
  Statistical Mechanics: Theory and Experiment} {\bfseries 2014} (2014) 06011}
  [\href{https://arxiv.org/abs/1309.4593}{{\ttfamily 1309.4593}}].

\bibitem{Foda:2015nfk}
O.~Foda and K.~Zarembo, \emph{{Overlaps of partial N\'eel states and Bethe
  states}}, \href{https://doi.org/10.1088/1742-5468/2016/02/023107}{\emph{J.
  Stat. Mech.} {\bfseries 1602} (2016) 023107}
  [\href{https://arxiv.org/abs/1512.02533}{{\ttfamily 1512.02533}}].

\bibitem{Piroli:2017sei}
L.~Piroli, B.~Pozsgay and E.~Vernier, \emph{{What is an integrable quench?}},
  \href{https://doi.org/10.1016/j.nuclphysb.2017.10.012}{\emph{Nucl. Phys. B}
  {\bfseries 925} (2017) 362}
  [\href{https://arxiv.org/abs/1709.04796}{{\ttfamily 1709.04796}}].

\bibitem{Pozsgay:2018}
B.~{Pozsgay}, \emph{{Overlaps with arbitrary two-site states in the XXZ spin
  chain}}, \href{https://doi.org/10.1088/1742-5468/aabbe1}{\emph{Journal of
  Statistical Mechanics: Theory and Experiment} {\bfseries 5} (2018) 053103}
  [\href{https://arxiv.org/abs/1801.03838}{{\ttfamily 1801.03838}}].

\bibitem{deLeeuw:2019ebw}
M.~De~Leeuw, T.~Gombor, C.~Kristjansen, G.~Linardopoulos and B.~Pozsgay,
  \emph{{Spin Chain Overlaps and the Twisted Yangian}},
  \href{https://doi.org/10.1007/JHEP01(2020)176}{\emph{JHEP} {\bfseries 01}
  (2020) 176} [\href{https://arxiv.org/abs/1912.09338}{{\ttfamily
  1912.09338}}].

\bibitem{Pozsgay:2019}
B.~{Pozsgay}, L.~{Piroli} and E.~{Vernier}, \emph{{Integrable Matrix Product
  States from boundary integrability}},
  \href{https://doi.org/10.21468/SciPostPhys.6.5.062}{\emph{SciPost Physics}
  {\bfseries 6} (2019) 062} [\href{https://arxiv.org/abs/1812.11094}{{\ttfamily
  1812.11094}}].

\bibitem{Jiang:2020sdw}
Y.~Jiang and B.~Pozsgay, \emph{{On exact overlaps in integrable spin chains}},
  \href{https://doi.org/10.1007/JHEP06(2020)022}{\emph{JHEP} {\bfseries 06}
  (2020) 022} [\href{https://arxiv.org/abs/2002.12065}{{\ttfamily
  2002.12065}}].

\bibitem{Chen:2020xel}
H.-H. Chen, \emph{{Exact overlaps in the Lieb-Liniger model from coordinate
  Bethe ansatz}},
  \href{https://doi.org/10.1016/j.physletb.2020.135631}{\emph{Phys. Lett. B}
  {\bfseries 808} (2020) 135631}
  [\href{https://arxiv.org/abs/2003.02711}{{\ttfamily 2003.02711}}].

\bibitem{deLeeuw:2015hxa}
M.~de~Leeuw, C.~Kristjansen and K.~Zarembo, \emph{{One-point Functions in
  Defect CFT and Integrability}},
  \href{https://doi.org/10.1007/JHEP08(2015)098}{\emph{JHEP} {\bfseries 08}
  (2015) 098} [\href{https://arxiv.org/abs/1506.06958}{{\ttfamily
  1506.06958}}].

\bibitem{Komatsu:2020sup}
S.~Komatsu and Y.~Wang, \emph{{Non-perturbative defect one-point functions in
  planar $\mathcal{N}=4$ super-Yang-Mills}},
  \href{https://doi.org/10.1016/j.nuclphysb.2020.115120}{\emph{Nucl. Phys. B}
  {\bfseries 958} (2020) 115120}
  [\href{https://arxiv.org/abs/2004.09514}{{\ttfamily 2004.09514}}].

\bibitem{Buhl-Mortensen:2015gfd}
I.~Buhl-Mortensen, M.~de~Leeuw, C.~Kristjansen and K.~Zarembo, \emph{{One-point
  Functions in AdS/dCFT from Matrix Product States}},
  \href{https://doi.org/10.1007/JHEP02(2016)052}{\emph{JHEP} {\bfseries 02}
  (2016) 052} [\href{https://arxiv.org/abs/1512.02532}{{\ttfamily
  1512.02532}}].

\bibitem{deLeeuw:2016umh}
M.~de~Leeuw, C.~Kristjansen and S.~Mori, \emph{{AdS/dCFT one-point functions of
  the SU(3) sector}},
  \href{https://doi.org/10.1016/j.physletb.2016.10.044}{\emph{Phys. Lett. B}
  {\bfseries 763} (2016) 197}
  [\href{https://arxiv.org/abs/1607.03123}{{\ttfamily 1607.03123}}].

\bibitem{Buhl-Mortensen:2017ind}
I.~Buhl-Mortensen, M.~de~Leeuw, A.~C. Ipsen, C.~Kristjansen and M.~Wilhelm,
  \emph{{Asymptotic One-Point Functions in Gauge-String Duality with Defects}},
  \href{https://doi.org/10.1103/PhysRevLett.119.261604}{\emph{Phys. Rev. Lett.}
  {\bfseries 119} (2017) 261604}
  [\href{https://arxiv.org/abs/1704.07386}{{\ttfamily 1704.07386}}].

\bibitem{deLeeuw:2018mkd}
M.~De~Leeuw, C.~Kristjansen and G.~Linardopoulos, \emph{{Scalar one-point
  functions and matrix product states of AdS/dCFT}},
  \href{https://doi.org/10.1016/j.physletb.2018.03.083}{\emph{Phys. Lett. B}
  {\bfseries 781} (2018) 238}
  [\href{https://arxiv.org/abs/1802.01598}{{\ttfamily 1802.01598}}].

\bibitem{Gombor:2020kgu}
T.~Gombor and Z.~Bajnok, \emph{{Boundary states, overlaps, nesting and
  bootstrapping AdS/dCFT}},
  \href{https://doi.org/10.1007/JHEP10(2020)123}{\emph{JHEP} {\bfseries 10}
  (2020) 123} [\href{https://arxiv.org/abs/2004.11329}{{\ttfamily
  2004.11329}}].

\bibitem{Gombor:2020auk}
T.~Gombor and Z.~Bajnok, \emph{{Boundary state bootstrap and asymptotic
  overlaps in AdS/dCFT}},  \href{https://arxiv.org/abs/2006.16151}{{\ttfamily
  2006.16151}}.

\bibitem{Kristjansen:2020mhn}
C.~Kristjansen, D.~M\"uller and K.~Zarembo, \emph{{Integrable boundary states
  in D3-D5 dCFT: beyond scalars}},
  \href{https://doi.org/10.1007/JHEP08(2020)103}{\emph{JHEP} {\bfseries 08}
  (2020) 103} [\href{https://arxiv.org/abs/2005.01392}{{\ttfamily
  2005.01392}}].

\bibitem{Kristjansen:2020vbe}
C.~Kristjansen, D.~M\"uller and K.~Zarembo, \emph{{Overlaps and Fermionic
  Dualities for Integrable Super Spin Chains}},
  \href{https://arxiv.org/abs/2011.12192}{{\ttfamily 2011.12192}}.

\bibitem{Chen:2019gsb}
G.~Chen, R.~de~Mello~Koch, M.~Kim and H.~J. Van~Zyl, \emph{{Absorption of
  closed strings by giant gravitons}},
  \href{https://doi.org/10.1007/JHEP10(2019)133}{\emph{JHEP} {\bfseries 10}
  (2019) 133} [\href{https://arxiv.org/abs/1908.03553}{{\ttfamily
  1908.03553}}].

\bibitem{Berenstein:2006qk}
D.~Berenstein, D.~H. Correa and S.~E. Vazquez, \emph{{A Study of open strings
  ending on giant gravitons, spin chains and integrability}},
  \href{https://doi.org/10.1088/1126-6708/2006/09/065}{\emph{JHEP} {\bfseries
  09} (2006) 065} [\href{https://arxiv.org/abs/hep-th/0604123}{{\ttfamily
  hep-th/0604123}}].

\bibitem{Ciavarella:2010tp}
A.~Ciavarella, \emph{{Giant magnons and non-maximal giant gravitons}},
  \href{https://doi.org/10.1007/JHEP01(2011)040}{\emph{JHEP} {\bfseries 01}
  (2011) 040} [\href{https://arxiv.org/abs/1011.1440}{{\ttfamily 1011.1440}}].

\bibitem{Linardopoulos:2021rfq}
G.~Linardopoulos and K.~Zarembo, \emph{{String integrability of defect CFT and
  dynamical reflection matrices}},
  \href{https://arxiv.org/abs/2102.12381}{{\ttfamily 2102.12381}}.

\bibitem{Dekel:2011ja}
A.~Dekel and Y.~Oz, \emph{{Integrability of Green-Schwarz Sigma Models with
  Boundaries}}, \href{https://doi.org/10.1007/JHEP08(2011)004}{\emph{JHEP}
  {\bfseries 08} (2011) 004} [\href{https://arxiv.org/abs/1106.3446}{{\ttfamily
  1106.3446}}].

\bibitem{Young:2014lka}
D.~Young, \emph{{ABJ(M) Chiral Primary Three-Point Function at Two-loops}},
  \href{https://doi.org/10.1007/JHEP07(2014)120}{\emph{JHEP} {\bfseries 07}
  (2014) 120} [\href{https://arxiv.org/abs/1404.1117}{{\ttfamily 1404.1117}}].

\bibitem{Young:2014sia}
D.~Young, \emph{{An Extremal Chiral Primary Three-Point Function at Two-loops
  in ABJ(M)}}, \href{https://doi.org/10.1007/JHEP12(2014)141}{\emph{JHEP}
  {\bfseries 12} (2014) 141} [\href{https://arxiv.org/abs/1411.0626}{{\ttfamily
  1411.0626}}].

\bibitem{Bianchi:2020cfn}
M.~S. Bianchi, \emph{{On three-point functions in ABJM and the latitude Wilson
  loop}}, \href{https://doi.org/10.1007/JHEP10(2020)075}{\emph{JHEP} {\bfseries
  10} (2020) 075} [\href{https://arxiv.org/abs/2005.09522}{{\ttfamily
  2005.09522}}].

\bibitem{Baggio:2012rr}
M.~Baggio, J.~de~Boer and K.~Papadodimas, \emph{{A non-renormalization theorem
  for chiral primary 3-point functions}},
  \href{https://doi.org/10.1007/JHEP07(2012)137}{\emph{JHEP} {\bfseries 07}
  (2012) 137} [\href{https://arxiv.org/abs/1203.1036}{{\ttfamily 1203.1036}}].

\bibitem{Dedushenko:2016jxl}
M.~Dedushenko, S.~S. Pufu and R.~Yacoby, \emph{{A one-dimensional theory for
  Higgs branch operators}},
  \href{https://doi.org/10.1007/JHEP03(2018)138}{\emph{JHEP} {\bfseries 03}
  (2018) 138} [\href{https://arxiv.org/abs/1610.00740}{{\ttfamily
  1610.00740}}].

\bibitem{Dedushenko:2017avn}
M.~Dedushenko, Y.~Fan, S.~S. Pufu and R.~Yacoby, \emph{{Coulomb Branch
  Operators and Mirror Symmetry in Three Dimensions}},
  \href{https://doi.org/10.1007/JHEP04(2018)037}{\emph{JHEP} {\bfseries 04}
  (2018) 037} [\href{https://arxiv.org/abs/1712.09384}{{\ttfamily
  1712.09384}}].

\bibitem{Dedushenko:2018icp}
M.~Dedushenko, Y.~Fan, S.~S. Pufu and R.~Yacoby, \emph{{Coulomb Branch
  Quantization and Abelianized Monopole Bubbling}},
  \href{https://doi.org/10.1007/JHEP10(2019)179}{\emph{JHEP} {\bfseries 10}
  (2019) 179} [\href{https://arxiv.org/abs/1812.08788}{{\ttfamily
  1812.08788}}].

\bibitem{Mezei:2017kmw}
M.~Mezei, S.~S. Pufu and Y.~Wang, \emph{{A 2d/1d Holographic Duality}},
  \href{https://arxiv.org/abs/1703.08749}{{\ttfamily 1703.08749}}.

\bibitem{Gaiotto:2020vqj}
D.~Gaiotto and J.~Abajian, \emph{{Twisted M2 brane holography and sphere
  correlation functions}},  \href{https://arxiv.org/abs/2004.13810}{{\ttfamily
  2004.13810}}.

\bibitem{Chester:2018aca}
S.~M. Chester, S.~S. Pufu and X.~Yin, \emph{{The M-Theory S-Matrix From ABJM:
  Beyond 11D Supergravity}},
  \href{https://doi.org/10.1007/JHEP08(2018)115}{\emph{JHEP} {\bfseries 08}
  (2018) 115} [\href{https://arxiv.org/abs/1804.00949}{{\ttfamily
  1804.00949}}].

\bibitem{Binder:2019mpb}
D.~J. Binder, S.~M. Chester and S.~S. Pufu, \emph{{AdS$_{4}$/CFT$_{3}$ from
  weak to strong string coupling}},
  \href{https://doi.org/10.1007/JHEP01(2020)034}{\emph{JHEP} {\bfseries 01}
  (2020) 034} [\href{https://arxiv.org/abs/1906.07195}{{\ttfamily
  1906.07195}}].

\bibitem{Chester:2020jay}
S.~M. Chester, R.~R. Kalloor and A.~Sharon, \emph{{3d $ \mathcal{N} $ = 4 OPE
  coefficients from Fermi gas}},
  \href{https://doi.org/10.1007/JHEP07(2020)041}{\emph{JHEP} {\bfseries 07}
  (2020) 041} [\href{https://arxiv.org/abs/2004.13603}{{\ttfamily
  2004.13603}}].

\bibitem{KW:2021}
S.~Komatsu and Y.~Wang, \emph{{to appear}}, .

\bibitem{Chen:2019kgc}
G.~Chen, R.~De~Mello~Koch, M.~Kim and H.~J. Van~Zyl, \emph{{Structure constants
  of heavy operators in ABJM and ABJ theory}},
  \href{https://doi.org/10.1103/PhysRevD.100.086019}{\emph{Phys. Rev. D}
  {\bfseries 100} (2019) 086019}
  [\href{https://arxiv.org/abs/1909.03215}{{\ttfamily 1909.03215}}].

\bibitem{Benna:2008zy}
M.~Benna, I.~Klebanov, T.~Klose and M.~Smedback, \emph{{Superconformal
  Chern-Simons Theories and AdS(4)/CFT(3) Correspondence}},
  \href{https://doi.org/10.1088/1126-6708/2008/09/072}{\emph{JHEP} {\bfseries
  09} (2008) 072} [\href{https://arxiv.org/abs/0806.1519}{{\ttfamily
  0806.1519}}].

\bibitem{Klose:2010ki}
T.~Klose, \emph{{Review of AdS/CFT Integrability, Chapter IV.3: N=6
  Chern-Simons and Strings on AdS4xCP3}},
  \href{https://doi.org/10.1007/s11005-011-0520-y}{\emph{Lett. Math. Phys.}
  {\bfseries 99} (2012) 401} [\href{https://arxiv.org/abs/1012.3999}{{\ttfamily
  1012.3999}}].

\bibitem{Escobedo:2010xs}
J.~Escobedo, N.~Gromov, A.~Sever and P.~Vieira, \emph{{Tailoring Three-Point
  Functions and Integrability}},
  \href{https://doi.org/10.1007/JHEP09(2011)028}{\emph{JHEP} {\bfseries 09}
  (2011) 028} [\href{https://arxiv.org/abs/1012.2475}{{\ttfamily 1012.2475}}].

\bibitem{Dey:2011ea}
T.~K. Dey, \emph{{Exact Large $R$-charge Correlators in ABJM Theory}},
  \href{https://doi.org/10.1007/JHEP08(2011)066}{\emph{JHEP} {\bfseries 08}
  (2011) 066} [\href{https://arxiv.org/abs/1105.0218}{{\ttfamily 1105.0218}}].

\bibitem{Chakrabortty:2011fd}
S.~Chakrabortty and T.~K. Dey, \emph{{Correlators of Giant Gravitons from dual
  ABJ(M) Theory}}, \href{https://doi.org/10.1007/JHEP03(2012)062}{\emph{JHEP}
  {\bfseries 03} (2012) 062} [\href{https://arxiv.org/abs/1112.6299}{{\ttfamily
  1112.6299}}].

\bibitem{Giovannoni:2011pn}
D.~Giovannoni, J.~Murugan and A.~Prinsloo, \emph{{The Giant graviton on
  $AdS_{4} x CP^{3}$ - another step towards the emergence of geometry}},
  \href{https://doi.org/10.1007/JHEP12(2011)003}{\emph{JHEP} {\bfseries 12}
  (2011) 003} [\href{https://arxiv.org/abs/1108.3084}{{\ttfamily 1108.3084}}].

\bibitem{Hirano:2012vz}
S.~Hirano, C.~Kristjansen and D.~Young, \emph{{Giant Gravitons on $AdS_4 \times
  \mathbb{C}P^3$ and their Holographic Three-point Functions}},
  \href{https://doi.org/10.1007/JHEP07(2012)006}{\emph{JHEP} {\bfseries 07}
  (2012) 006} [\href{https://arxiv.org/abs/1205.1959}{{\ttfamily 1205.1959}}].

\bibitem{Corley:2001zk}
S.~Corley, A.~Jevicki and S.~Ramgoolam, \emph{{Exact correlators of giant
  gravitons from dual N=4 SYM theory}},
  \href{https://doi.org/10.4310/ATMP.2001.v5.n4.a6}{\emph{Adv. Theor. Math.
  Phys.} {\bfseries 5} (2002) 809}
  [\href{https://arxiv.org/abs/hep-th/0111222}{{\ttfamily hep-th/0111222}}].

\bibitem{Caputa:2012dg}
P.~Caputa and B.~A.~E. Mohammed, \emph{{From Schurs to Giants in ABJ(M)}},
  \href{https://doi.org/10.1007/JHEP01(2013)055}{\emph{JHEP} {\bfseries 01}
  (2013) 055} [\href{https://arxiv.org/abs/1210.7705}{{\ttfamily 1210.7705}}].

\bibitem{Berenstein:2008dc}
D.~Berenstein and D.~Trancanelli, \emph{{Three-dimensional N=6 SCFT's and their
  membrane dynamics}},
  \href{https://doi.org/10.1103/PhysRevD.78.106009}{\emph{Phys. Rev. D}
  {\bfseries 78} (2008) 106009}
  [\href{https://arxiv.org/abs/0808.2503}{{\ttfamily 0808.2503}}].

\bibitem{Nishioka:2008ib}
T.~Nishioka and T.~Takayanagi, \emph{{Fuzzy Ring from M2-brane Giant Torus}},
  \href{https://doi.org/10.1088/1126-6708/2008/10/082}{\emph{JHEP} {\bfseries
  10} (2008) 082} [\href{https://arxiv.org/abs/0808.2691}{{\ttfamily
  0808.2691}}].

\bibitem{Hamilton:2009iv}
A.~Hamilton, J.~Murugan, A.~Prinsloo and M.~Strydom, \emph{{A Note on dual
  giant gravitons in AdS(4) x CP**3}},
  \href{https://doi.org/10.1088/1126-6708/2009/04/132}{\emph{JHEP} {\bfseries
  04} (2009) 132} [\href{https://arxiv.org/abs/0901.0009}{{\ttfamily
  0901.0009}}].

\bibitem{Murugan:2011zd}
J.~Murugan and A.~Prinsloo, \emph{{ABJM Dibaryon Spectroscopy}},
  \href{https://doi.org/10.1007/JHEP05(2011)129}{\emph{JHEP} {\bfseries 05}
  (2011) 129} [\href{https://arxiv.org/abs/1103.1163}{{\ttfamily 1103.1163}}].

\bibitem{Gutierrez:2010bb}
N.~Gutierrez, Y.~Lozano and D.~Rodriguez-Gomez, \emph{{Charged particle-like
  branes in ABJM}}, \href{https://doi.org/10.1007/JHEP09(2010)101}{\emph{JHEP}
  {\bfseries 09} (2010) 101} [\href{https://arxiv.org/abs/1004.2826}{{\ttfamily
  1004.2826}}].

\bibitem{Lozano:2011dd}
Y.~Lozano, M.~Picos, K.~Sfetsos and K.~Siampos, \emph{{ABJM Baryon Stability
  and Myers effect}},
  \href{https://doi.org/10.1007/JHEP07(2011)032}{\emph{JHEP} {\bfseries 07}
  (2011) 032} [\href{https://arxiv.org/abs/1105.0939}{{\ttfamily 1105.0939}}].

\bibitem{Herrero:2011bk}
M.~Herrero, Y.~Lozano and M.~Picos, \emph{{Dielectric 5-Branes and Giant
  Gravitons in ABJM}},
  \href{https://doi.org/10.1007/JHEP08(2011)132}{\emph{JHEP} {\bfseries 08}
  (2011) 132} [\href{https://arxiv.org/abs/1107.5475}{{\ttfamily 1107.5475}}].

\bibitem{Lozano:2013ota}
Y.~Lozano, J.~Murugan and A.~Prinsloo, \emph{{A giant graviton genealogy}},
  \href{https://doi.org/10.1007/JHEP08(2013)109}{\emph{JHEP} {\bfseries 08}
  (2013) 109} [\href{https://arxiv.org/abs/1305.6932}{{\ttfamily 1305.6932}}].

\bibitem{Liendo:2015cgi}
P.~Liendo, C.~Meneghelli and V.~Mitev, \emph{{On Correlation Functions of BPS
  Operators in 3d ${\mathcal{N}}$ = 6 Superconformal Theories}},
  \href{https://doi.org/10.1007/s00220-016-2715-7}{\emph{Commun. Math. Phys.}
  {\bfseries 350} (2017) 387}
  [\href{https://arxiv.org/abs/1512.06072}{{\ttfamily 1512.06072}}].

\bibitem{Bastianelli:1999en}
F.~Bastianelli and R.~Zucchini, \emph{{Three point functions of chiral primary
  operators in d = 3, N=8 and d = 6, N=(2,0) SCFT at large N}},
  \href{https://doi.org/10.1016/S0370-2693(99)01179-X}{\emph{Phys. Lett. B}
  {\bfseries 467} (1999) 61}
  [\href{https://arxiv.org/abs/hep-th/9907047}{{\ttfamily hep-th/9907047}}].

\bibitem{Gromov:2014eha}
N.~Gromov and G.~Sizov, \emph{{Exact Slope and Interpolating Functions in N=6
  Supersymmetric Chern-Simons Theory}},
  \href{https://doi.org/10.1103/PhysRevLett.113.121601}{\emph{Phys. Rev. Lett.}
  {\bfseries 113} (2014) 121601}
  [\href{https://arxiv.org/abs/1403.1894}{{\ttfamily 1403.1894}}].

\bibitem{Drukker:2009sf}
N.~Drukker and J.~Plefka, \emph{{Superprotected n-point correlation functions
  of local operators in N=4 super Yang-Mills}},
  \href{https://doi.org/10.1088/1126-6708/2009/04/052}{\emph{JHEP} {\bfseries
  04} (2009) 052} [\href{https://arxiv.org/abs/0901.3653}{{\ttfamily
  0901.3653}}].

\bibitem{Gorini:2020new}
N.~Gorini, L.~Griguolo, L.~Guerrini, S.~Penati, D.~Seminara and P.~Soresina,
  \emph{{The topological line of ABJ(M) theory}},
  \href{https://arxiv.org/abs/2012.11613}{{\ttfamily 2012.11613}}.

\bibitem{OSPS:2016}
O.~Ohlsson~Sax, R.~Pereira and A.~Sfondrini, \emph{{unpublished}}, .

\bibitem{Pereira:2017unx}
R.~Pereira, \emph{{Correlation Functions in Integrable Theories: From weak to
  strong coupling}}, Ph.D. thesis, Uppsala U., 2017.

\bibitem{Kazama:2014sxa}
Y.~Kazama, S.~Komatsu and T.~Nishimura, \emph{{Novel construction and the
  monodromy relation for three-point functions at weak coupling}},
  \href{https://doi.org/10.1007/JHEP01(2015)095}{\emph{JHEP} {\bfseries 01}
  (2015) 095} [\href{https://arxiv.org/abs/1410.8533}{{\ttfamily 1410.8533}}].

\bibitem{Vescovi:2021fjf}
E.~Vescovi, \emph{{On the four-point function of determinant operators in
  $\mathcal{N}=4$ SYM}},  \href{https://arxiv.org/abs/2101.05117}{{\ttfamily
  2101.05117}}.

\bibitem{Ghoshal:1993tm}
S.~Ghoshal and A.~B. Zamolodchikov, \emph{{Boundary S matrix and boundary state
  in two-dimensional integrable quantum field theory}},
  \href{https://doi.org/10.1142/S0217751X94001552}{\emph{Int. J. Mod. Phys. A}
  {\bfseries 9} (1994) 3841}
  [\href{https://arxiv.org/abs/hep-th/9306002}{{\ttfamily hep-th/9306002}}].

\bibitem{Gromov:2012uv}
N.~Gromov and P.~Vieira, \emph{{Tailoring Three-Point Functions and
  Integrability IV. Theta-morphism}},
  \href{https://doi.org/10.1007/JHEP04(2014)068}{\emph{JHEP} {\bfseries 04}
  (2014) 068} [\href{https://arxiv.org/abs/1205.5288}{{\ttfamily 1205.5288}}].

\bibitem{vladimirov1986proof}
A.~A. Vladimirov, \emph{Proof of the invariance of the bethe-ansatz solutions
  under complex conjugation}, {\emph{Theoretical and Mathematical Physics}
  {\bfseries 66} (1986) 102}.

\bibitem{Caetano:2020dyp}
J.~Caetano and S.~Komatsu, \emph{{Functional equations and separation of
  variables for exact $g$-function}},
  \href{https://doi.org/10.1007/JHEP09(2020)180}{\emph{JHEP} {\bfseries 09}
  (2020) 180} [\href{https://arxiv.org/abs/2004.05071}{{\ttfamily
  2004.05071}}].

\bibitem{Cavaglia:2021mft}
A.~Cavagli\`a, N.~Gromov and F.~Levkovich-Maslyuk, \emph{{Separation of
  Variables in AdS/CFT: Functional Approach for the Fishnet CFT}},
  \href{https://arxiv.org/abs/2103.15800}{{\ttfamily 2103.15800}}.

\bibitem{Gombor:2021uxz}
T.~Gombor and B.~Pozsgay, \emph{{On factorized overlaps: Algebraic Bethe
  Ansatz, twists, and Separation of Variables}},
  \href{https://arxiv.org/abs/2101.10354}{{\ttfamily 2101.10354}}.

\bibitem{Bissi:2012ff}
A.~Bissi, C.~Kristjansen, A.~Martirosyan and M.~Orselli, \emph{{On Three-point
  Functions in the $AdS_4/CFT_3$ Correspondence}},
  \href{https://doi.org/10.1007/JHEP01(2013)137}{\emph{JHEP} {\bfseries 01}
  (2013) 137} [\href{https://arxiv.org/abs/1211.1359}{{\ttfamily 1211.1359}}].

\bibitem{Koch:2016qkp}
R.~de~Mello~Koch and H.~J.~R. van Zyl, \emph{{Inelastic Magnon Scattering}},
  \href{https://doi.org/10.1016/j.physletb.2017.02.056}{\emph{Phys. Lett. B}
  {\bfseries 768} (2017) 187}
  [\href{https://arxiv.org/abs/1603.06414}{{\ttfamily 1603.06414}}].

\bibitem{deMelloKoch:2018tlb}
R.~de~Mello~Koch, M.~Kim and H.~J. Zyl, \emph{{Integrable Subsectors from
  Holography}}, \href{https://doi.org/10.1007/JHEP05(2018)198}{\emph{JHEP}
  {\bfseries 05} (2018) 198}
  [\href{https://arxiv.org/abs/1802.01367}{{\ttfamily 1802.01367}}].

\bibitem{Chen:2018sbp}
H.-H. Chen, H.~Ouyang and J.-B. Wu, \emph{{Open Spin Chains from Determinant
  Like Operators in ABJM Theory}},
  \href{https://doi.org/10.1103/PhysRevD.98.106012}{\emph{Phys. Rev. D}
  {\bfseries 98} (2018) 106012}
  [\href{https://arxiv.org/abs/1809.09941}{{\ttfamily 1809.09941}}].

\bibitem{Bai:2019soy}
N.~Bai, H.-H. Chen, H.~Ouyang and J.-B. Wu, \emph{{Two-Loop Integrability of
  ABJM Open Spin Chain from Giant Graviton}},
  \href{https://doi.org/10.1007/JHEP03(2019)193}{\emph{JHEP} {\bfseries 03}
  (2019) 193} [\href{https://arxiv.org/abs/1901.03949}{{\ttfamily
  1901.03949}}].

\bibitem{Chen:2019igg}
H.-H. Chen, \emph{{Asymptotic Bethe ansatz of ABJM open spin chain from giant
  graviton}}, \href{https://doi.org/10.1007/JHEP08(2019)109}{\emph{JHEP}
  {\bfseries 08} (2019) 109}
  [\href{https://arxiv.org/abs/1906.09886}{{\ttfamily 1906.09886}}].

\bibitem{Gromov:2007ky}
N.~Gromov and P.~Vieira, \emph{{Complete 1-loop test of AdS/CFT}},
  \href{https://doi.org/10.1088/1126-6708/2008/04/046}{\emph{JHEP} {\bfseries
  04} (2008) 046} [\href{https://arxiv.org/abs/0709.3487}{{\ttfamily
  0709.3487}}].

\bibitem{Marboe:2016yyn}
C.~Marboe and D.~Volin, \emph{{Fast analytic solver of rational Bethe
  equations}}, \href{https://doi.org/10.1088/1751-8121/aa6b88}{\emph{J. Phys.
  A} {\bfseries 50} (2017) 204002}
  [\href{https://arxiv.org/abs/1608.06504}{{\ttfamily 1608.06504}}].

\bibitem{Kulish:1983rd}
P.~Kulish and N.~Reshetikhin, \emph{{DIAGONALIZATION OF GL(N) INVARIANT
  TRANSFER MATRICES AND QUANTUM N WAVE SYSTEM (LEE MODEL)}},
  \href{https://doi.org/10.1088/0305-4470/16/16/001}{\emph{J. Phys. A}
  {\bfseries 16} (1983) L591}.

\bibitem{Slavnov:2019hdn}
N.~Slavnov, \emph{{Introduction to the nested algebraic Bethe ansatz}},  in
  \emph{{Les Houches Summer School}: {Structures in local quantum field
  theory}}, 11, 2019, \href{https://arxiv.org/abs/1911.12811}{{\ttfamily
  1911.12811}}, \href{https://doi.org/10.21468/SciPostPhysLectNotes.19}{DOI}.

\end{thebibliography}
\end{document}